\documentclass[aps,prd,10pt,superscriptaddress,tightline,nofootinbib,numerical,floatfix,twocolumn]{revtex4-2}
\usepackage{stackengine}
\usepackage{bm}
\usepackage{braket}
\usepackage{color}
\usepackage{lipsum}
\usepackage{dcolumn}
\usepackage{epsfig}
\usepackage[font={small}]{caption}
\usepackage{float}
\usepackage{subcaption}
\usepackage[T1]{fontenc}
\usepackage{graphicx}
\usepackage{lipsum}
\usepackage[colorlinks]{hyperref}
\AtBeginDocument{
	\hypersetup{
		urlcolor=black,
		citecolor=green,
		linkcolor=blue,
	}%
}
\usepackage{mathtools}
\usepackage{placeins}
\usepackage{xcolor}
\usepackage{amsmath}
\usepackage{cleveref}
\usepackage{amssymb}
\usepackage{amsbsy}
\usepackage{amsthm}
\usepackage{tabularx}
\usepackage{multirow}
\usepackage{romannum}
\usepackage{ulem}
\usepackage[toc,title,page]{appendix}

\AtBeginDocument{\pagenumbering{arabic}}

\begin{document}

	\title{Catapulting towards massive and large spatial quantum superposition  }
	
	\newcommand{\affone}{Centre for Quantum Computation and Communication Technology, School of Mathematics and Physics, University of Queensland, Brisbane, Queensland 4072, Australia}
	\newcommand{\afftwo}{Department of Physics and Astronomy, University College London, Gower Street, WC1E 6BT London, United Kingdom.}
	\newcommand{\affthree}{Van Swinderen Institute, University of Groningen, 9747 AG Groningen, The Netherlands.}

	\author{Run Zhou}
	\affiliation{\affthree}
	\author{Ryan J. Marshman}
	\affiliation{\affone}
	\author{Sougato Bose}
	\affiliation{\afftwo}

	\author{Anupam Mazumdar}
	\affiliation{\affthree}

	\date{\today}

	\begin{abstract}
		Large spatial quantum superposition of size ${\cal O}(1-10)~{\rm \mu \text{m}}$ for mass $m \sim 10^{-17}-10^{-14}$~kg is required to probe the foundations of quantum mechanics and testing classical and quantum nature of gravity via entanglement in a laboratory.  In this paper, we will show that it is possible to accelerate the two spin states of a macroscopic nano-crystal sourced by the inhomogeneous nonlinear magnetic field in the Stern-Gerlach type setup. We will assume that the electronic spin can be embedded at the centre of the nano-crystal, such as the nitrogen-vacancy (NV) centre of diamond. Our analysis will be generic to any dopant or any material. We will show that we can create a desired superposition size within $1-2$ seconds by catapulting the trajectories of the two spin states with a modest magnetic field gradient and then recombine the trajectories for a coherent interference. We will show the demanding nature of the precision required in the magnetic field  to recover $99\%$ spin coherence confidence level at the moment of interference.
	\end{abstract}
		\maketitle
		
	\section{Introduction}
	
Gravity is one of the weakest interactions of nature, and it is very special because of its universality. Unlike any other known interactions, it is not yet clear whether gravity respects the rules of quantum mechanics at a microscopic level~\cite{Kiefer}. No experimental proof validates whether the gravitational interaction is indeed quantum. The spacetime we have witnessed so far in gravitational experiments is extremely classical without any hint of quantum-ness~\cite{Will:2014kxa}.

The conventional wisdom is that the gravitational effects will become important only when we approach the Planckian length or the time scale, making it extremely challenging to test the quantum nature of gravity in a laboratory. Furthermore, tests from the cosmological perturbations in the cosmic microwave background radiation~\cite{venin}, or the positive detection of the primordial gravitational waves~\cite{Ashoorioon:2012kh} do not confirm the quantum nature of gravity or, as a matter of fact, any other
astrophysical tests~\cite{Addazi}. They all have many astrophysical uncertainties, making it extremely challenging to conclude the true nature of gravity. Also, the feeble nature of the gravitational interaction makes it extremely hard to detect the graviton as an individual quanta~\cite{dyson}.

Despite all these challenges, there is one hope for gravity. The gravitational interaction is a long-range interaction like in the case of quantum electrodynamics. Hence, it gives us a unique possibility to test its quantum properties in the infrared.

Recently, a tabletop experiment has been proposed to explore such quantum origin of gravity with the help of quantum superposition and quantum entanglement~\cite{Bose:2017nin, Marletto}. The protocol known as the quantum gravity induced entanglement of masses (QGEM) is based on the quantum interaction of gravity with the quantum state of matter to generate the entanglement. The latter is purely a quantum observable and has no classical analogue. If gravity is indeed quantum, it will entangle the two masses in quantum spatial superpositions~\cite{Marshman:2019sne,Bose:2022uxe}. In the canonical approach to quantum gravity, the gravitational interaction is being mediated by the hypothetical massless spin-2 graviton, whose quantum properties can be studied~\cite{Bose:2022uxe}, see also for the path integral approach~\cite{Bose:2017nin,Christodoulou:2022vte}, and the Arnowitt-Desse-Meissner (ADM) approach~\cite{Danielson:2021egj}. The critical point to note here is that a creation of a spatial superposition is governed by its own dynamical degree of freedom, which conserves the equations of motion governed by the electromagnetic properties of the material. To understand the theoretical aspects of the entanglement we will always need to consider the dynamical aspects of two masses, see Ref.~\cite{Bose:2022uxe}.

A large spatial superposition for a massive object tests the foundations of quantum mechanics~\cite{Leggett,Arndt2014}, equivalence principle of gravity~\cite{Barker:2022mdz,Bose:2022czr}, falsifies spontaneous collapse mechanisms~\cite{Penrose,Bassi}, and places bounds on decoherence mechanisms~\cite{Joos:1984uk,Schlosshauer}. Furthermore, as an application of a massive quantum interferometer, we can use them as a quantum sensor~\cite{Toros:2020dbf}, and probe very high-frequency gravitational waves~\cite{Marshman:2018upe}.

To realise some of these ambitious experiments, especially QGEM, we will require a large spatially localised state of superposition 
$\Delta Z \sim {\cal O}(10-100)~{\rm \mu \text{m}}$ for large masses of order $m\sim 10^{-15}-10^{-14}$~kg~\cite{Bose:2017nin,vandeKamp:2020rqh}. We are assuming that the superposition is in the $z$-direction. These requirements are far beyond the scales achieved to date in any laboratory (e.g., macromolecules $m \sim 10^{-22}$~kg over $\Delta Z \sim 0.25~{\rm \mu \text{m}}$, or atoms of mass $m\sim 10^{-25}$~kg over $\Delta Z~\sim 0.5$~m~\cite{Arndt1999,Nimmrichter2013,Arndt2014,Kovachy2015,Yaakov2019}).

Despite numerous challenges, there are already physical schemes to obtain small $\Delta Z$ and $m$~\cite{Bose,Blencowe,Bouwmeester, Sekatski,Romero-Isart,Cirac,Khalili,Scala,Wan,Bateman,Yin,Pino,Clarke,Ringbauer,Khosla,Kaltenbaek,Romero-Isart-2017,Pedernales,Hogan,Machluf,Margalit,Mann,Wood:2021vpp}, and there are
arguments presented on how to achieve large superposition in a vacuum by using the Stern-Gerlach principle in presence of a magnetic field gradient~\cite{Pedernales,Wood:2021vpp,Marshman:2021wyk}. Based on these ideas, a feasibility experiment has been performed with the help of atoms, showing that such a Stern-Gerlach Interferometer (SGI) for massive objects can indeed be realisable~\cite{Margalit}. Of course, we will now need to increase the mass by nearly $6-7$ orders of magnitude, which will pose a serious technical, if not a fundamental, challenge.

In this paper, we aim to improve the existing mechanism for creating a large spatial superposition~\cite{Bose:2017nin,Pedernales,Wood:2021vpp,Marshman:2021wyk}. In our current work we will 
consider the effect of spatially dependent non-linear magnetic field to create the superposition as opposed to 
the spatially dependent linear magnetic field in Ref.\cite{Marshman:2021wyk}. We will utilise the non-linear magnetic field profile  to further increase the spatial superposition size. We will concentrate on the one-dimensional interferometer, which avoids the issues related to the two-dimensional SGI~\cite{Paraniak}. We will be focusing on applying a much lower magnetic field gradient {\it first} described in the original paper of QGEM~\cite{Bose:2017nin}. We are assuming that the electronic spin can be embedded at the centre of the nano-crystal, such as in the case of a nitrogen-vacancy of diamond's  (NV) centre. Our discussions will be generic to any dopant and material, but for the illustration, we will use the material properties similar to that of the diamond. Also, we will avoid the low region of the magnetic field for the Majorana spin-flip~\cite{Majorana,Inguscio}, discussed below.

Typically, we will require the experimental configurations with a magnetic field which originates in a single current-carrying wire or a permanent magnet, where the magnetic field goes as $|B|\propto 1/z$, where $z $ is the distance from the current source. The magnetic field can then be expanded around a small region. Such configurations were considered in~\cite{Marshman:2021wyk}. However, as we will see below, if we consider the nonlinear part of the magnetic field dependence, we can generate an even larger superposition size at a shorter time scale. Indeed, a detailed discussion of obtaining such a magnetic field profile will require separate consideration, such as a quadrupole field from coils in an anti-Helmholtz configuration. 

We will further assume that we can achieve the required level of internal cooling of the nano-crystal and the external cooling for maintaining the coherence of the spin for $1-2$ seconds; see the bounds on ambient temperatures in~\cite{Bose:2017nin,vandeKamp:2020rqh,Tilly:2021qef}. We will also assume that the crystal is ideal; in this respect, we are assuming that the impurities are very small, such that the spin coherence can be maintained. The internal cooling for the crystal will also suppress the phonon vibration sufficiently to maintain the spin coherence~\cite{Bose:2017nin}. Given all these effects are under control, we will ask how large superposition size can we achieve for masses $10^{-17}, 10^{-16}, 10^{-15}$~kg massive objects.

We will apply the inhomogeneous magnetic field profile and the bias magnetic field. We will consider the nonlinear dependence of the magnetic field in one-direction-$z$, without loss of any generality. In this regard, we will create the superposition primarily in the $z$-direction. We will first create a velocity difference between the two paths of the spins by 
creating anharmonic oscillations and create a sufficiently large velocity difference between the two paths to catapult the trajectories as far as possible to create a large $\Delta Z$. Then we will bring the trajectories back to interfere the two paths, and study the spin coherence~\cite{Englert1988,Schwinger1988}. 

While creating the spatial superposition, we will lose the spin coherence; therefore to create interference, we will need to ensure that the spin coherence is restored at the moment of interference. We will demand that the spin coherence be $99\%$, which will place a severe constraint on the two paths and hence any fluctuations they incur in creating the superposition. We will see that our analysis following~\cite{Englert1988,Schwinger1988} will put a stringent constraint on the magnetic field fluctuations, which we can tolerate. We will also assume that the entire setup is performed in a free-fall experiment, such that the Earth's gravitational acceleration can be negligible. The latter is necessary for avoiding any gravity-induced and relative acceleration noise; the details can be found in~\cite{Toros:2020dbf}. 

The paper is organised as follows. In section II, we will discuss the foundations of the SGI setup and discusses the nonlinear magnetic field profile and the constraints on the magnetic field. In section III, we will discuss various stages of the two trajectories for masses $10^{-17}, 10^{-16},10^{-15}$~kg. In section IV, we will discuss the scaling behaviour of the superposition size. In section V, we will discuss the constraint on the magnetic field fluctuation, which we can tolerate for the spin coherence, and in section VI, we will conclude our paper.

\section{Stern-Gerlach Interferometer}
	
We can write the Hamiltonian of the spin embedded in the nano-crystal as \cite{Loubser,Pedernales,Marshman:2021wyk}
	\begin{equation}\label{Hamitonian}
		H=\frac{\hat{\boldsymbol{p}}^{2}}{2 m}+\hbar D \hat{\boldsymbol{S}}^{2}-\frac{\chi_{m} m}{2 \mu_{0}} \boldsymbol{B}^{2}-\hat{\boldsymbol{\mu}}\cdot \boldsymbol{B},
	\end{equation}	
	where $m$ is the mass of the nano-crystal, $\hat{\boldsymbol{p}}$ and $\hat{\boldsymbol{S}}$ are momentum and spin operators, respectively. $D$ is the NV zero-field splitting. $\chi_{m}$ is the magnetic susceptibility. $\mu_{0}$ is the vacuum permeability, $\hat{\boldsymbol{\mu}}=-g\mu_{B}\hat{\boldsymbol{S}}$ is the spin magnetic moment, where $g\approx 2$ is the Land\`{e} g-factor, $\mu_{B}=\frac{e \hbar}{2 m_{e}}$ is the Bohr magneton, $e$ is the electron charge and $m_{e}$ is the electron rest mass. $\boldsymbol{B}$ is the magnetic field. We will assume that the spin is embedded in the centre of the nano-crystal. We will not consider the effects of external torque in this paper; we are assuming that we can engineer a situation so that the torque and the rotational effects of the mass are negligible or decoupled from the translation; for possible mechanisms to cool rotation, see~\cite{Kuhn,Schafer,Rudolph,Frimmer,Japha}. At this point, we also consider an idealised system with no impurities. Of course in reality, we will need to consider the impurities. However, for this toy model, we will not consider these effects here for the time being. Note that we are neglecting the gravitational potential here. We will be interested in experimenting with a free-fall setup to minimise gravity gradient noise and dephasing due to Earth's gravitational potential, see the discussion in Ref.~\cite{Toros:2020dbf}. 
	
	With these assumptions, the last two terms in Eq.(\ref{Hamitonian}) represent the potential energy
	\begin{align}\label{Potential}
		\hat{\boldsymbol{U}}=-\frac{\chi_{m} m}{2 \mu_{0}} \boldsymbol{B}^{2}-\hat{\boldsymbol{\mu}}\cdot \boldsymbol{B},
	\end{align}
	from which we can calculate the acceleration of the nano-crystal as
	\begin{align}\label{Acceleration1}
		\hat{\boldsymbol{a}}&=-\frac{1}{m}\nabla \hat{\boldsymbol{U}} \nonumber \\
		&=\frac{\chi_{m}}{2\mu_{0}}\nabla \boldsymbol{B}^{2}-\frac{g e \hbar}{2 m m_{e}}\nabla \hat{\boldsymbol{S}}\cdot \boldsymbol{B}.  
	\end{align}
	Eq.(\ref{Acceleration1}) shows that if the specific form of the magnetic field is determined, then the acceleration can be calculated to obtain the trajectory of the nano-crystal in the magnetic field. 
	
	We will assume that the magnetic field takes the following simple form
	\begin{align}\label{magneticfield}
		\boldsymbol{B}=(B_{0}+\eta z^{2}-\eta x^{2}) \hat{\boldsymbol{z}}-2\eta zx \hat{\boldsymbol{x}},
	\end{align}
	where $B_{0}$ is a fixed constant magnetic field, we will explain below why do we need this bias magnetic field to align the NV-spin in the $z$ direction. The 
	$\eta$ is a coefficient with the dimension $\text{T}~\text{m}^{-2}$, while $\eta z$ will determine the magnetic field gradient~\footnote{We can test that the magnetic field function satisfies Maxwell's equation ${\bf \nabla \cdot B=0}$ and ${\bf \nabla \times B=0}$ away from the source term.}. Here, we have assumed that at the initial moment the coordinate of the NV centre along the x-axis is zero ($x = 0$) and that the embedded spin is aligned in the z-direction ($S_{x} = 0$).
	
	Let's now calculate $\nabla \boldsymbol{B}^{2}=2(B_{0}+\eta z^{2})2\eta z\hat{\boldsymbol{z}}$, and $\nabla \hat{\boldsymbol{S}}\cdot \boldsymbol{B}=(2\eta z S_{z})\hat{\boldsymbol{z}}$, respectively, and then combine the results to get the expression for the acceleration.
	\begin{align}\label{Acceleration2}
		a_{z}=\left(\frac{\chi_{m}}{\mu_{0}}(B_{0}+\eta z^{2})2\eta z-S_{z}\frac{g e \hbar}{m m_{e}}\eta z\right)\hat{\boldsymbol{z}}.
	\end{align}
The initial state superposition is given by, $\left(\ket{\uparrow}_{z}+\ket{\downarrow}_{z}\right)/\sqrt{2}$,  the internal spin of the NV centre, and the z-direction for the wave packet separation.

The above Eq.(\ref{Acceleration2}) shows that the wave packet only separates in the z-direction and in the x coordinate of the NV centre the acceleration is always zero. The spin state in the $x$ and $y$ basis will experience a rapid Larmor precession, therefore the averaging the spin yields no net force along the $x$ axis in our case. This means that as long as $B_{0}$ in Eq.({\ref{magneticfield}}) is not zero, we can ensure that the spin direction is approximately aligned along the z-axis and avoid the Majorana spin flips, see~\cite{Marshman:2021wyk}. But this is an ideal situation. In the actual experiment, the spin would have a Larmor precession around the z-axis. The minimum allowable value of $B_{0}$ in Eq.({\ref{magneticfield}}) can be determined by both the Larmor precession frequency and the adiabatic condition of the frequency which forbids the particle motion along the $x$-axis, see for details~\cite{Marshman:2021wyk}. The Larmor precession frequency is given by
	\begin{align}\label{LarmorPrecession}
		\omega_{L}=\frac{g e}{2 m_{e}}|\boldsymbol{B(x,z)}|,
	\end{align}
    will require to satisfy the adiabatic condition $\dot{\omega_{L}}\ll \omega_{L}^{2}$~\cite{Marshman:2021wyk}.
    Combining Eq.(\ref{magneticfield}), Eq.(\ref{LarmorPrecession}), and the adiabaticity condition, we are able to obtain the minimum magnetic field, labeled $B_{min}$. The minimum Larmor precession frequency corresponding to the minimum magnetic field that satisfies the adiabatic condition is
    \begin{align}
    	\omega_{L}^{min}=\frac{g e}{2 m_{e}}B_{min}.
    \end{align}
    In this paper, we will set  $B_{0}\geq B_{min}$, and that $B_{0}$ is the minimum magnetic field experienced by the wave packet, so the adiabatic condition is always satisfied during the evolution of the wave packet. We will take $B_{0}\approx 5.7\times 10^{-4}\text{T}$, which ensure that adiabatic condition is always satisfied in this paper. 
    
    
\section{Catapulting trajectories}    
    
\begin{figure*}[htbp]
	\centering
	\begin{subfigure}{0.325\linewidth}
		\centering
		\includegraphics[width=0.9\linewidth]{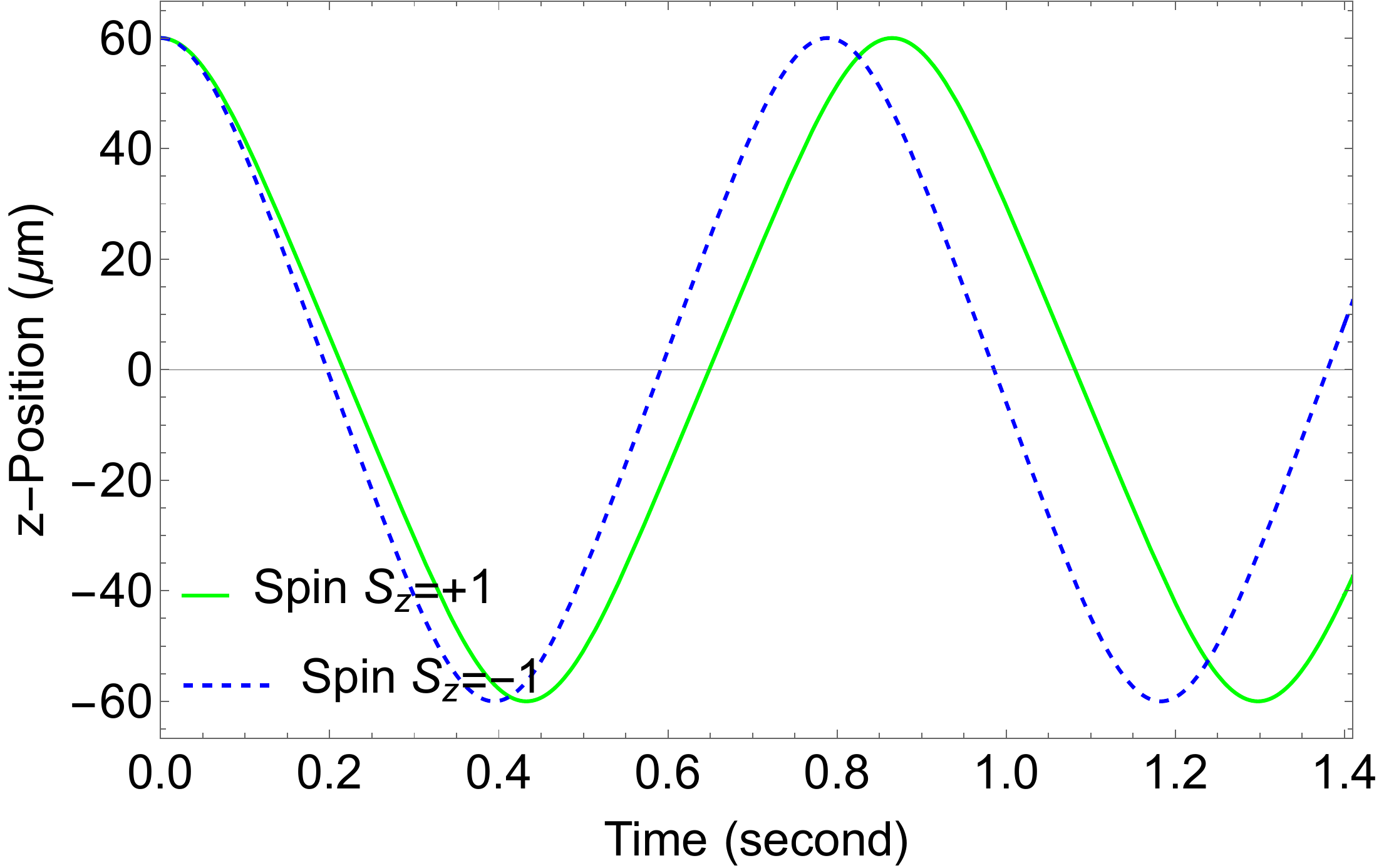}
		\caption{ }
	\end{subfigure}
	\centering
	\begin{subfigure}{0.325\linewidth}
		\centering
		\includegraphics[width=0.9\linewidth]{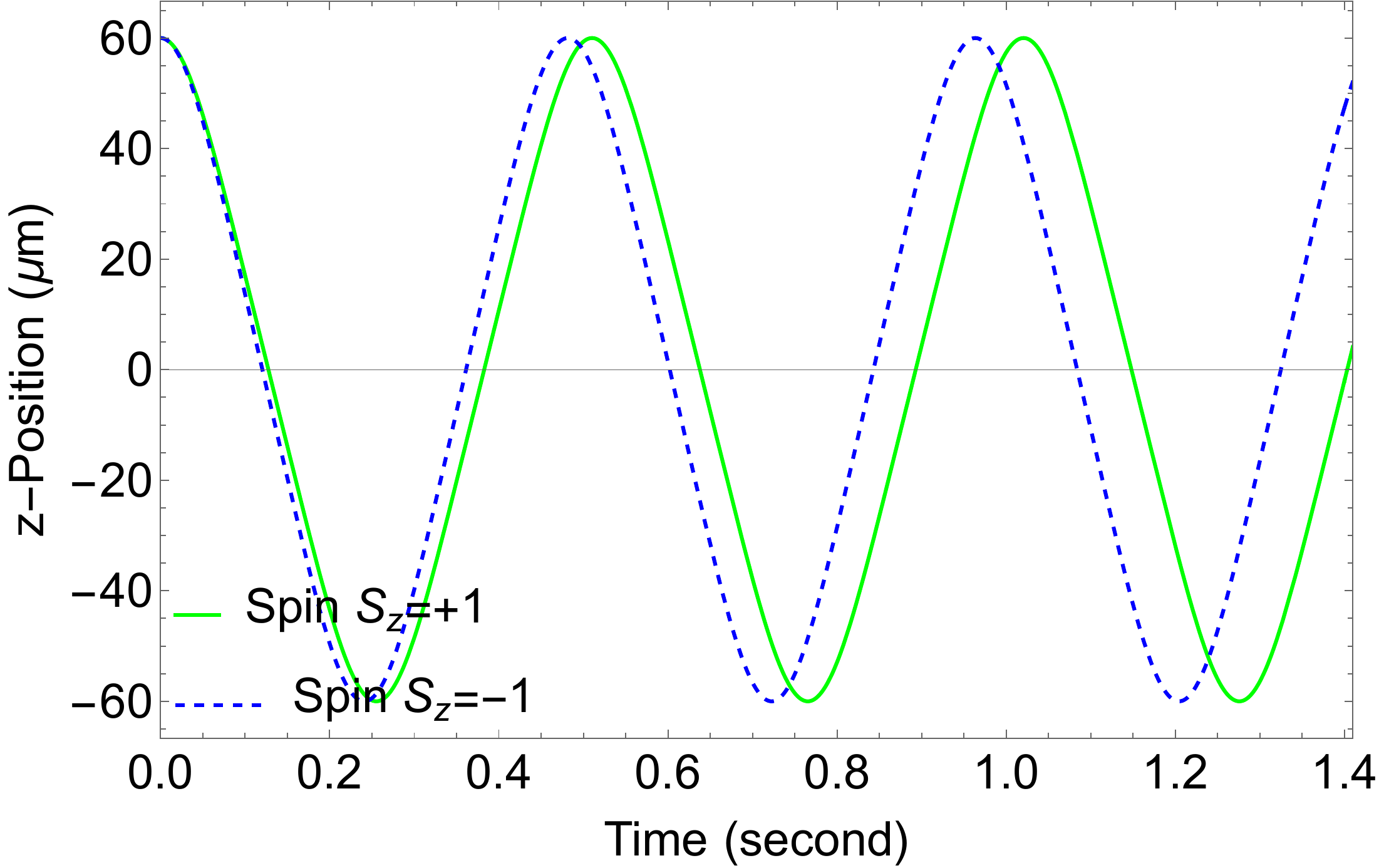}
		\caption{ }
	\end{subfigure}
	\centering
	\begin{subfigure}{0.325\linewidth}
		\centering
		\includegraphics[width=0.9\linewidth]{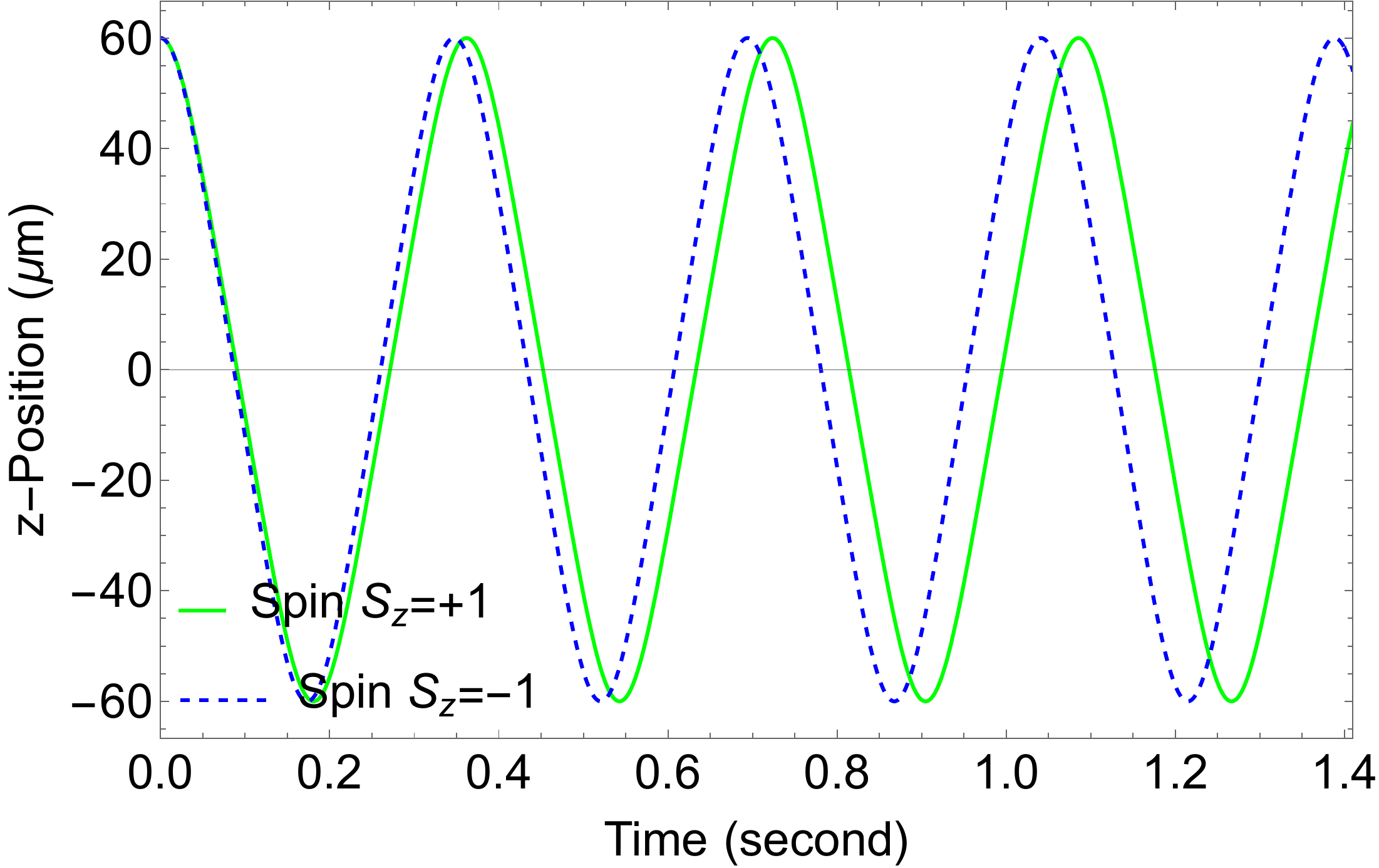}
		\caption{ }
	\end{subfigure}\\
\centering
\begin{subfigure}{0.325\linewidth}
	\centering
	\includegraphics[width=0.9\linewidth]{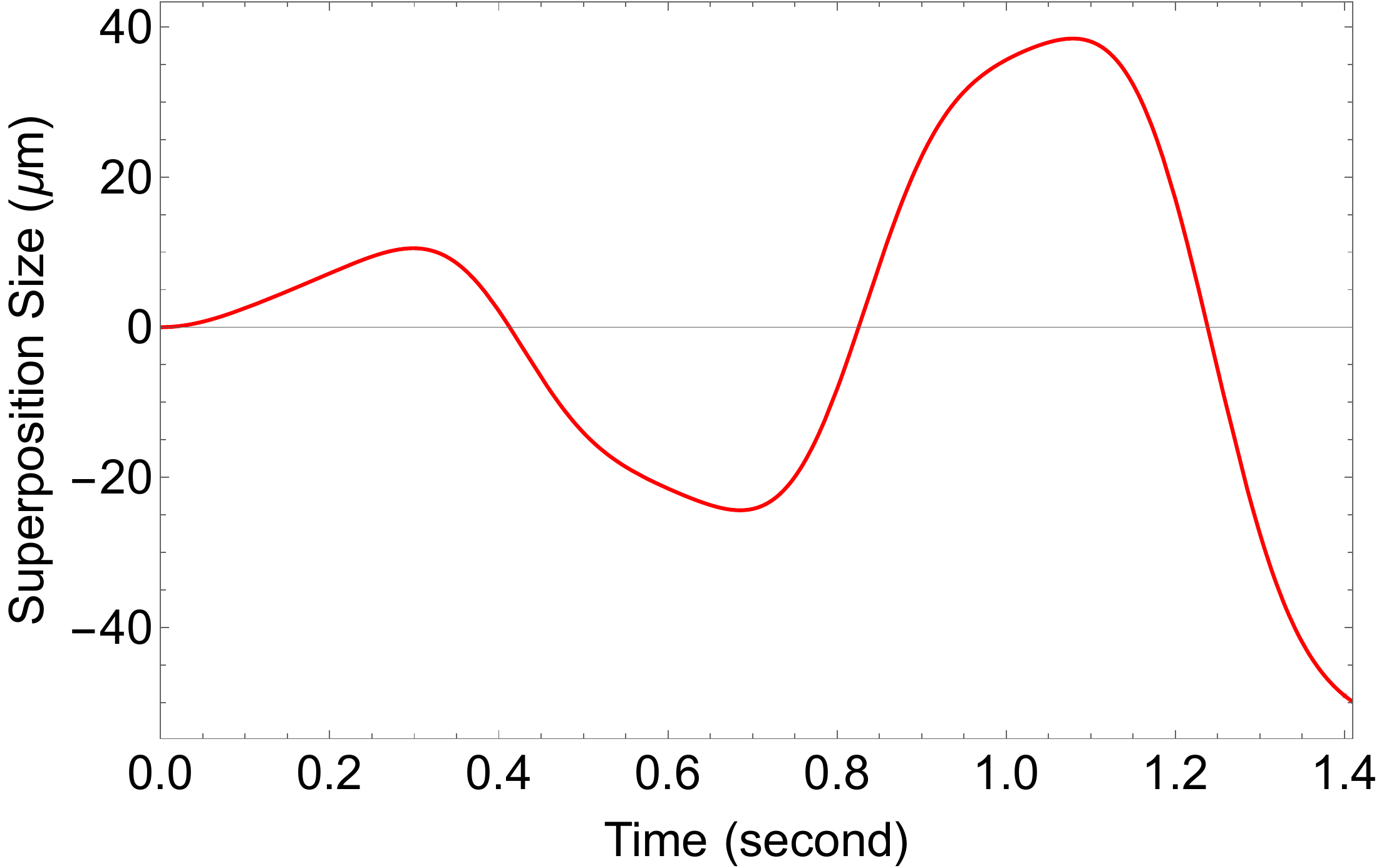}
	\caption{ }
\end{subfigure}
\centering
\begin{subfigure}{0.325\linewidth}
	\centering
	\includegraphics[width=0.9\linewidth]{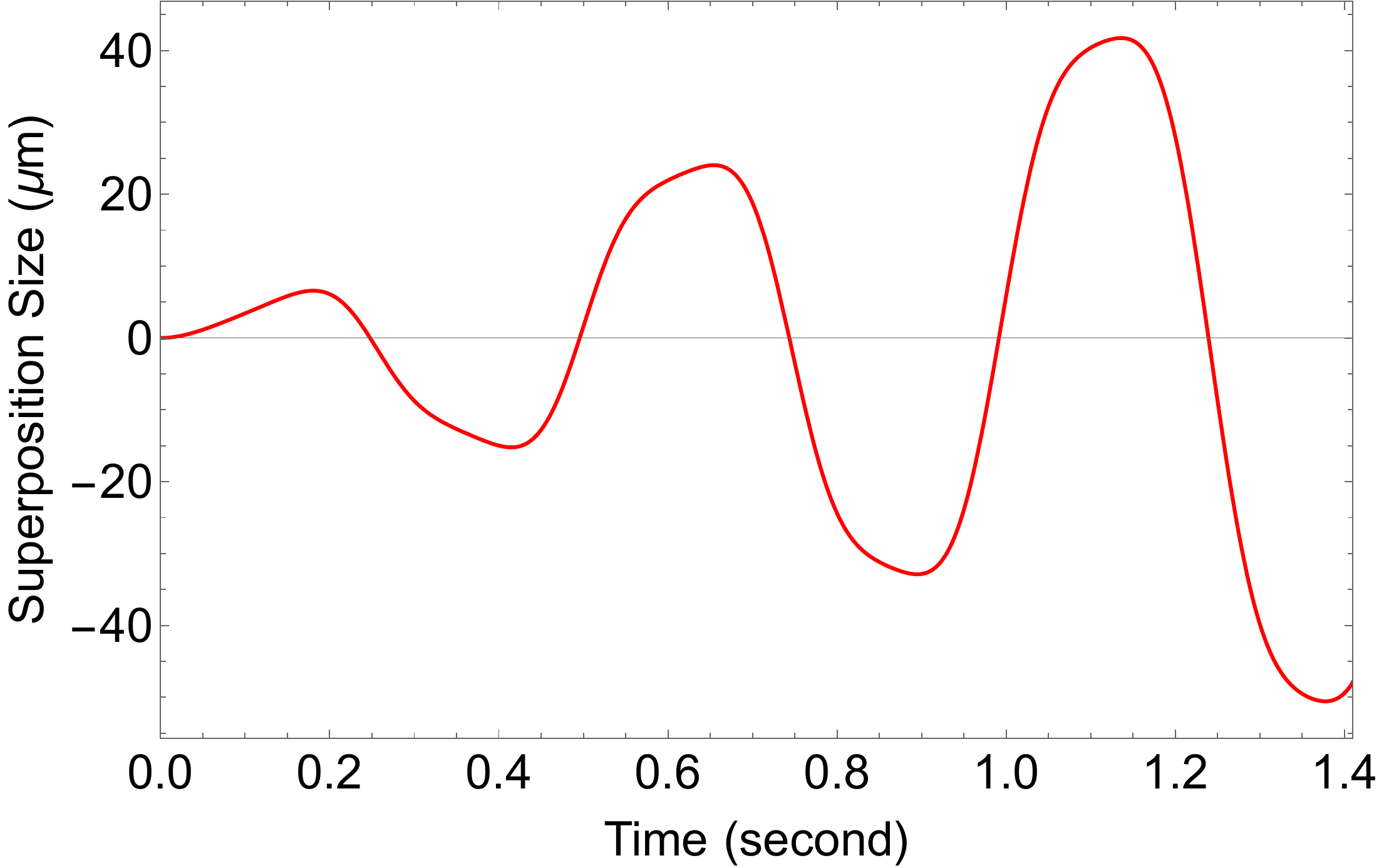}
	\caption{ }
\end{subfigure}
\centering
\begin{subfigure}{0.325\linewidth}
	\centering
	\includegraphics[width=0.9\linewidth]{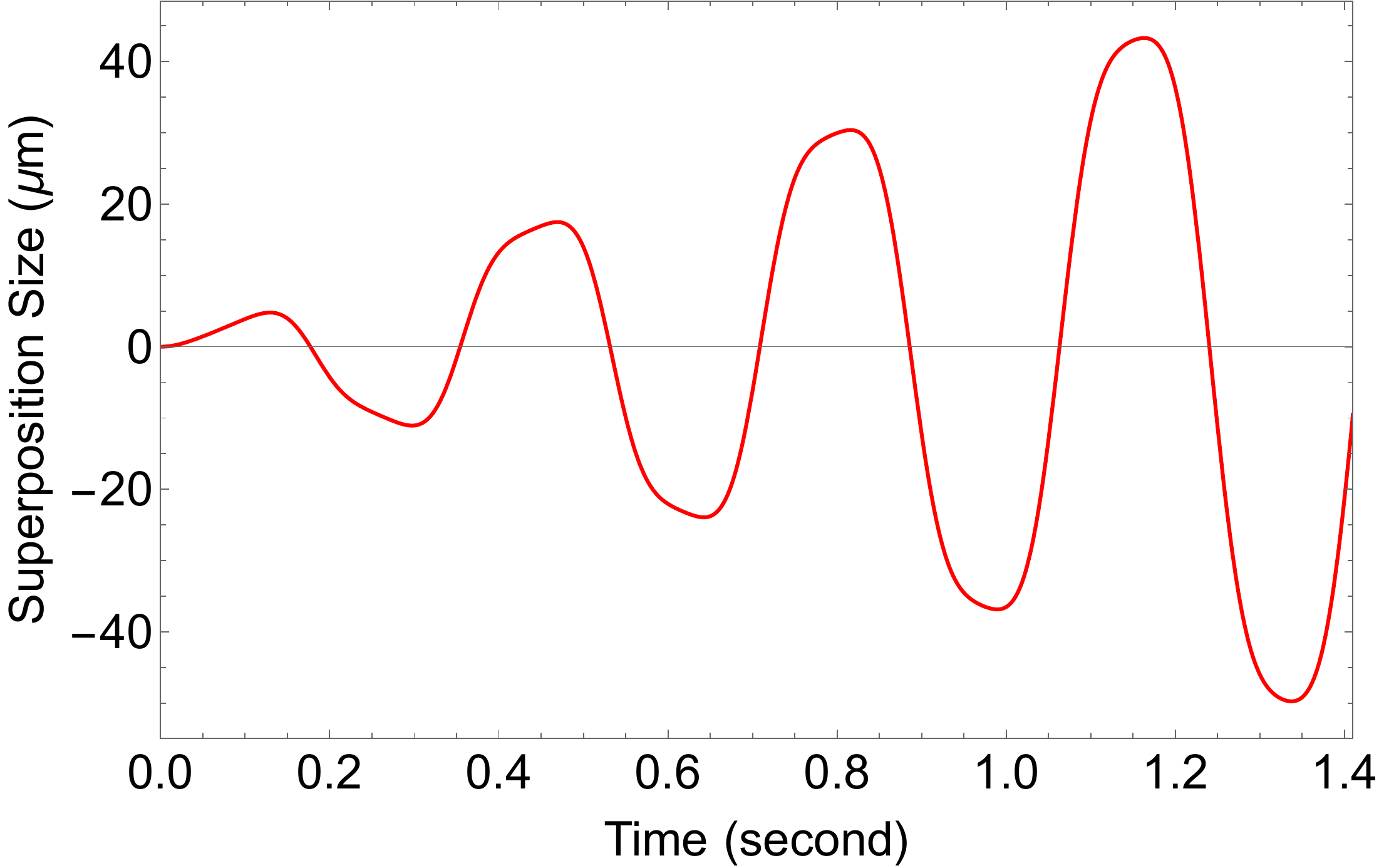}
	\caption{ }
\end{subfigure}
	\caption{Trajectories of the two wave packets under different magnetic field gradients and the corresponding superposition size. $\eta$ is a parameter associated with the magnetic field profile. With the increase of the value of $\eta$ (from left to right, they are $1.4\times 10^{6}~\text{T}~\text{m}^{-2}$, $2.4\times 10^{6}~\text{T}~\text{m}^{-2}$ and $3.4\times 10^{6}~\text{T}~\text{m}^{-2}$), the motional frequency of the wave packet increases accordingly, but the maximum superposition size that can be achieved over the same period of time remains almost unchanged (the maximum superposition size is about $40~{\rm \mu \text{m}}$ within $1.2$ seconds). The mass here is $10^{-17}~ \text{kg}$.}\label{IncreasingGradient}
	\end{figure*}

Note that  the difference between the two wave packet trajectories is mainly caused by the difference in the spin eigenvalues of the second term on the right in Eq.(\ref{Acceleration2}). We will expect to get a large superposition size by increasing the value of $\eta$. However, by increasing the value of $\eta$ will only increase the motional frequency of the wave packet, and will not directly to increase the superposition size, $\Delta Z$. Furthermore, it maintains the superposition size for a longer period.
We will clearly see this results in Fig.\ref{IncreasingGradient}. When we fix the mass of the nano-crystal and increase the value of $\eta$, we will find that the motional frequency of the wave packet increases with $\eta$ while the maximum superposition size remains almost unchanged, such that it can reach $\Delta Z \sim 40~{\rm \mu \text{m}}$ within $\tau \sim 1.2$ seconds for the case $m=10^{-17}~\text{kg}$. Similar results were found for the other two masses considered.

\begin{figure*}[htbp]
	\centering
	\begin{subfigure}{0.325\linewidth}
		\centering
		\includegraphics[width=0.9\linewidth]{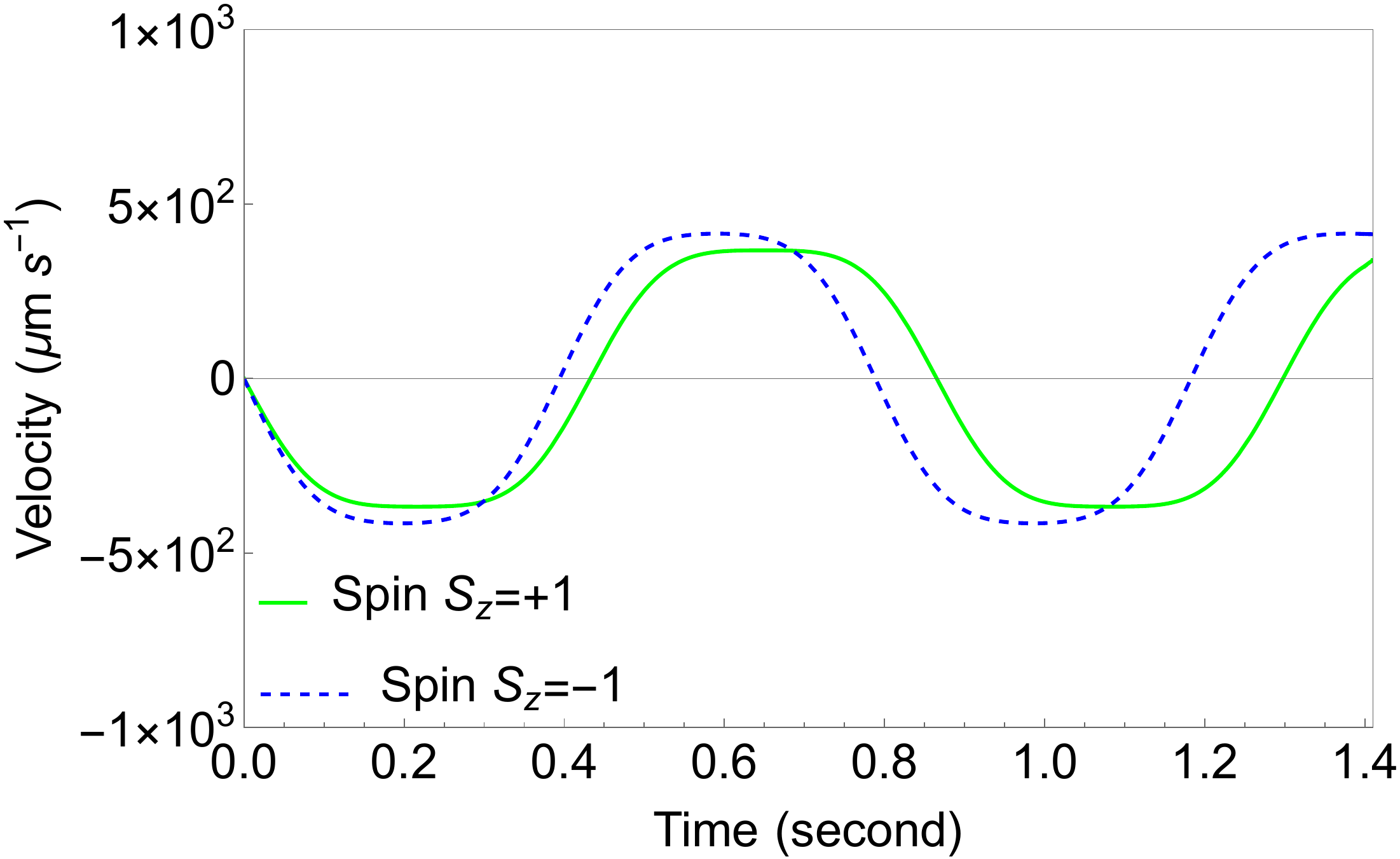}
		\caption{ }
	\end{subfigure}
	\centering
	\begin{subfigure}{0.325\linewidth}
		\centering
		\includegraphics[width=0.9\linewidth]{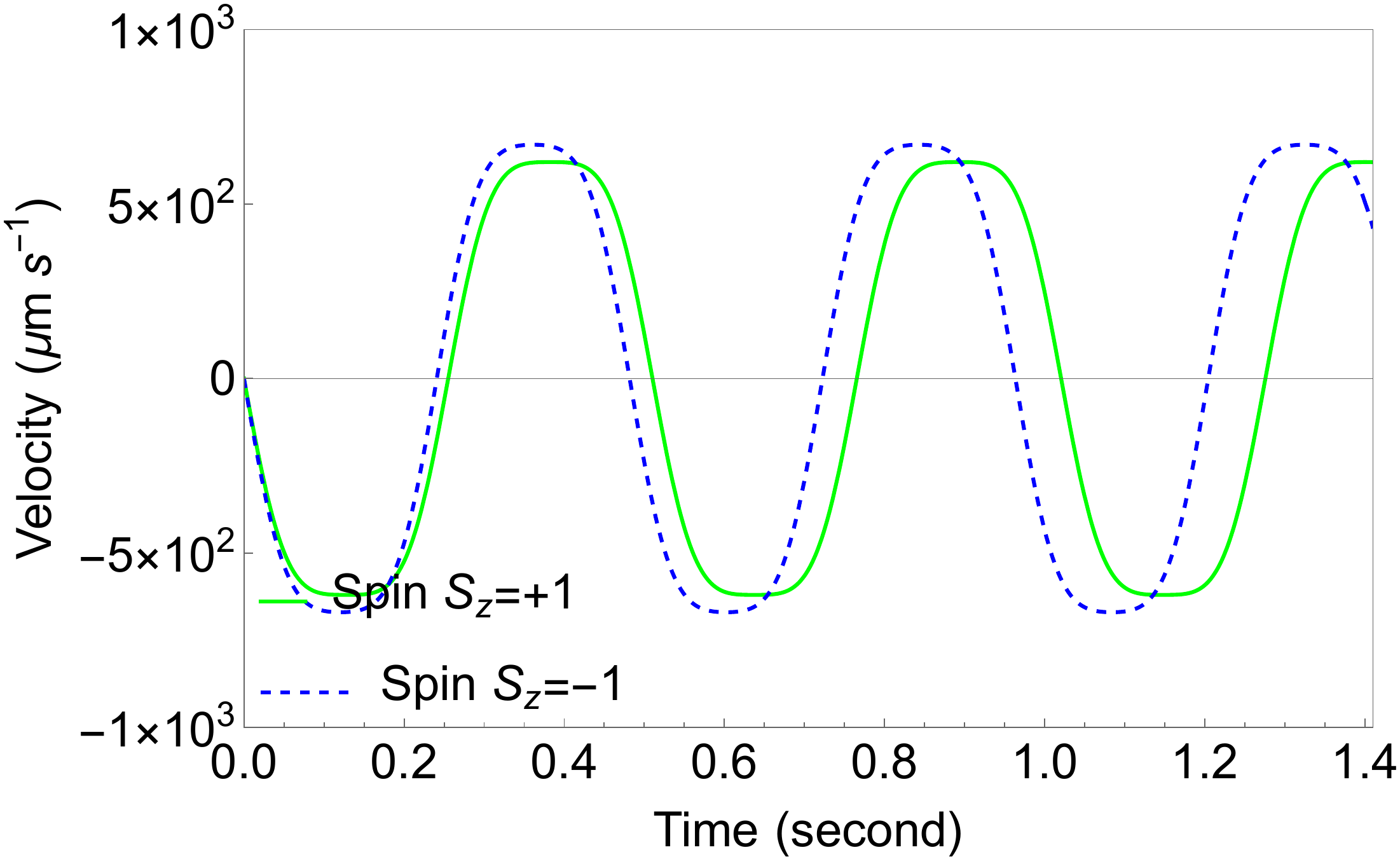}
		\caption{ }
	\end{subfigure}
	\centering
	\begin{subfigure}{0.325\linewidth}
		\centering
		\includegraphics[width=0.9\linewidth]{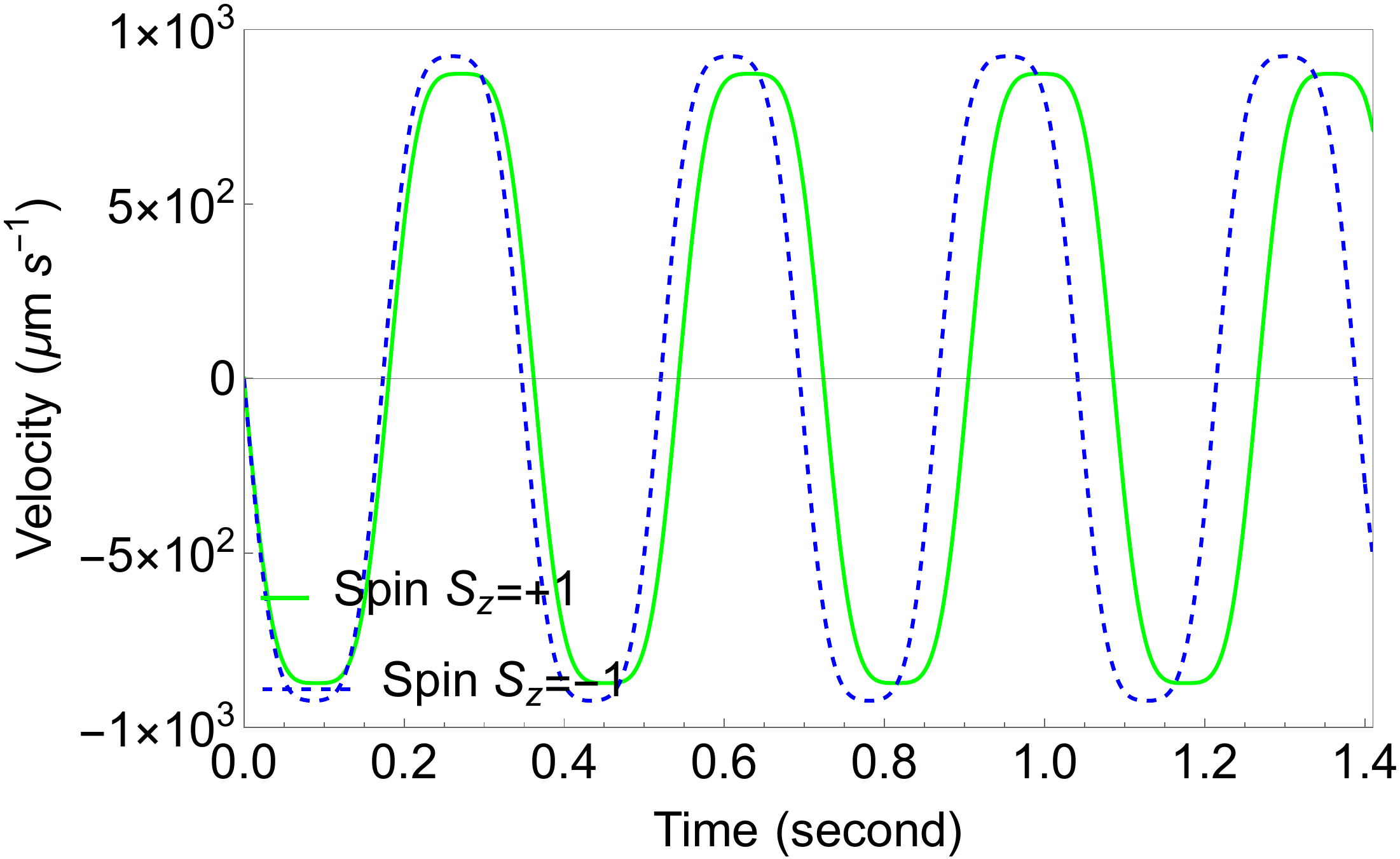}
		\caption{ }
	\end{subfigure}\\
	\centering
	\begin{subfigure}{0.325\linewidth}
		\centering
		\includegraphics[width=0.9\linewidth]{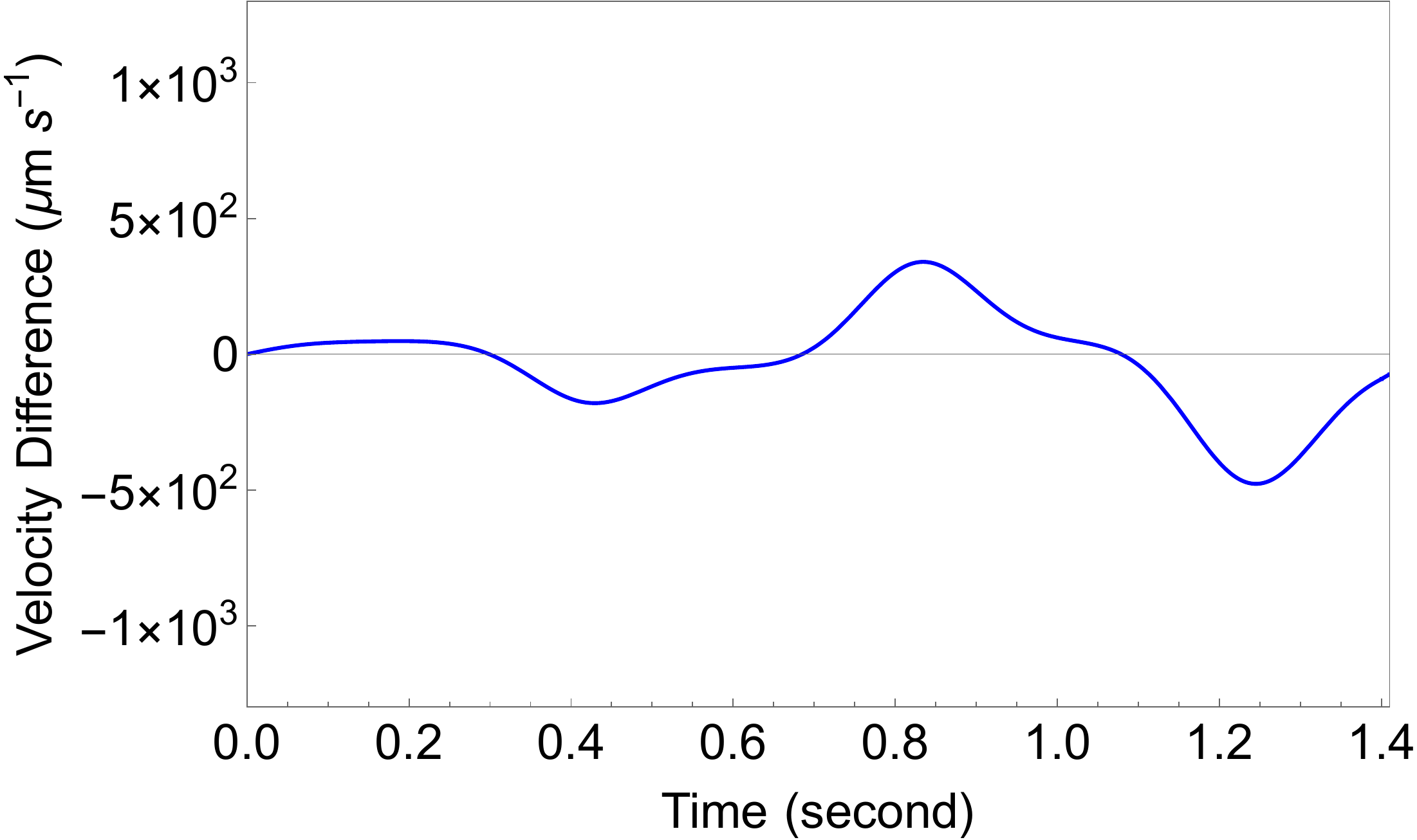}
		\caption{ }
	\end{subfigure}
	\centering
	\begin{subfigure}{0.325\linewidth}
		\centering
		\includegraphics[width=0.9\linewidth]{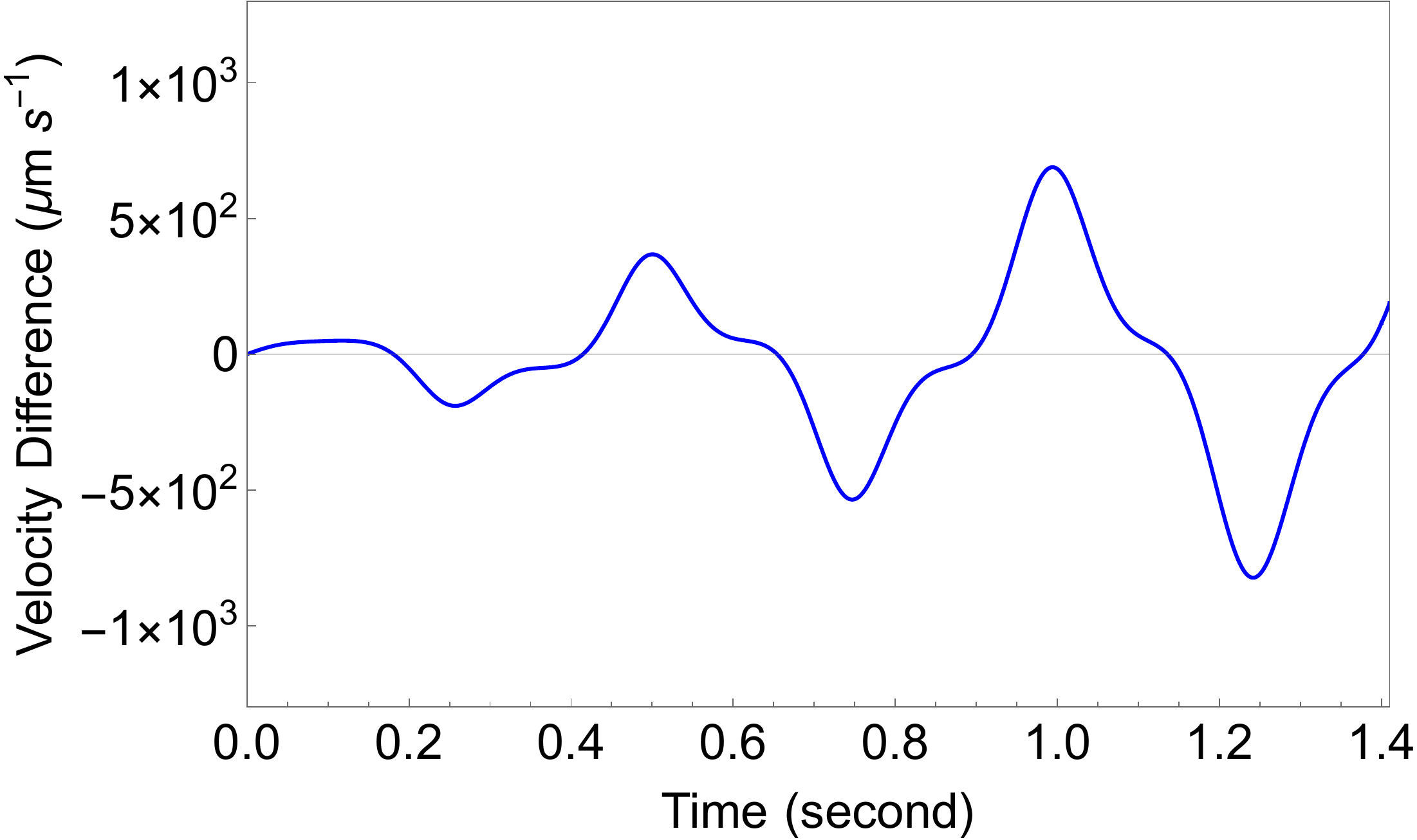}
		\caption{ }
	\end{subfigure}
	\centering
	\begin{subfigure}{0.325\linewidth}
		\centering
		\includegraphics[width=0.9\linewidth]{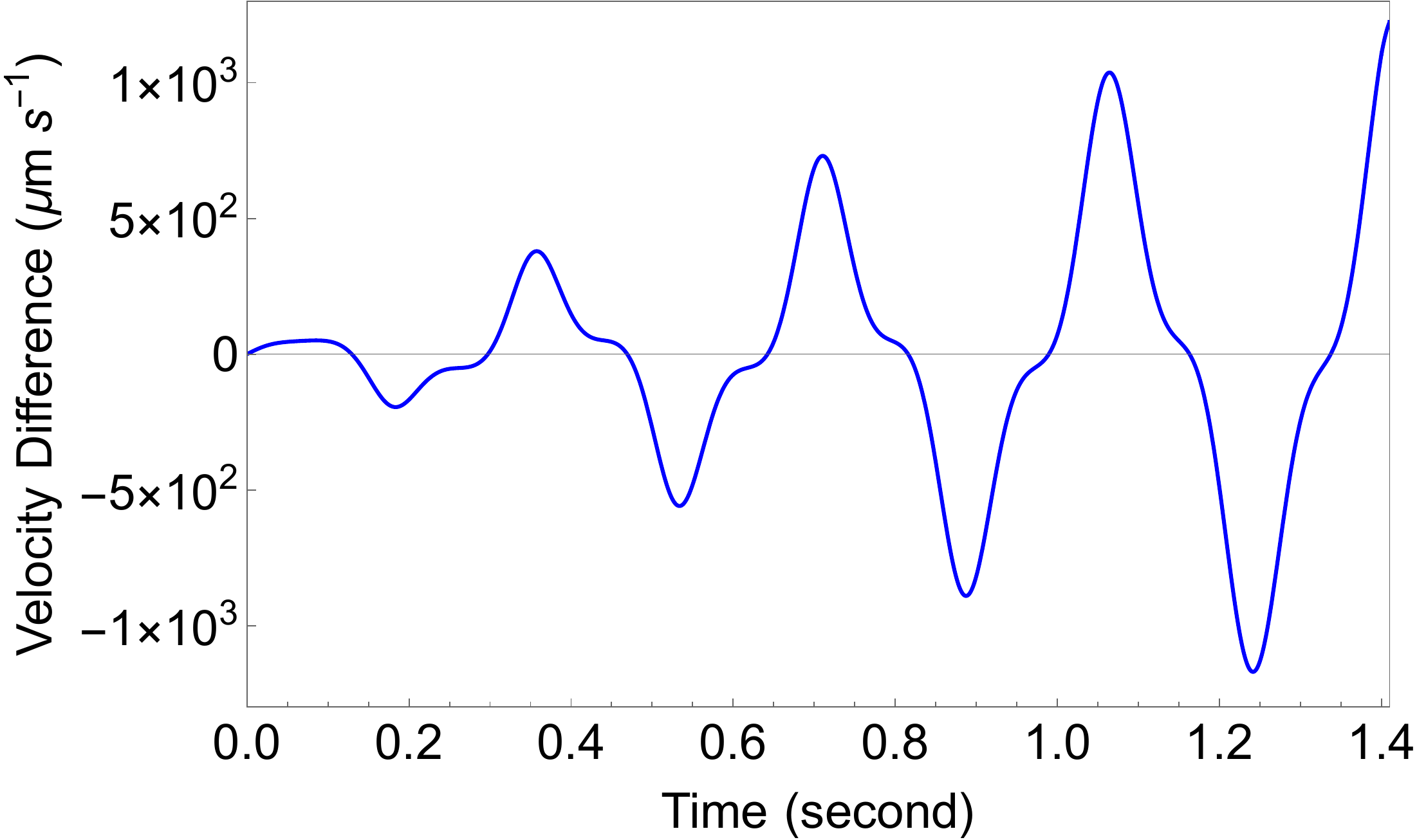}
		\caption{ }
	\end{subfigure}
	\caption{We have shown the velocity curve for the two wave packets under different $\eta$ and the corresponding velocity difference. By increasing the value of $\eta$ (from left to right, they are $1.4\times 10^{6}~\text{T}~\text{m}^{-2}$, $2.4\times 10^{6}~\text{T}~\text{m}^{-2}$ and $3.4\times 10^{6}~\text{T}~\text{m}^{-2}$), the maximum velocity difference that can be achieved over the same period of time also increases (around 400 $\mu \text{m}~\text{s}^{-1}$, 700 $\mu \text{m}~\text{s}^{-1}$, and 1100 $\mu \text{m}~\text{s}^{-1}$ in less than 1.2 $\text{s}$, respectively). The mass here is $10^{-17}~\text{kg}$.}\label{VelocityDifference1}
\end{figure*}


		\begin{figure*}
		\begin{subfigure}{0.325\linewidth}
			\centering
			\includegraphics[width=0.9\linewidth]{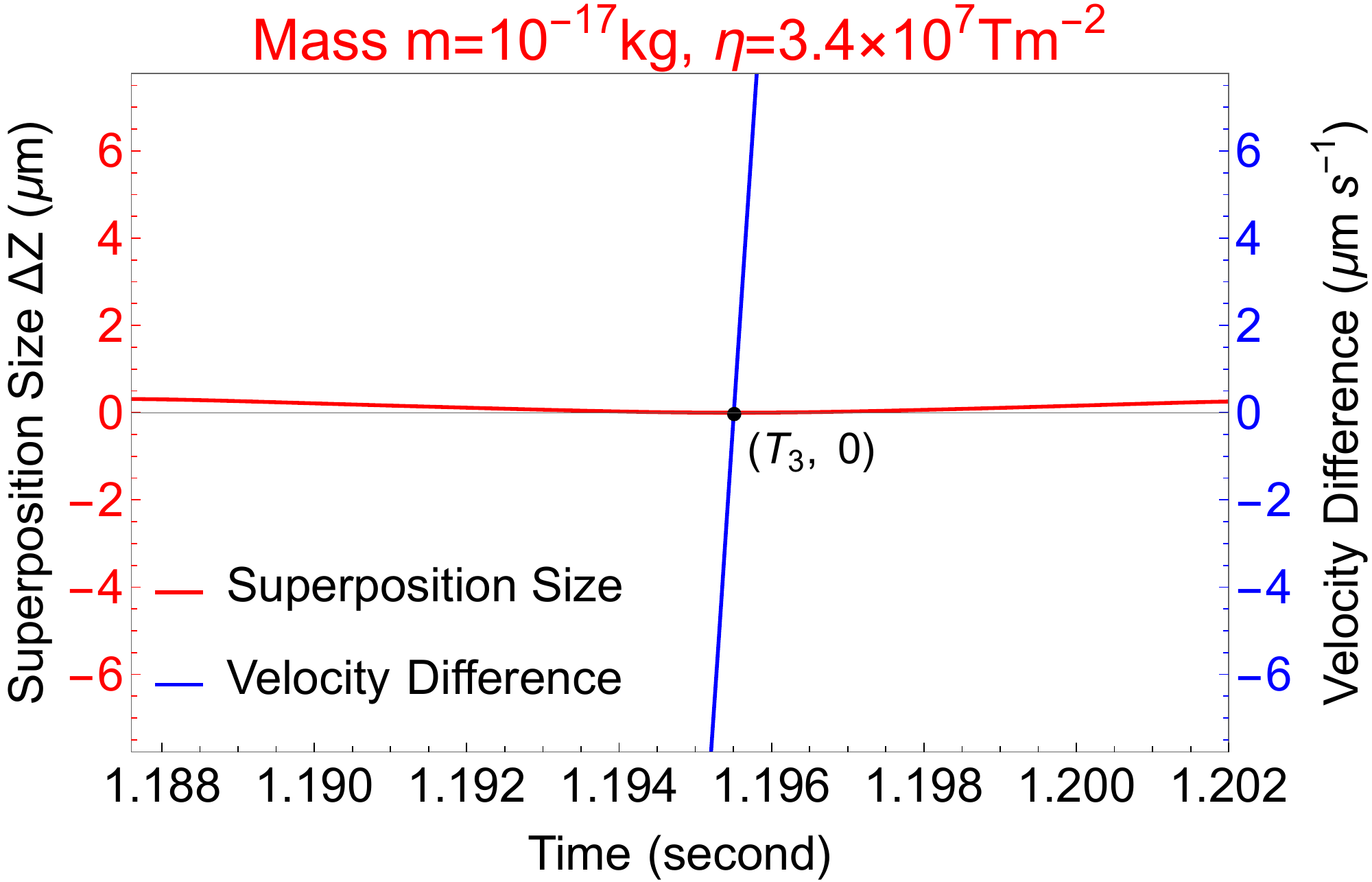}
			\caption{ }\label{Recombined-Mass17}
		\end{subfigure}
		\centering
		\begin{subfigure}{0.325\linewidth}
			\centering
			\includegraphics[width=0.9\linewidth]{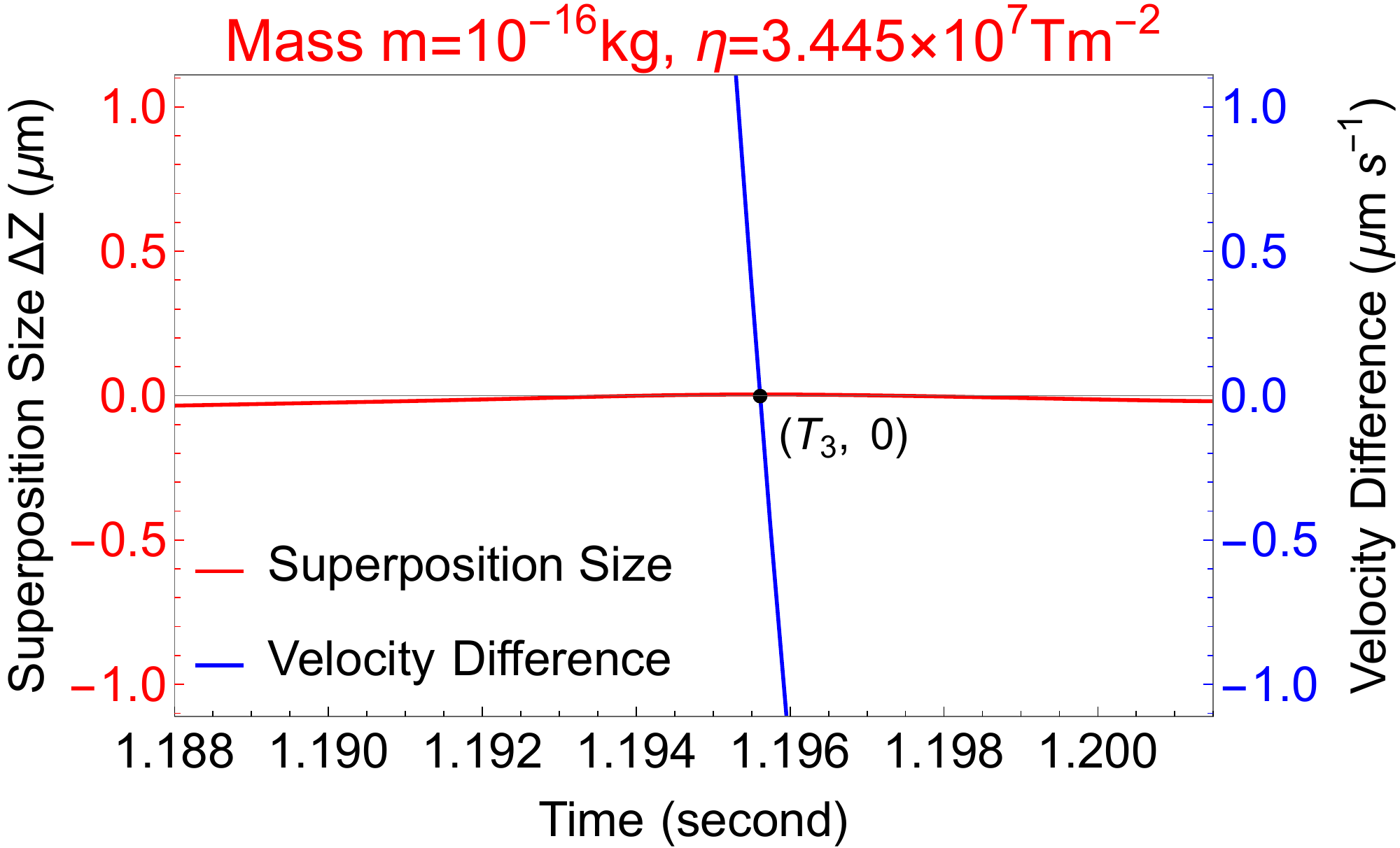}
			\caption{ }
		\end{subfigure}
		\centering
		\begin{subfigure}{0.325\linewidth}
			\centering
			\includegraphics[width=0.9\linewidth]{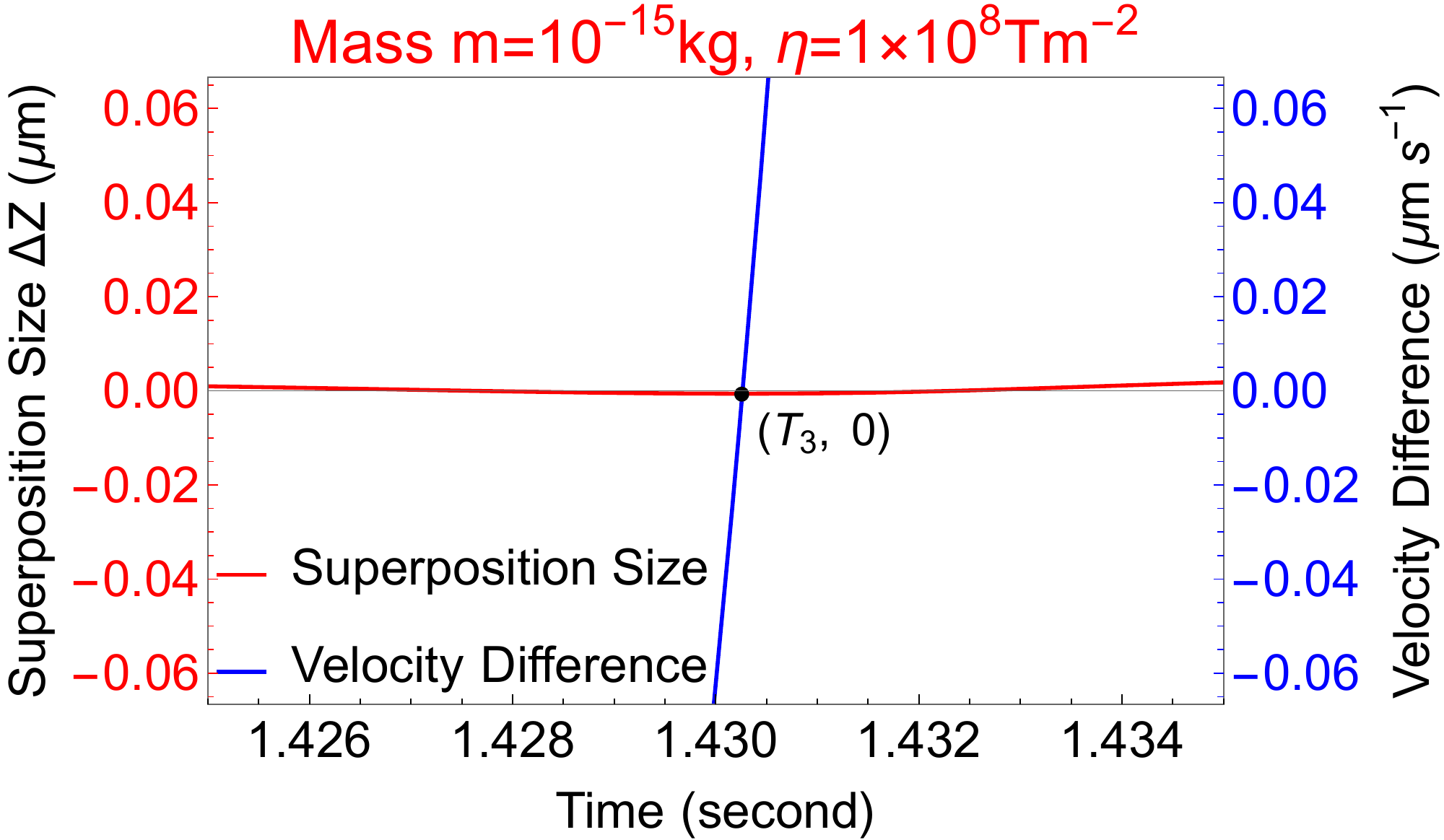}
			\caption{ }
		\end{subfigure}
		\caption{We have shown the velocity difference and the superposition size when the two wave packet's trajectories are closed for the interference for $m=10^{-17} ~\text{kg}$, $10^{-16} ~\text{kg}$, $10^{-15} ~\text{kg}$, respectively. Here we $T_{3}$ has different values for the different masses.}
	\end{figure*}


	Although the superposition size does not increase with $\eta$, we can increase the velocity difference between the two wave packets in a short time by increasing the value of $\eta$ (as shown in Fig.(\ref{VelocityDifference1})). 
	When there is a large velocity difference between the two wave packets and the spatial position coincides, we can adjust the magnetic field so that the two wave packets are located at the lowest point of the potential energy (that is  $z=0$). By doing so, we can catapult the two wave packets around in the magnetic field and achieve a large superposition size.
	 
\begin{figure*}[t!]
	\centering
	\begin{subfigure}{0.325\linewidth}
		\centering
		\includegraphics[width=0.9\linewidth]{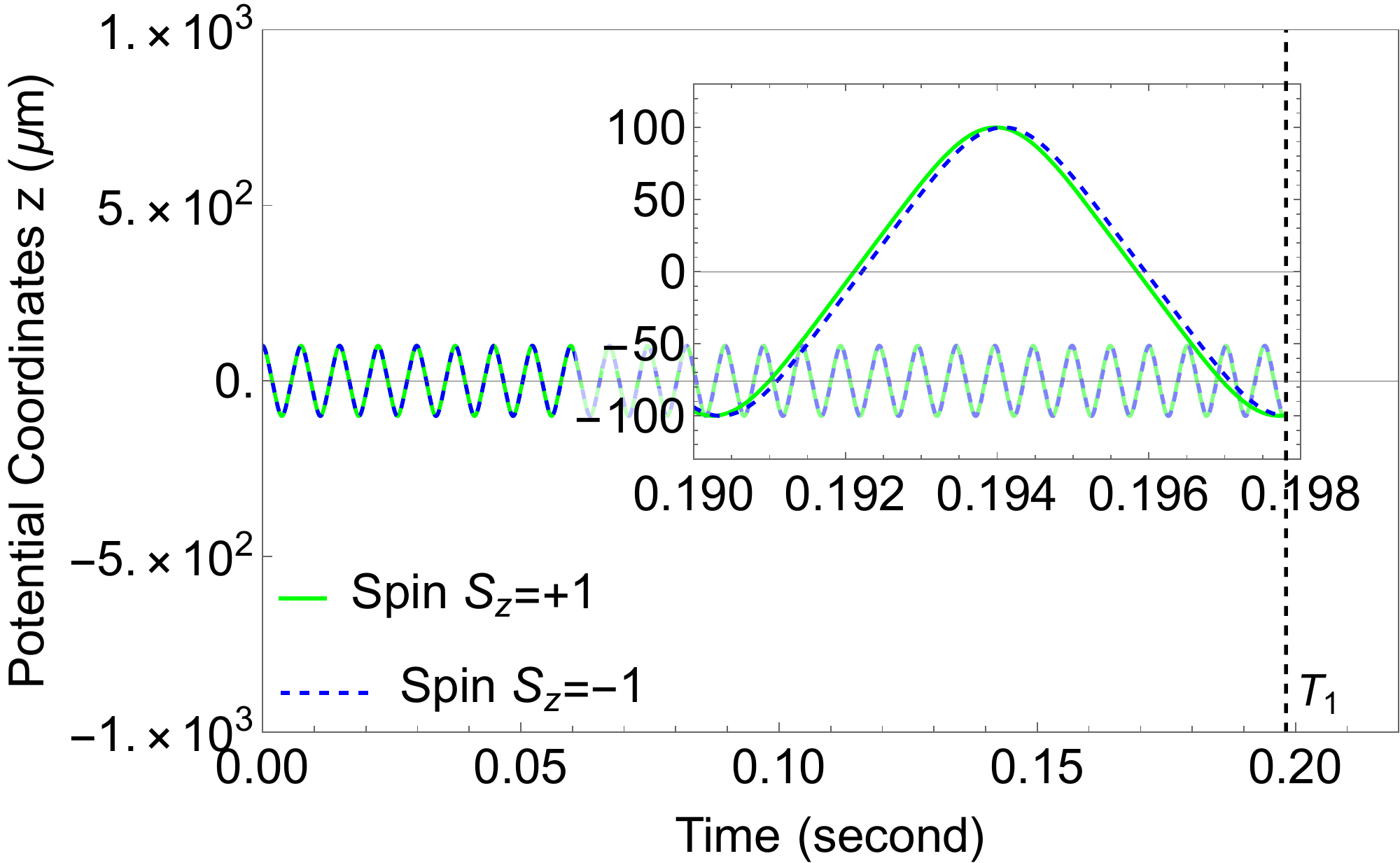}
		\caption{Potential coordinates - stage \Romannum{1}}\label{PotentialCoordinates-Mass17-1}
	\end{subfigure}
	\centering
	\begin{subfigure}{0.325\linewidth}
		\centering
		\includegraphics[width=0.9\linewidth]{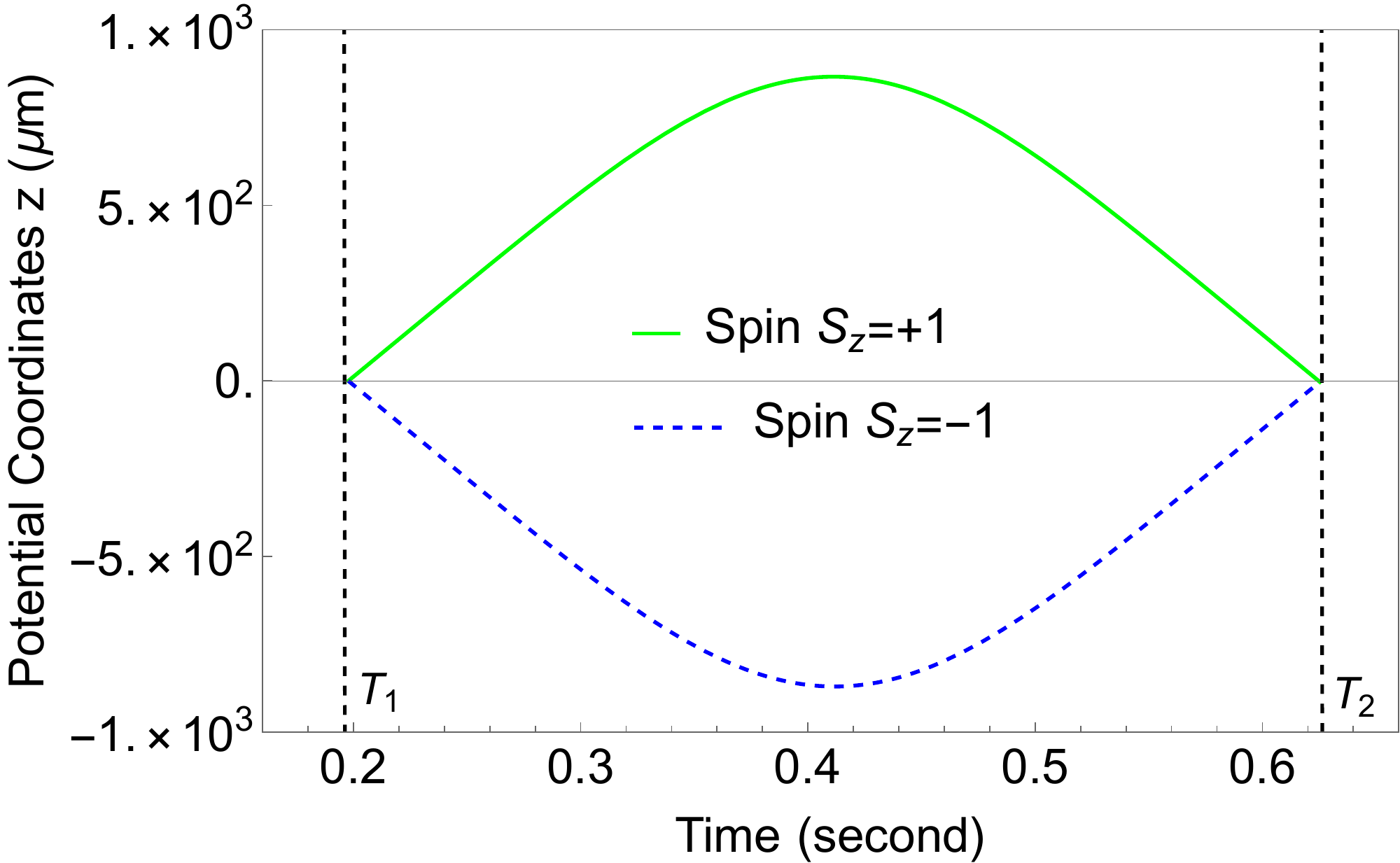}
		\caption{Potential coordinates - stage \Romannum{2}}\label{PotentialCoordinates-Mass17-2}
	\end{subfigure}
	\centering
	\begin{subfigure}{0.325\linewidth}
		\centering
		\includegraphics[width=0.9\linewidth]{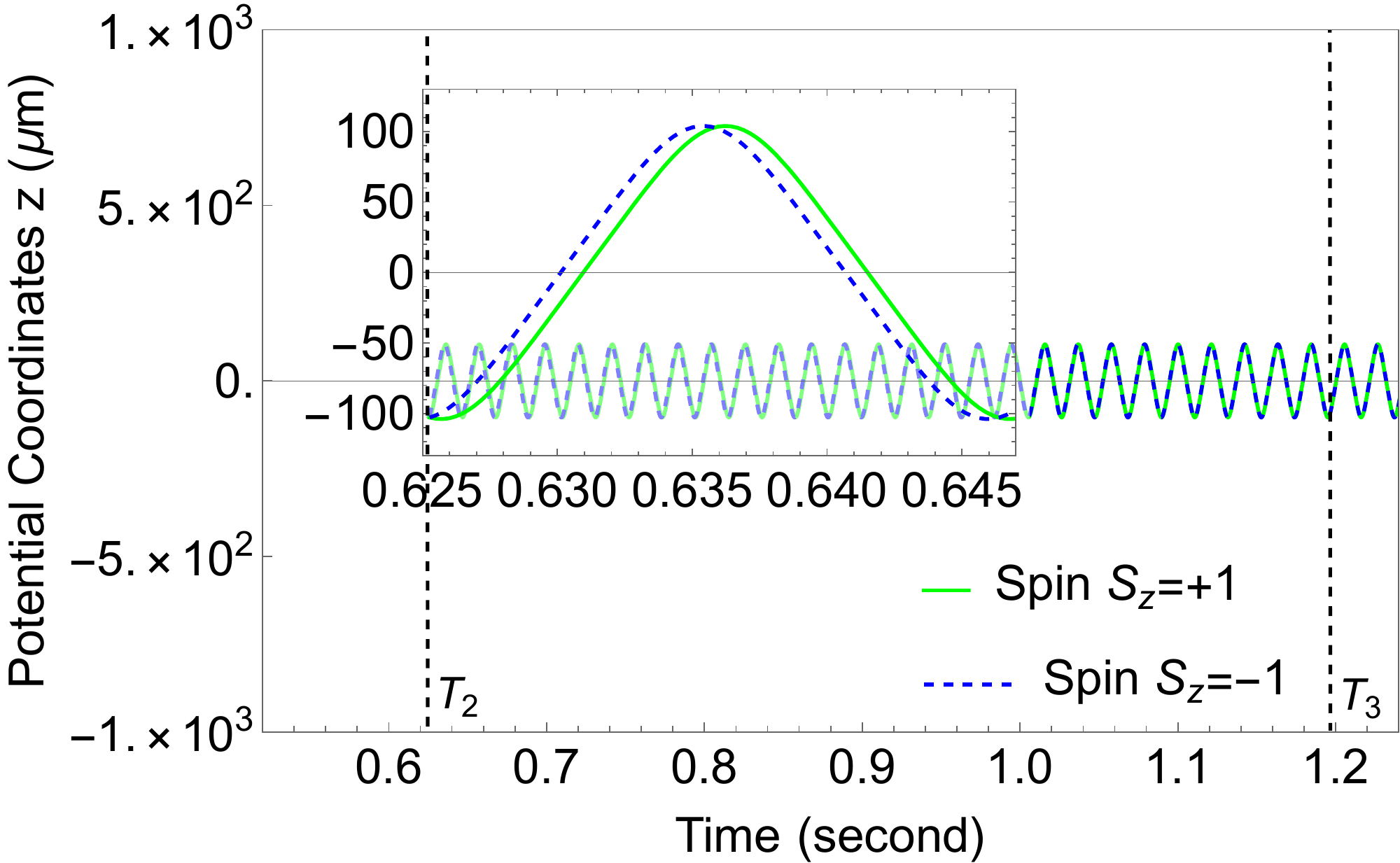}
		\caption{Potential coordinates - stage \Romannum{3}}
	\end{subfigure}\\
	\vspace{0.2cm}
    \centering
    \begin{subfigure}{0.325\linewidth}
	    \centering
    	\includegraphics[width=0.9\linewidth]{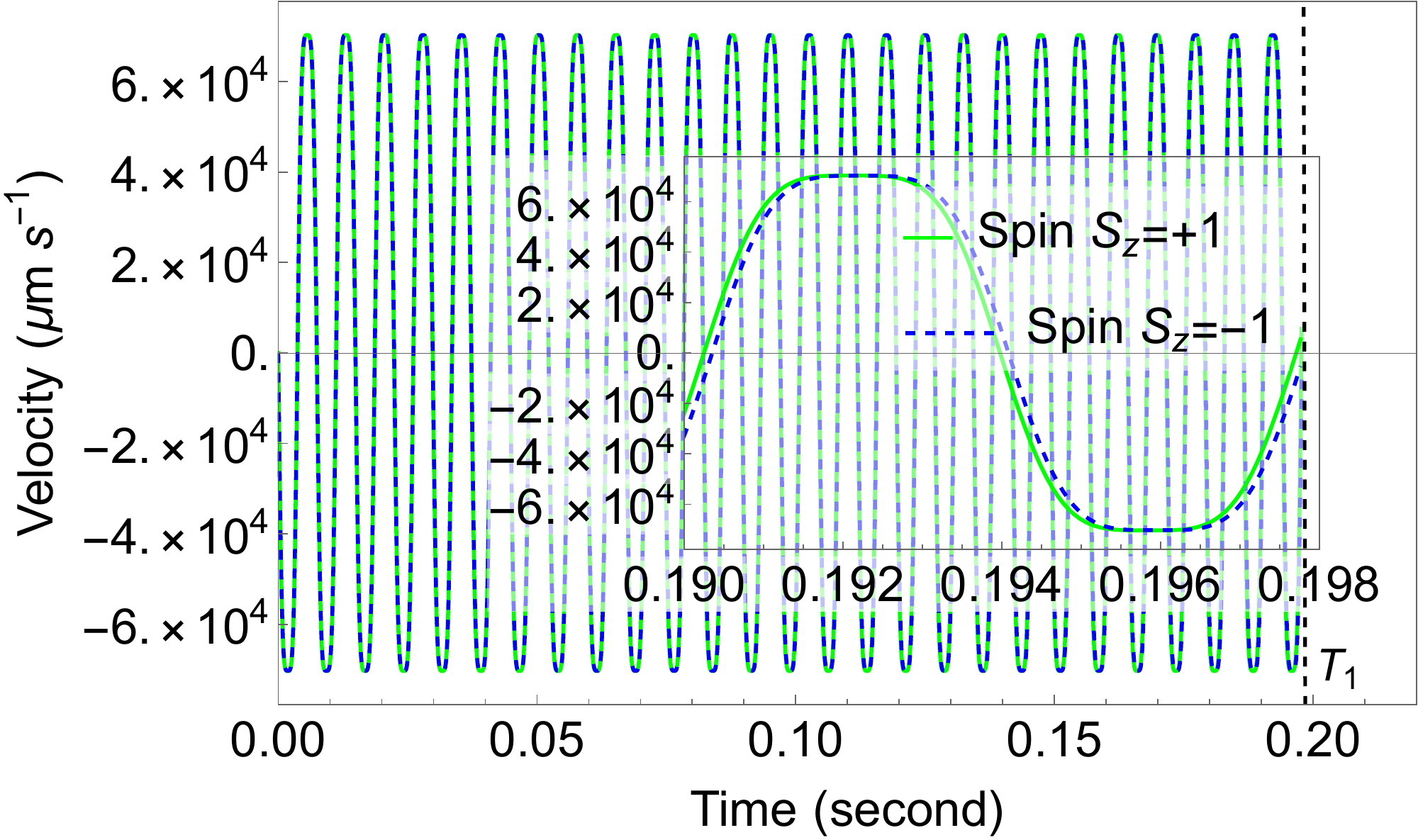}
	    \caption{Velocity - stage \Romannum{1}}\label{Velocity-Mass17-1}
    \end{subfigure}
    \centering
    \begin{subfigure}{0.325\linewidth}
	    \centering
	    \includegraphics[width=0.9\linewidth]{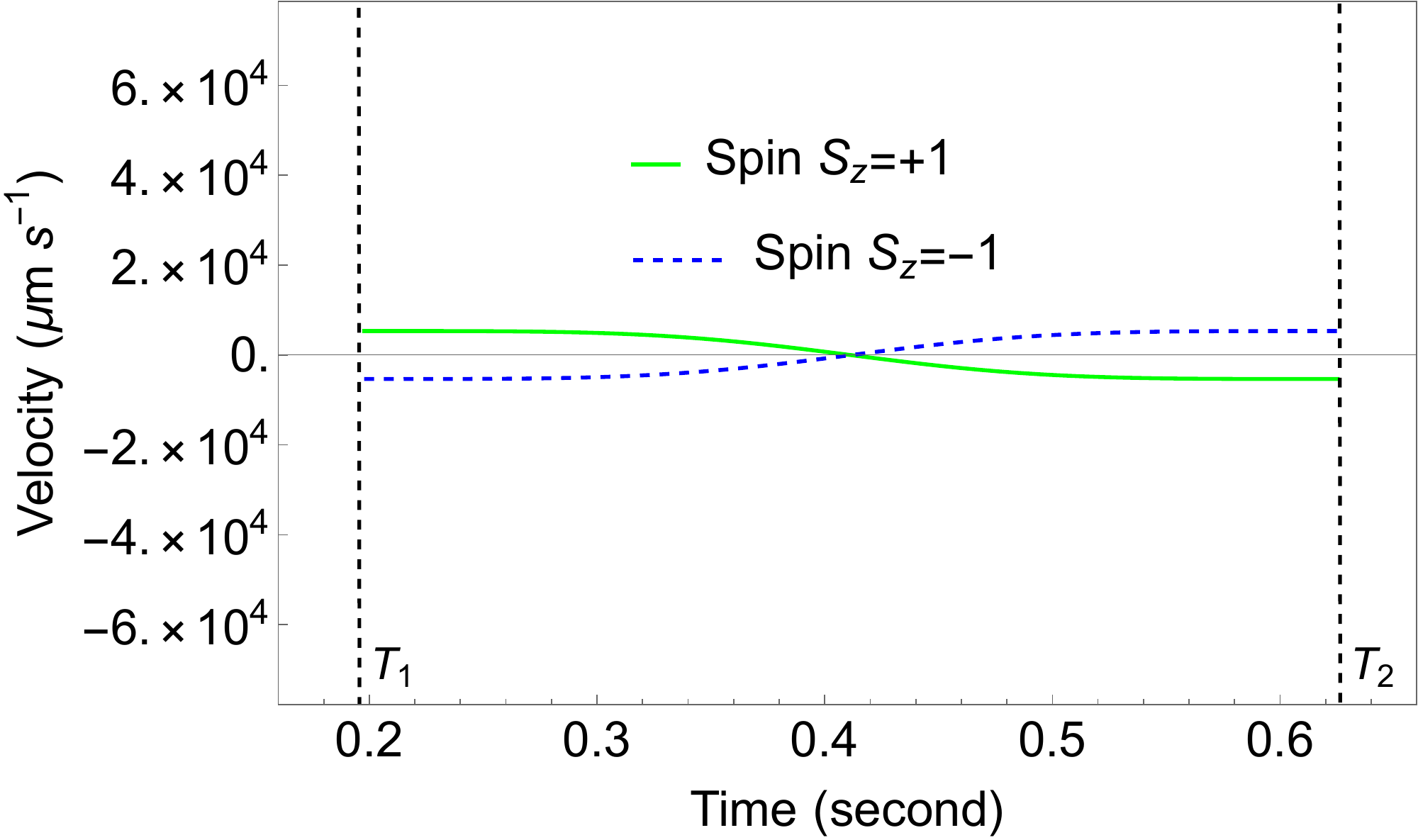}
	    \caption{Velocity - stage \Romannum{2}}
    \end{subfigure}
    \centering
    \begin{subfigure}{0.325\linewidth}
	    \centering
	    \includegraphics[width=0.9\linewidth]{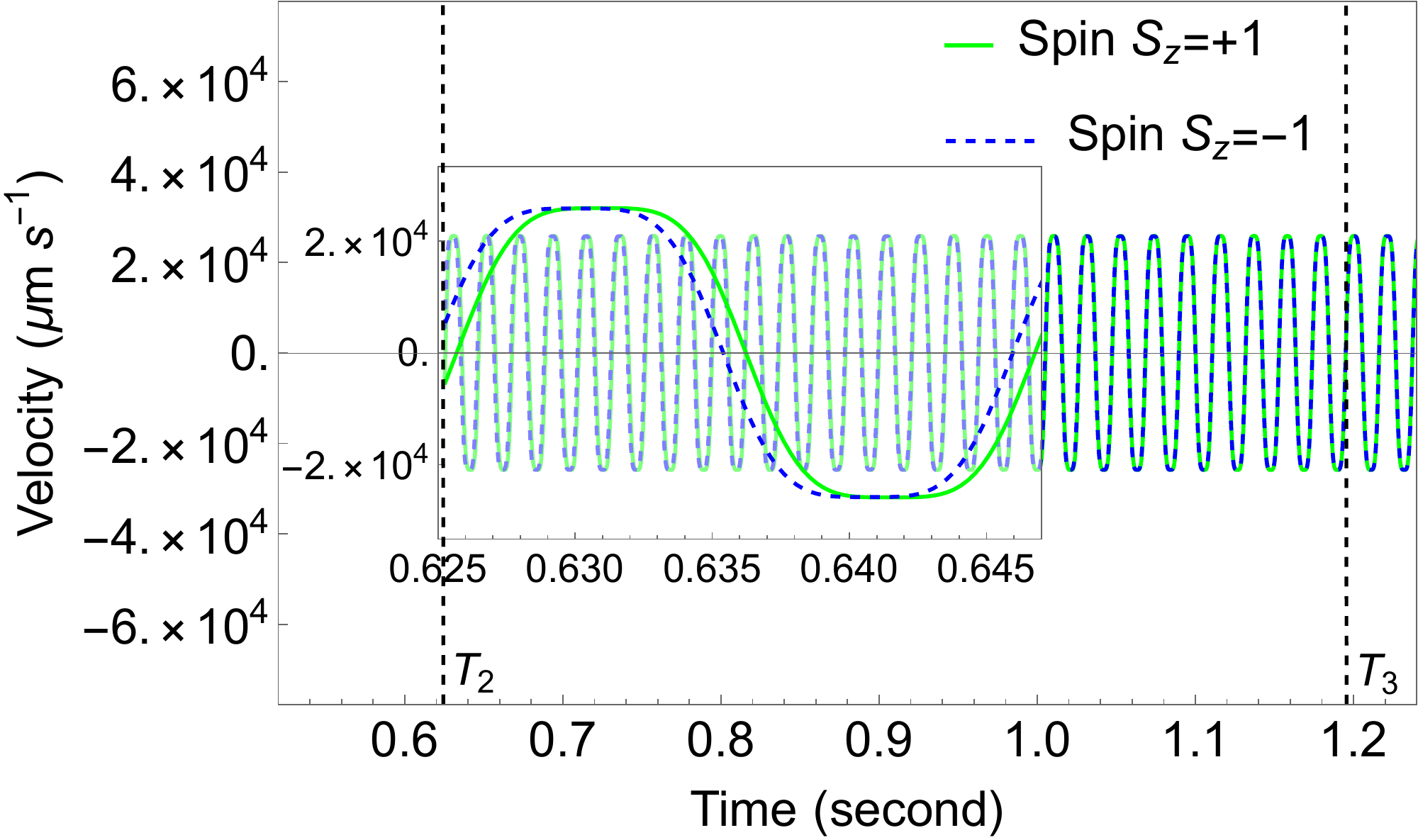}
	    \caption{Velocity - stage \Romannum{3}}
    \end{subfigure}\\
	\vspace{0.2cm}
    \centering
    \begin{subfigure}{0.325\linewidth}
    	\centering
    	\includegraphics[width=0.9\linewidth]{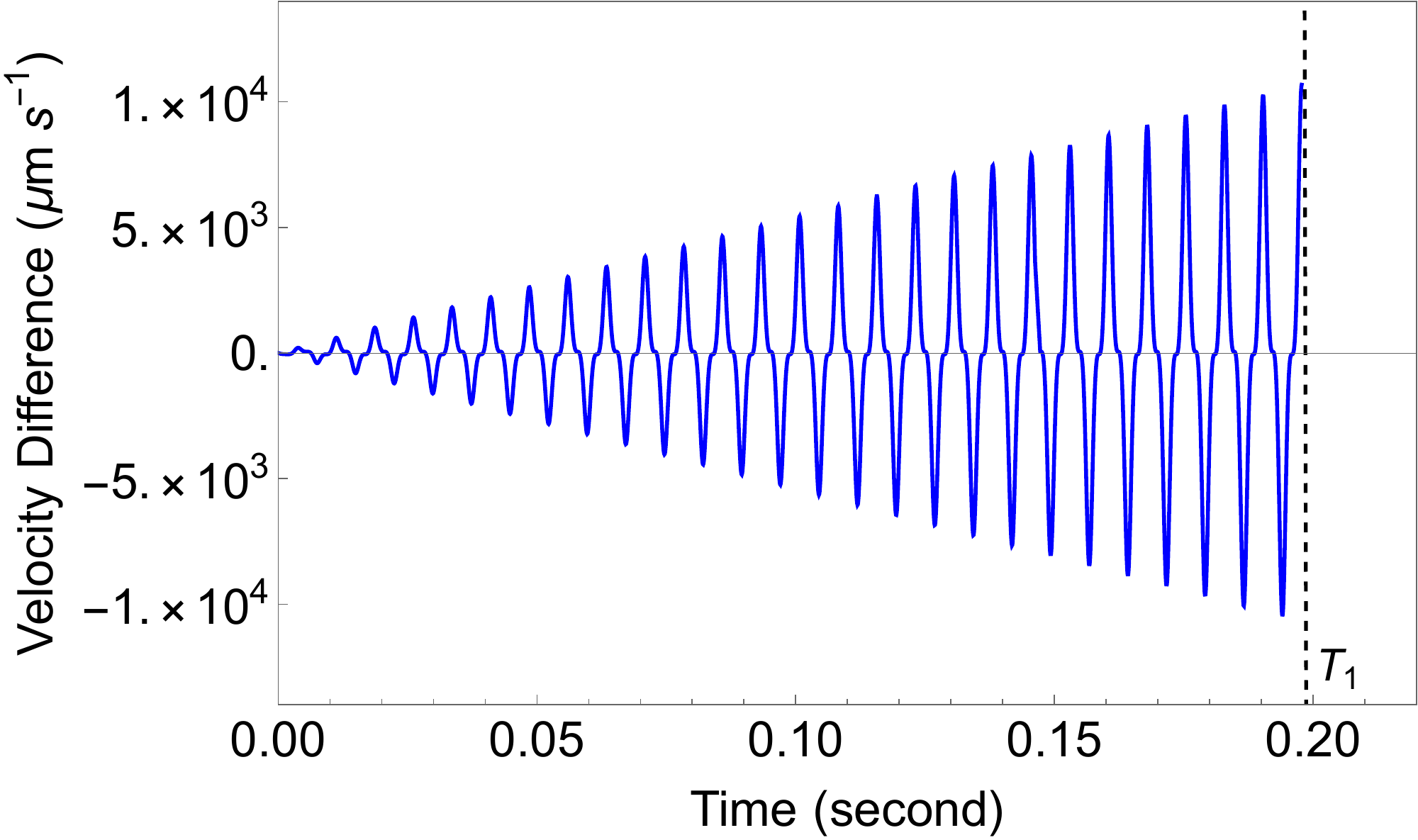}
    	\caption{Velocity difference - stage \Romannum{1}}\label{VelocityDifference-Mass17-1}
    \end{subfigure}
    \centering
    \begin{subfigure}{0.325\linewidth}
    	\centering
    	\includegraphics[width=0.9\linewidth]{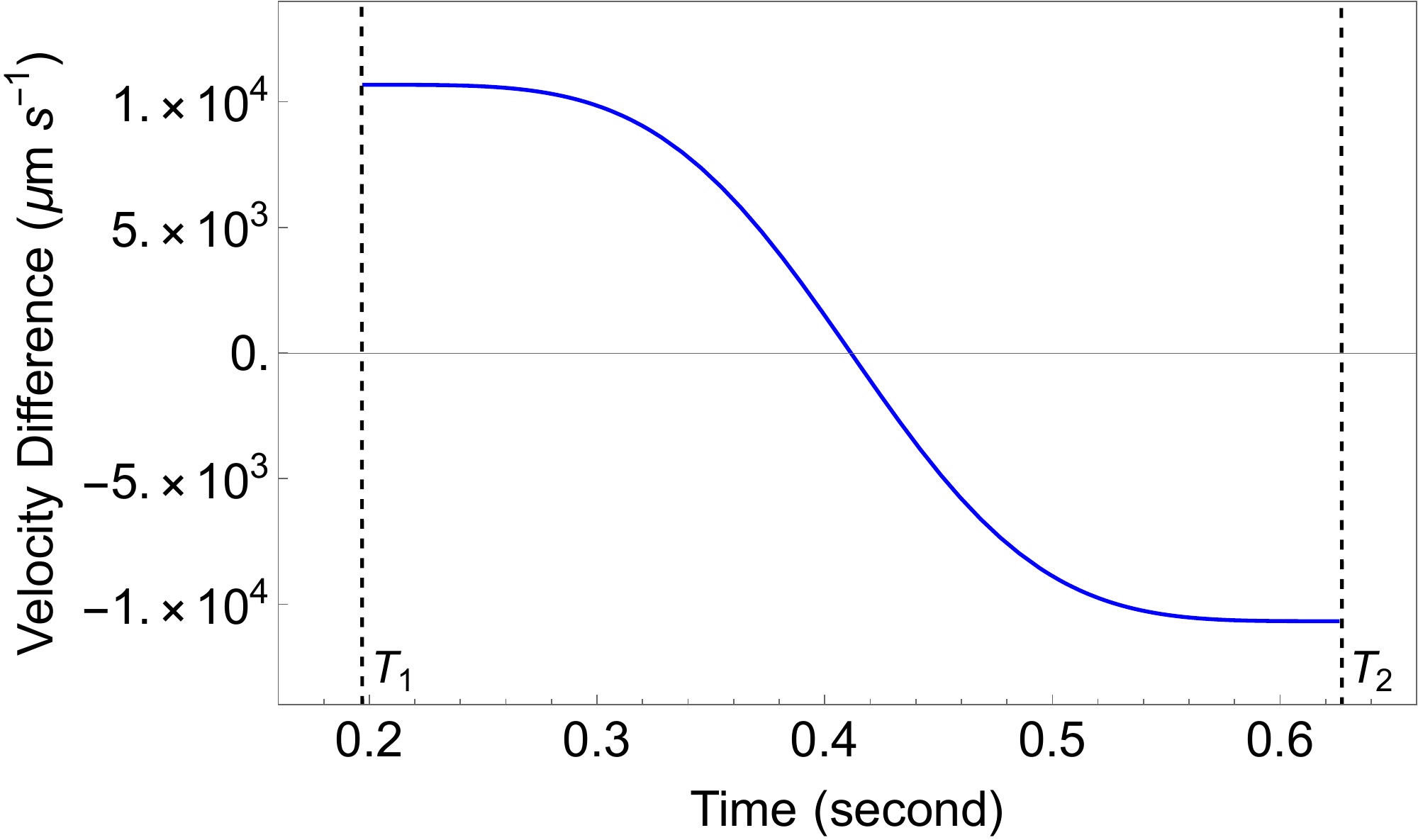}
    	\caption{Velocity difference - stage \Romannum{2}}
    \end{subfigure}
    \centering
    \begin{subfigure}{0.325\linewidth}
    	\centering
    	\includegraphics[width=0.9\linewidth]{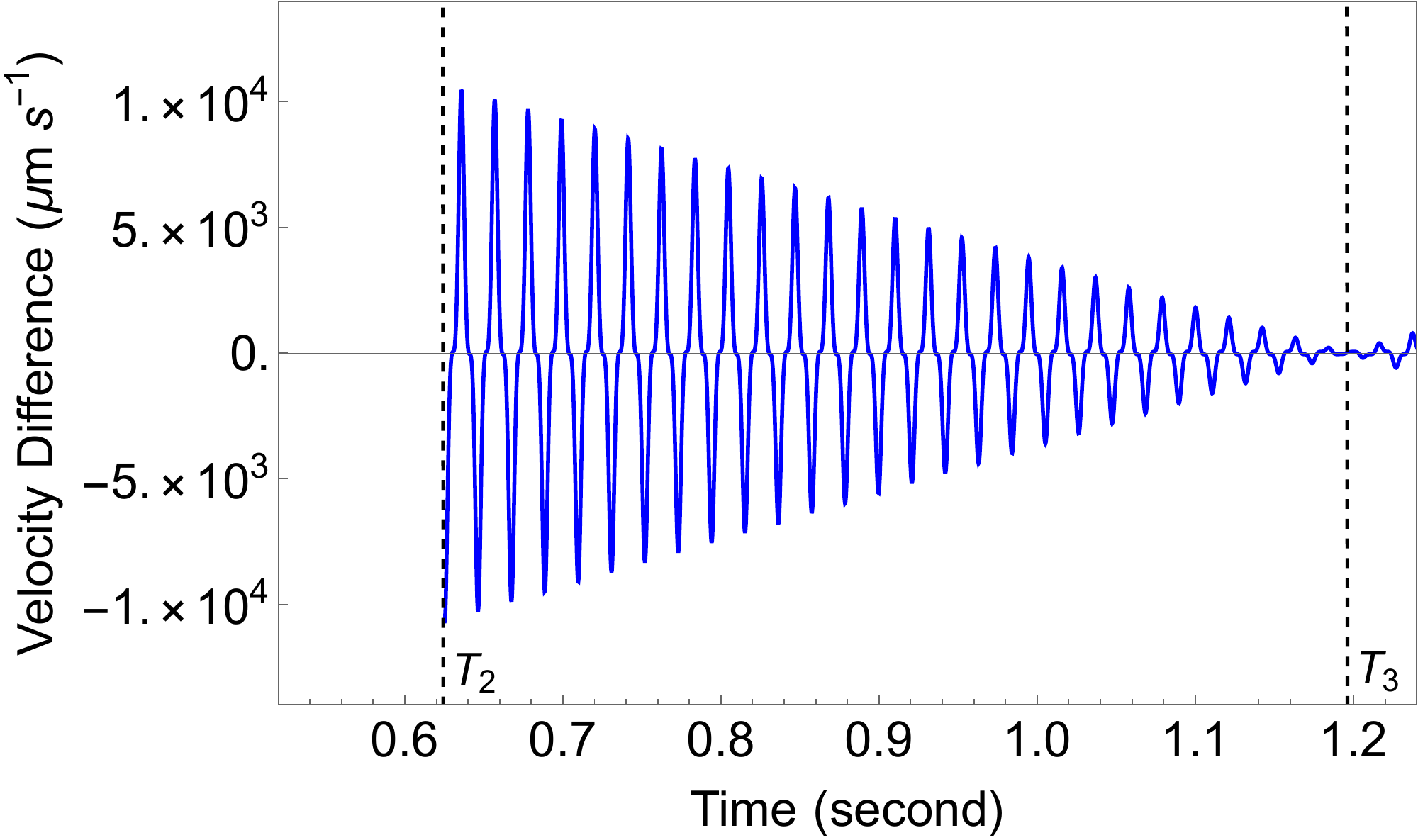}
    	\caption{Velocity difference - stage \Romannum{3}}\label{VelocityDifference-Mass17-3}
    \end{subfigure}\\
	\vspace{0.2cm}
    \centering
    \begin{subfigure}{0.325\linewidth}
    	\centering
    	\includegraphics[width=0.9\linewidth]{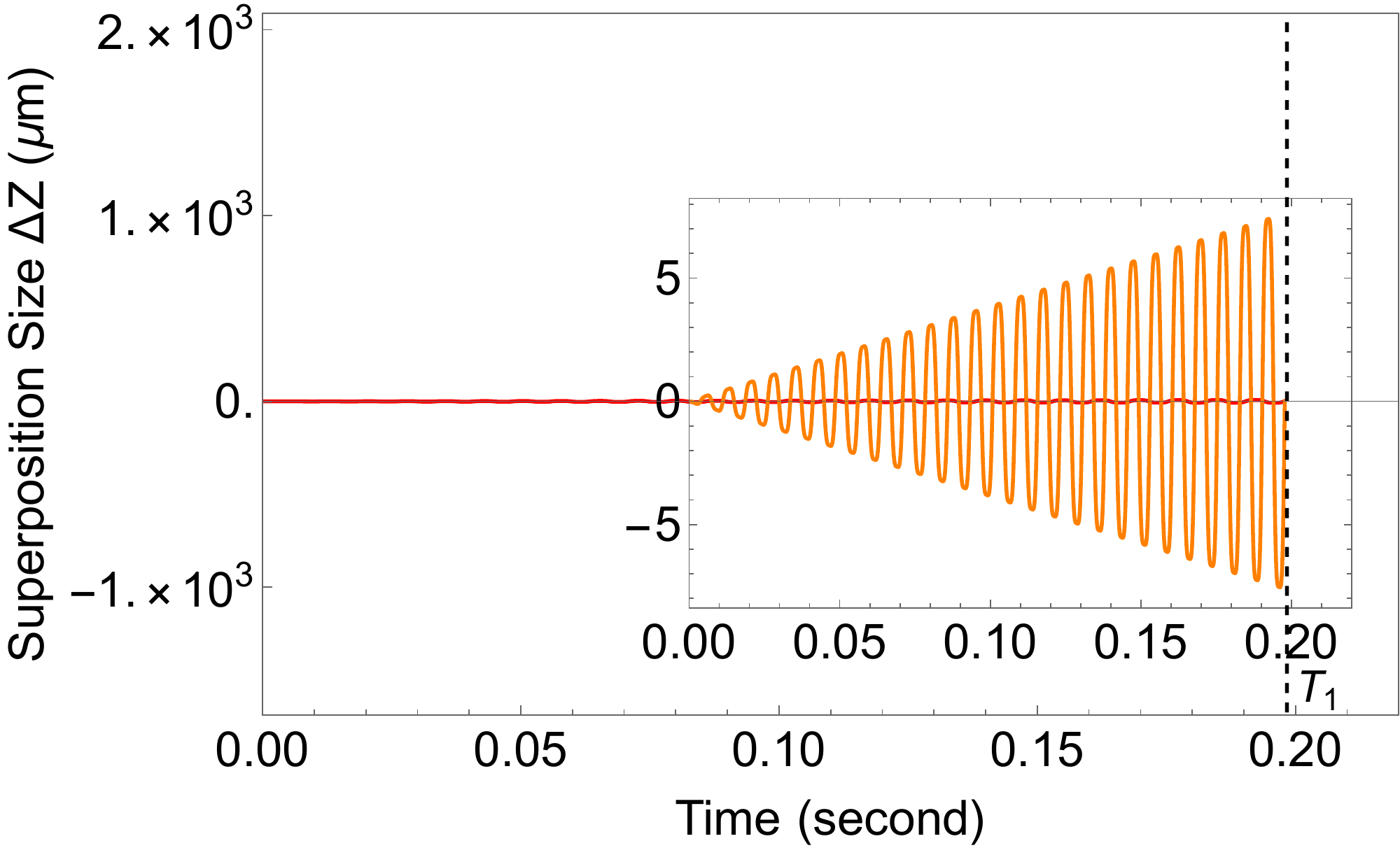}
    	\caption{Superposition size - stage \Romannum{1}}\label{SuperpositionSize-Mass17-1}
    \end{subfigure}
    \centering
    \begin{subfigure}{0.325\linewidth}
    	\centering
    	\includegraphics[width=0.9\linewidth]{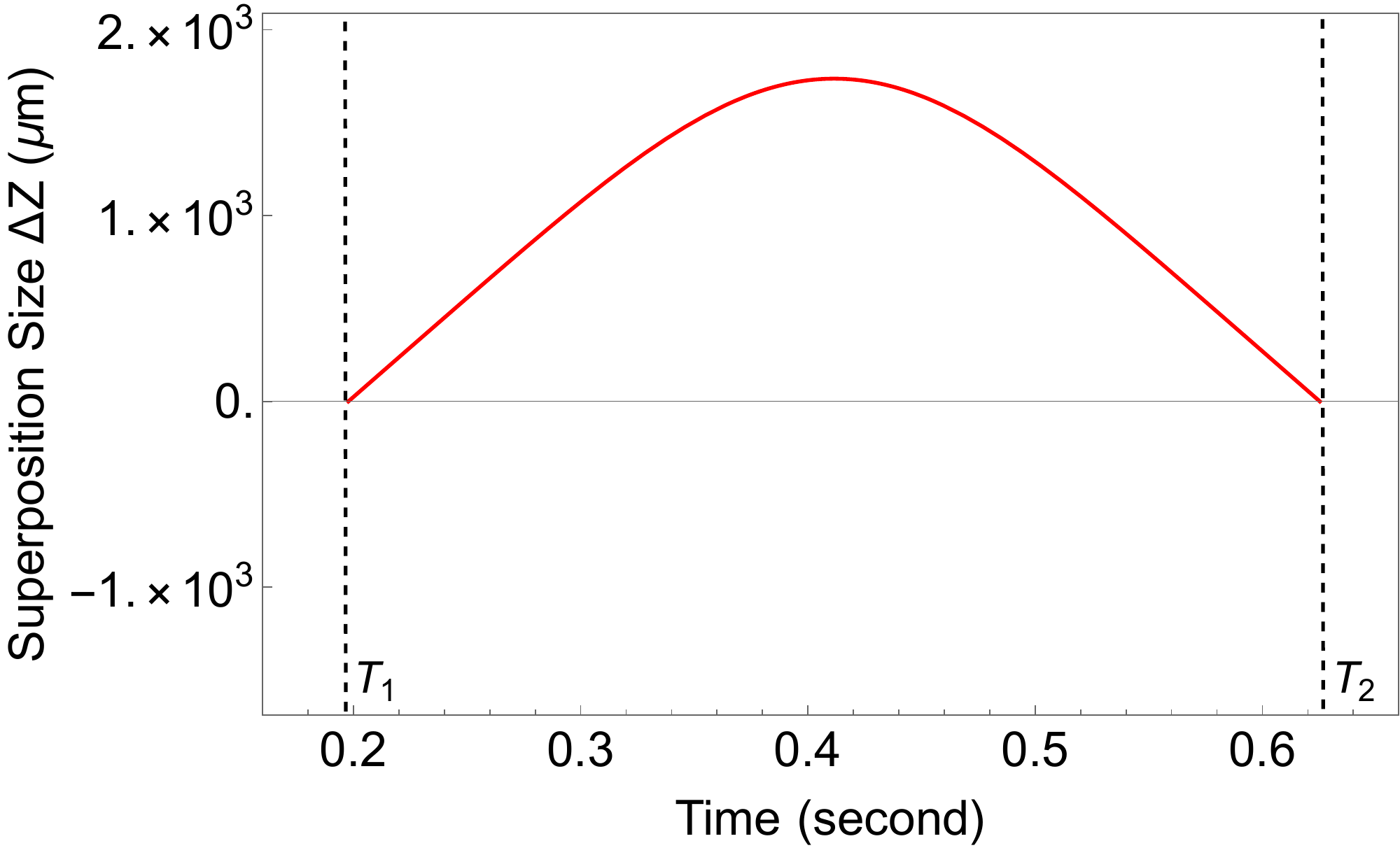}
    	\caption{Superposition size - stage \Romannum{2}}\label{SuperpositionSize-Mass17-2}
    \end{subfigure}
    \centering
    \begin{subfigure}{0.325\linewidth}
    	\centering
    	\includegraphics[width=0.9\linewidth]{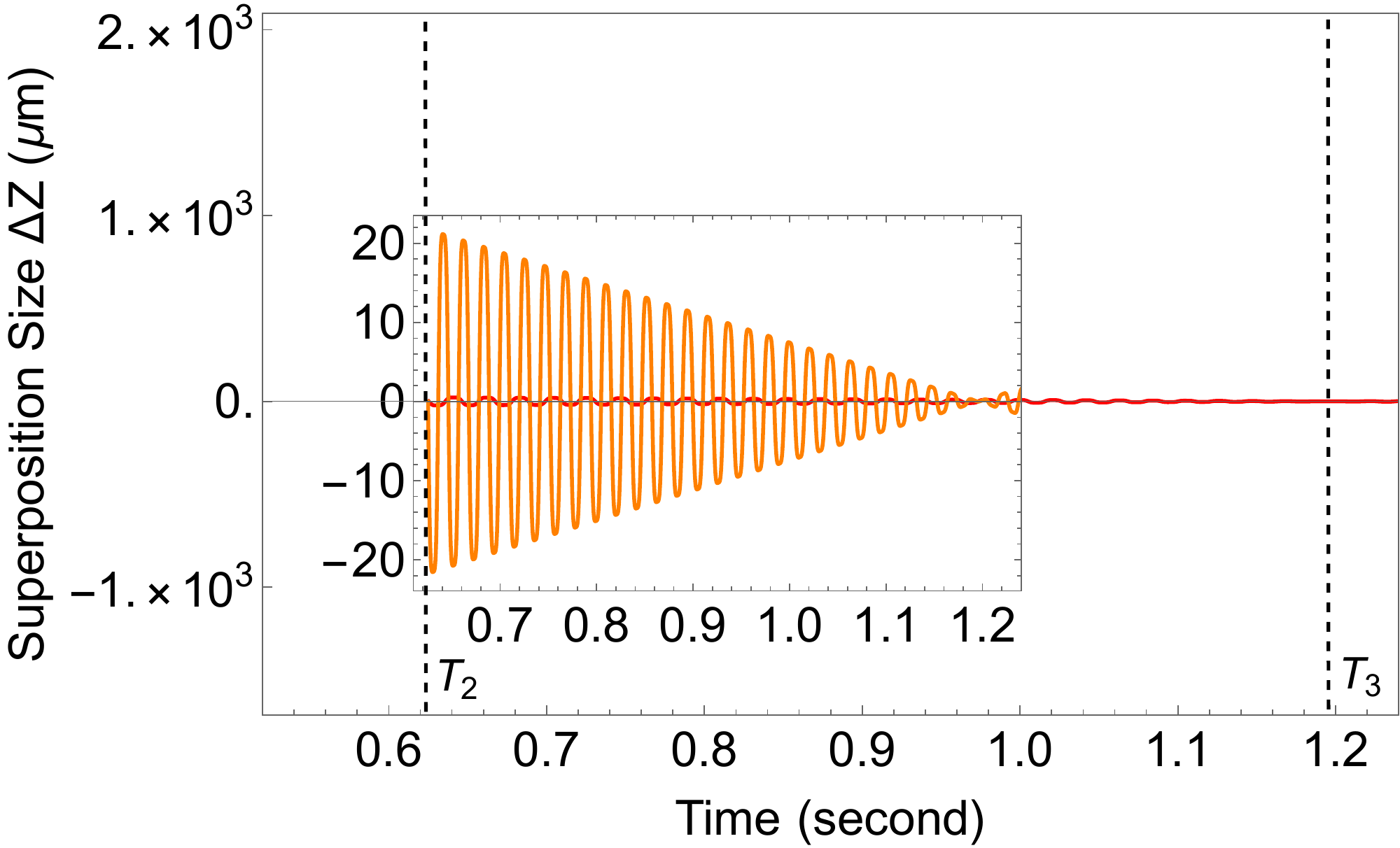}
    	\caption{Superposition size - stage \Romannum{3}}\label{SuperpositionSize-Mass17-3}
    \end{subfigure}
    	\caption{We show the dynamical aspects for the mass $m=10^{-17}$ kg, the magnetic field coordinates (potential coordinates) experienced, the velocities, the velocity differences, and the superposition size during the three experimental stages. We set different values of $\eta$ and the initial position of the wave packet in the magnetic field at different stages. Stage I, $\eta=1\times 10^{8}~\text{T}~\text{m}^{-2}$, with an initial coordinate $z=100~\mu \text{m}$. Stage II, $\eta=1\times 10^{5} ~\text{T}~\text{m}^{-2}$, with an initial coordinate $z=0~\mu \text{m}$. Stage III, $\eta=3.4\times 10^{7} ~\text{T}~\text{m}^{-2}$, with an initial coordinate $z=-102.8~\mu \text{m}$. The initial coordinates here refer to the  initialization coordinates of the NV center in the magnetic field at different experimental stages. Times $T_{1}$ and $T_{2}$ are determined by constraining the moment when the superposition size is zero (with an accuracy of $10^{-6}~\mu \text{m}$). Time $T_{3}$ is the moment when the velocity difference between two wave packets and the superposition size are zero.}\label{M17-FiveTypesFigures}
\end{figure*}

	We will now discuss the trajectories of the wave packets. Let us first consider the case where the nano-crystal has a mass of $10^{-17}$ kg. Similar analysis will arise for all the other masses under consideration. We will discuss the implementation in three stages. The purpose of the first stage is to obtain a large velocity difference between the two trajectories in a short time (around, say  $0.2$~s) by applying  ($\eta = 1\times 10^{8} ~\text{T}~\text{m}^{-2}$)~\footnote{The actual magnetic field gradient is small, as   ${\partial B}/\partial z\sim \eta z$, and $z\ll 1$m.  For  $z\sim 100{\rm \mu m}$, the maximum magnetic field gradient will be then ${\partial B}/\partial z\sim \eta z\sim 10^{4}~\text{T}~\text{m}^{-2}$. These values of the magnetic field gradient can be achievable in a labortaory~\cite{Machluf,Modena}.}. The purpose of the second stage is to generate, and then close, a large spatial superposition of the two trajectories. To do this we decrease the acceleration of the wave packets by decreasing the value of $\eta$. By adjusting the value of $\eta$ to an appropriate value, we can get a large superposition size in a relatively short coherent time scale (about $\sim 1$~s)~\cite{Bar-Gill,Abobeih}. When the spatial positions of the two wave packets coincide again, we begin the third stage. We then adjust the magnetic field gradient and the position of the particle in the potential energy. Doing so carefully will close the
	interferometer, bringing the spatial and momentum differences to become zero in both arms. Fig.(\ref{M17-FiveTypesFigures}) shows these steps graphically. The behaviour of the nano-crystals in the magnetic field for masses of $10^{-16} ~\text{kg}$ and $10^{-15} ~\text{kg}$ is very similar to that of $10^{-17} ~\text{kg}$, and we have included the numerical results of both the cases in the Appendix (\ref{CatapultingProcess}). We can also consider these stages in more detail
	\begin{itemize}
		\item Initialization, $t = 0$: The wave packet enters the inhomogeneous magnetic field region. The initial velocity along the $z$-direction is $\dot z(0)= 0$, and the initial position $z_{0}=0$. The parameter  $\eta= 1\times 10^{8}~ \text{T}~\text{m}^{-2}$. 
		
		\item Stage-\Romannum{1}, $0< t <T_{1}$: The two wave packets oscillate rapidly in the magnetic field, and the velocity difference between them grows larger and larger, see (Fig.(\ref{PotentialCoordinates-Mass17-1}) and Fig.(\ref{VelocityDifference-Mass17-1})). The magnetic field gradient is $\eta z \approx -1\times 10^{4}~\text{T}\text{m}^{-1}$ at $t=T_{1}$, where $\eta=1\times 10^{8}~\text{T}~\text{m}^{-2}, z\approx -1\times 10^{-4}~\text{m}$.
		
		\item At $t=T_{1}$: Since the difference in the change in the spatial position  (Fig.(\ref{SuperpositionSize-Mass17-1})) between the two wave packets is opposite to that of the difference in the velocity, it is possible to find a moment when the velocity difference is large enough and the superposition size is 0. This moment is marked $T_{1}$. At time $T_{1}$, if we can adjust the magnetic field such that the coordinate of the two wave packets in the magnetic field is $z= 0$, see (Fig.(\ref{PotentialCoordinates-Mass17-2})), where we find the value of $\eta\sim 1\times 10^{5}~\text{T}~\text{m}^{-2}$, and $\partial B_z=\eta z=0$ at  $z=0$.
		
		\item Stage-\Romannum{2}, $T_{1}< t< T_{2}$: In this stage, the two wave packets have different initial velocities at the new initial potential position, which is equivalent to ejecting the two wave packets away from each other. Moreover, due to the reduction in the magnetic field gradient, the spatial position difference between the two wave packets can be significantly increased, see (Fig.(\ref{SuperpositionSize-Mass17-2})).
		
		\item At $t=T_{2}$: The two wave packets meet again after a half-time period of motion, and we mark the time of their meeting as $T_{2}$, see (Fig.(\ref{PotentialCoordinates-Mass17-2})). At time $T_{2}$, we need to adjust the coordinates of the wave packet in the magnetic field and select an appropriate value of $\eta$, so that the trajectory of the two wave packets can be closed in a relatively short time (about $\sim 1$ s). At time $T_{2}$, the magnetic field of the wave packet is adjusted to $z_{T_{2}}=-102.8~ \mu \text{m}$. Parameter $\eta$ is adjusted to $3.4\times 10^{7} ~\text{T}~\text{m}^{-2}$.
		
		\item Stage-\Romannum{3}, $T_{2}< t< T_{3}$: The two wave packets still oscillate rapidly in the magnetic field, but their spatial position difference and velocity difference will be smaller and smaller, see (Fig.(\ref{VelocityDifference-Mass17-3}) and Fig.(\ref{SuperpositionSize-Mass17-3})).
		
		\item At $t=T_{3}$: The trajectories of the two wave packets are closed. The superposition size is zero and the velocity difference is zero, see (Fig.(\ref{Recombined-Mass17})). We will analyse this case separately in the context of spin coherence.
		
	\end{itemize}

In this experimental setup, the magnetic field was changed twice, but we did not consider the effect of the change in the magnetic field here. Note that the magnetic field changes in a short time, so it will only cause a small disturbance to the wave packet trajectory, but we are here taking this effect to be negligible. In fact, we can introduce switching functions of the magnetic field, for example, as discussed in \cite{Marshman:2021wyk}, and our main results will not be adversely affected. It is worth noting that the parameter $\eta$ used in stage \Romannum{1} is as high as $10^{8}~\text{T}~\text{m}^{-2}$, the corresponding maximum magnetic field gradient is $\partial B_z\sim 10^{4}~\text{T}\text{m}^{-1}$ (which can be achieved in the laboratory \cite{Mamin2012}), since the maximum magnetic field experienced by the wave packet is only $100~{\rm \mu \text{m}}$, see (Fig.(\ref{PotentialCoordinates-Mass17-1}), Fig.(\ref{PotentialCoordinates-Mass16-1}), Fig.(\ref{PotentialCoordinates-Mass15-1})) for different masses.

Since we initialize the coordinates of the NV centre in the magnetic field in stage $\text{\Romannum{2}}$ and stage $\text{\Romannum{3}}$, the coordinates of the NV centre in the magnetic field are inconsistent with their spatial coordinates in the latter two stages. To avoid confusion, unless otherwise specified, the coordinates mentioned in this paper refer to the coordinates of the NV centre in the magnetic field.

We have set the value of $\eta$ at the first stage to be $\eta \sim 1\times 10^{8}~\text{T}~\text{m}^{-2}$, which limits the time of the first stage to about $0.2$ s,  we set the limit the time of the second stage to about $0.5$ s, and require the trajectories of the two wave packets to be closed within $1.5$ s, then we can get the motion of the wave packets and the superposition size with different masses as shown in Figs.(\ref{M17-FiveTypesFigures}), (\ref{M16-FiveTypesFigures}) and (\ref{M15-FiveTypesFigures}).



\begin{figure*}
	\begin{subfigure}{0.325\linewidth}
	\includegraphics[width=0.9\linewidth]{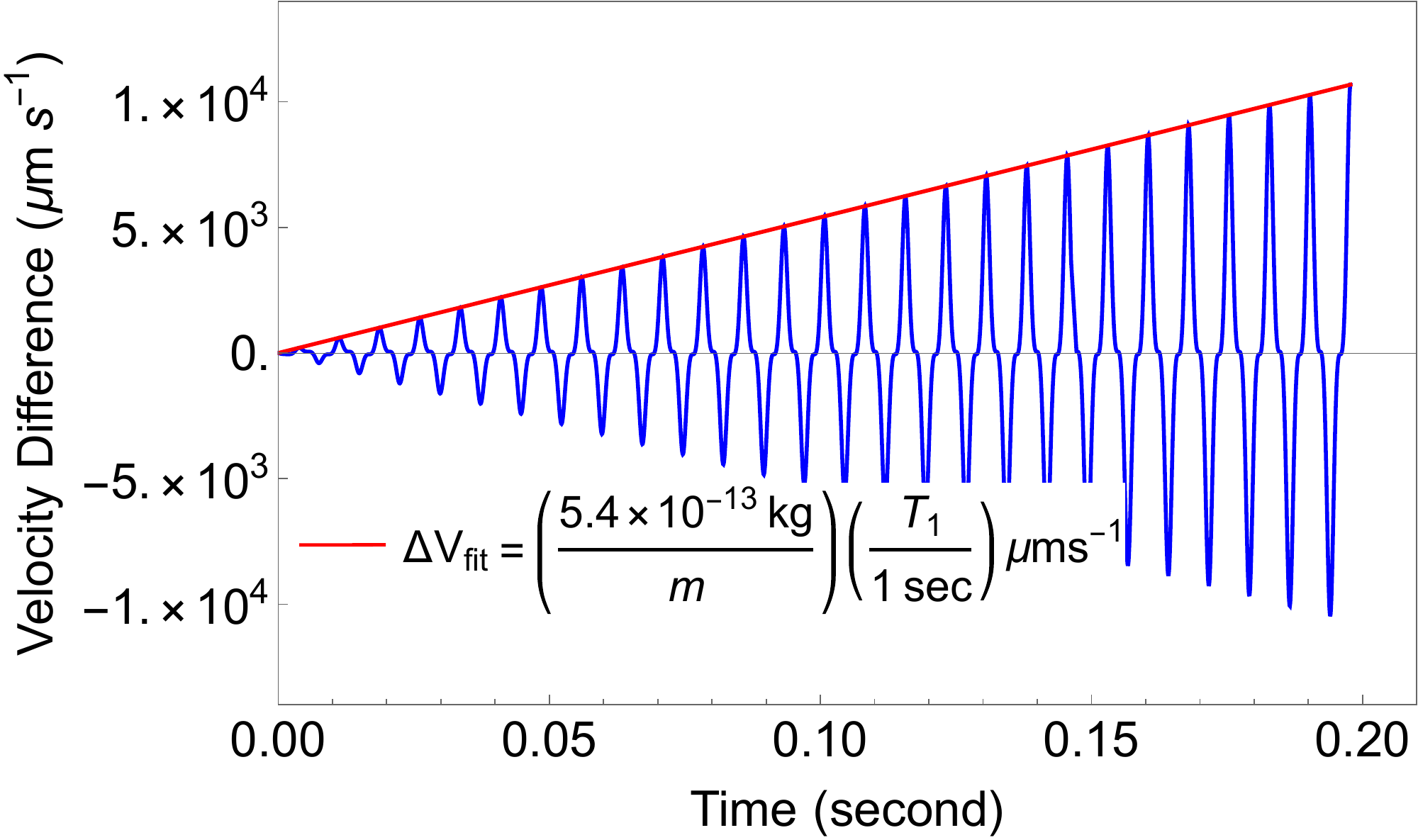}\label{VelocityDifferenceFitM17}
	\end{subfigure}
	\begin{subfigure}{0.325\linewidth}
	\includegraphics[width=0.9\linewidth]{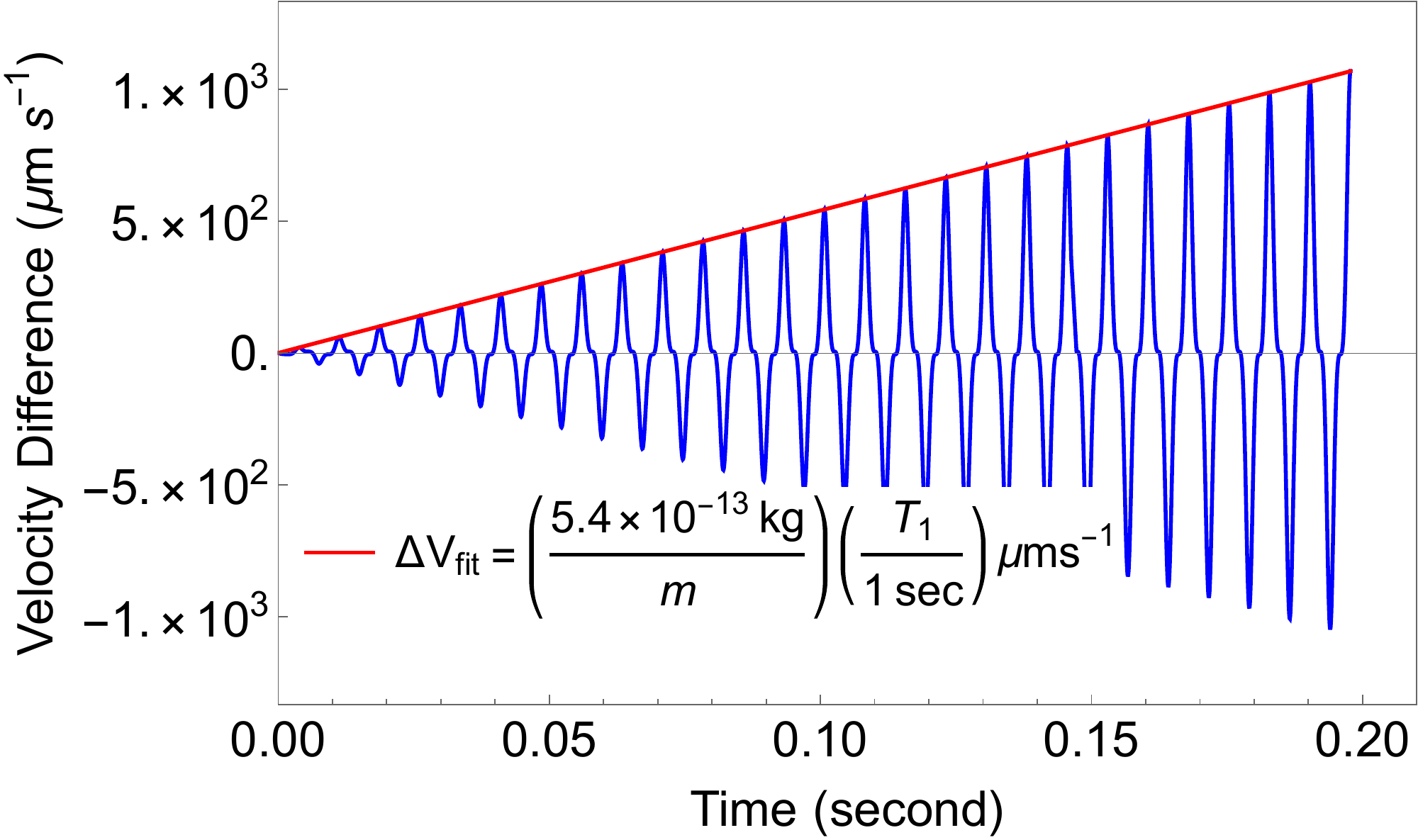}
	\end{subfigure}
	\begin{subfigure}{0.325\linewidth}
	\includegraphics[width=0.9\linewidth]{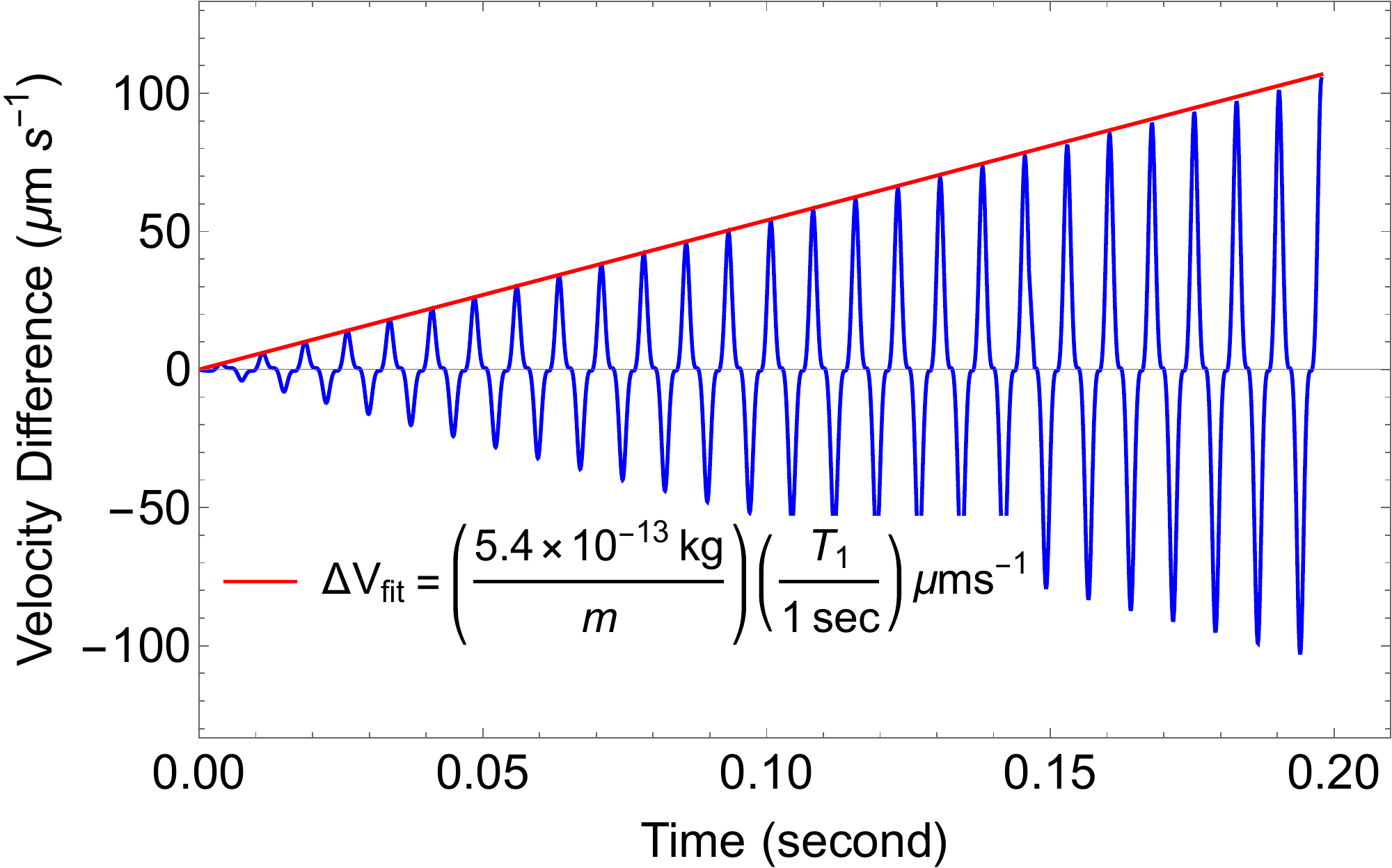}
	\end{subfigure}
	\caption{The linear fitting of the velocity difference for stage $\text{\Romannum{1}}$ under different masses. The red solid line is a linear fitting of the maximum velocity difference. $T_{1}$ is a variable here, representing the end time of the stage $\text{\Romannum{1}}$. The masses from left to right are $10^{-17}$ $\text{kg}$, $10^{-16}$ $\text{kg}$ and $10^{-15}$ $\text{kg}$ respectively. We have set $\eta=1\times 10^{8}$ $\text{T}~\text{m}^{-2}$.}\label{VelocityDifferenceFit}
\end{figure*}

\begin{figure}[t!]
	\centering
	\includegraphics[scale=0.3]{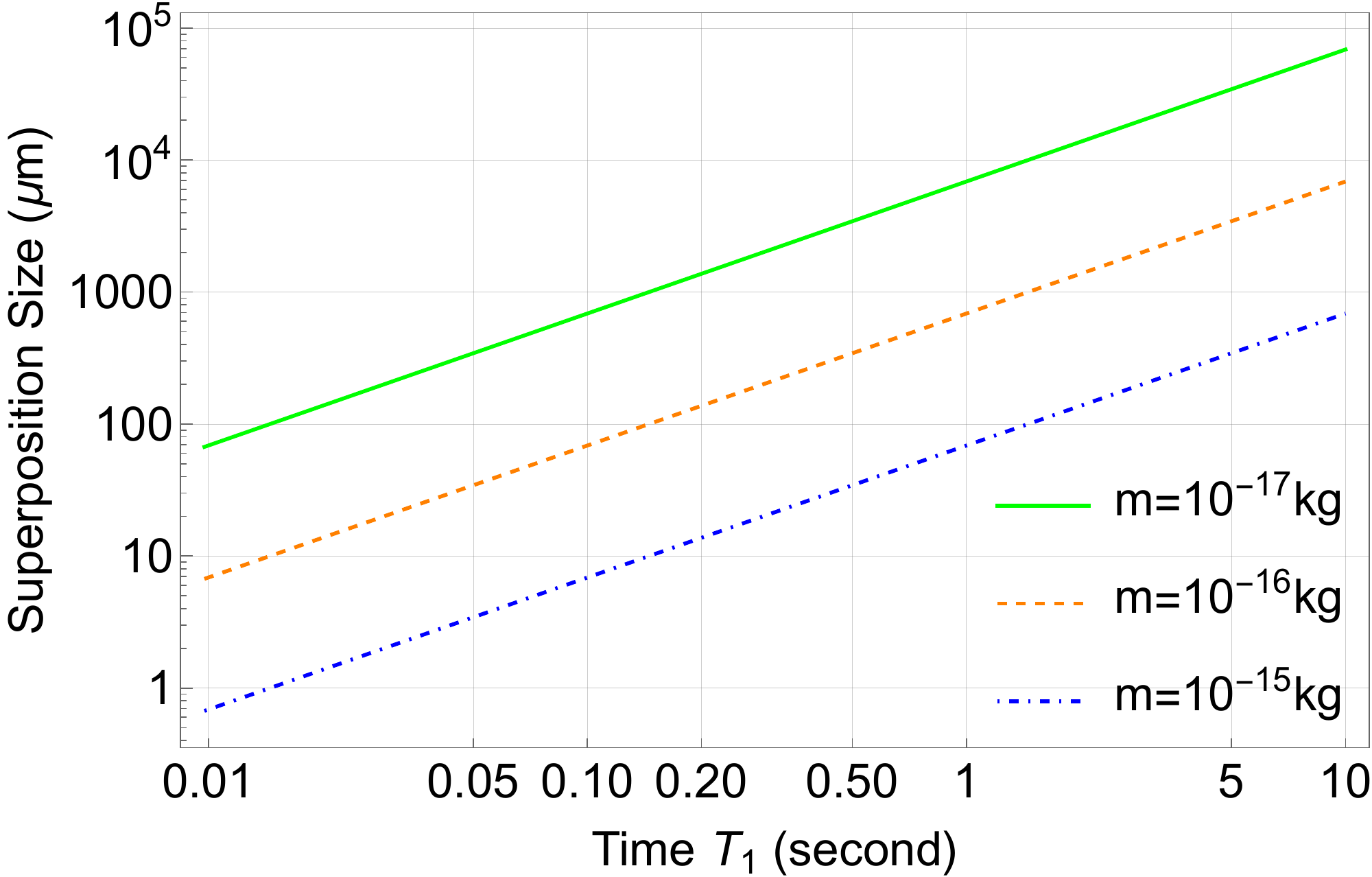}
	\caption{The scaling behavior of the superposed size (stage \Romannum{2}) under different masses obtained by catapulting the trajectories. The maximum superposition size we can achieve is inversely proportional to the mass of the nano-crystal. Here $\sqrt{A}=\pi/0.4$ Hz.}\label{ScalingBehavior1}
\end{figure}

\section{Scaling behaviour}

	We can use Eq.(\ref{Acceleration2}) to roughly analyse the motion of the wave packet for any masses. By substituting the values of each physical quantity into Eq.(\ref{Acceleration2}), it can be shown that when the mass is greater than $m\geq 10^{-17}$ kg, the motion of the wave packet is mainly dominated by $\eta z^{2}$ in the first term on the right-hand side of the equation. This term has nothing to do with mass; that is, as long as the values of $\eta$ and initial position are determined, the maximum velocity of the wave packet can be determined. This result can be seen in Fig.(\ref{Velocity-Mass17-1}), (\ref{Velocity-Mass16-1}) and (\ref{Velocity-Mass15-1}). The velocity difference between the two wave packets is caused by the second term on the right of Eq.(\ref{Acceleration2}). The value of this term is inversely proportional to the mass. Since the size of the velocity difference determines the superposition size, the maximum superposition size we can achieve should also be inversely proportional to the mass, i.e. $\Delta Z \sim 1/m$. This is result also borne out from our numerical results, as shown in Fig.(\ref{SuperpositionSize-Mass17-2}), (\ref{SuperpositionSize-Mass16-2}) and (\ref{SuperpositionSize-Mass15-2}). 
		
	It should be noted that in order to compare the behaviour of the wave packets with different masses, the values of $\eta$ and initial position are chosen properly at different stages of the experiment, but this does not mean that the wave packets with different masses we can only take these parameter values. For example, if we don't limit the time of the second stage of the experiment to about $0.5$ seconds, we can choose a smaller value of $\eta$ and get a larger superposition size.

	We first perform a linear fitting of the velocity difference in stage $\text{\Romannum{1}}$. The fitting formula is
\begin{equation}\label{VFit}
	\Delta V_{\text{fit}}=\left(\frac{5.4\times 10^{-13}~\text{kg}}{m}\right)\left(\frac{T_{1}}{1\, \text{sec}}\right)10^{-6}~\text{m}~s^{-1},
\end{equation}
where $\Delta V_{\text{fit}}$ is the maximum velocity difference reached in stage $\text{\Romannum{1}}$. $T_{1}$ is a variable here, representing the end time of the stage $\text{\Romannum{1}}$. The values of mass m are $10^{-17}~\text{kg}$, $10^{-16}~\text{kg}$ and $10^{-15}~\text{kg}$ respectively. The linear fitting results are shown in Fig.(\ref{VelocityDifferenceFit}).

Next, we take the velocity difference obtained in stage $\text{\Romannum{1}}$ as the initial velocity of the wave packet to study the trajectory of the wave packet in stage $\text{\Romannum{2}}$. In Sec.(\ref{SpinCoherence}) and Appendix (\ref{MagneticFieldControl}) we have discussed the fitting of wave packet trajectories. Now we only need to move the simple harmonic motion (Eq.(\ref{MotionEquationofSpatial})) by the $-\pi /2$ phase to fit the trajectory of the wave packet with the initial velocity $V_{\text{fit}}$ and initial position $z=0$. The fitting formula of wave packet trajectory in stage $\text{\Romannum{2}}$ is

\begin{equation}\label{TrajectoryFit}
	z(t)=\Delta Z_{0}\cos(\sqrt{A}t-\frac{\pi}{2}),
\end{equation}
where $\Delta Z_{0}$ and $\sqrt{A}$ are the amplitude and frequency of the wave packet motion, respectively. The specific expression of $A$ is found later, in Eq.(\ref{frequency}). Combining Eq.(\ref{VFit}) and Eq.(\ref{TrajectoryFit}), gives the amplitude, and thus maximum superposition size, as
\begin{equation}\label{amplitude}
	\Delta Z_{0}=\frac{\Delta V_{\text{fit}}}{\sqrt{A}}=\left(\frac{5.4\times 10^{-13} \text{kg}}{m}\right)\left(\frac{1\, \text{Hz}}{\sqrt{A}}\right)\left(\frac{T_{1}}{1\, \text{sec}}\right)10^{-6}\text{m}.
\end{equation}
For the sake of discussion, let $\sqrt{A}=(\pi/0.4)$Hz. That is, the half period of wave packet motion is 0.4 s, which is consistent with the time set in stage $\text{\Romannum{2}}$ of the experiment in this paper. It can be seen from Eq.(\ref{amplitude}) that the maximum superposition size of the ejection trajectory is inversely proportional to the mass of the nano-crystal. The scaling behaviour of the superimposed size is shown in Fig.(\ref{ScalingBehavior1}).

This formula is similar to Eq.(11) in~\cite{Marshman:2021wyk}. In Ref.~\cite{Marshman:2021wyk}, we only considered the magnetic field gradient, so there is only one parameter in the equation. In our case, we need to consider the initial velocity of the wave packet in addition to the gradient, so we need to add a parameter $T_1$, which represents the initial velocity of the wave packet. If we fix the value of $T_1$, that is, the magnitude of the initial velocity, then our
expression is the same as Eq.(11) of \cite{Marshman:2021wyk}. However, since we have two parameters to play with, we obtain a larger size of superposition compared to \cite{Marshman:2021wyk}. For instance, for $m=10^{-15}~\text{kg}$, we can obtain $\Delta Z= 16~{\rm \mu \text{m}}$ with our current proposal in total time of flight roughly $1.4$ seconds.
 In \cite{Marshman:2021wyk}, we had obtained $\Delta Z=0.11~{\rm \mu \text{m}}$ for the same time period.

When discussing the scaling behaviour of the superposition size by numerically fitting the velocity difference in stage $\text{\Romannum{1}}$ and the trajectory of the wave packet in stage $\text{\Romannum{2}}$, we should make the following points

\begin{itemize}
	\item The maximum velocity difference achieved in stage $\text{\Romannum{1}}$ is determined by the gradient parameter $\eta$ and the initial position of the nano-crystal of the stage $\text{\Romannum{1}}$. The greater the value of $\eta$ and initial position, the greater the maximum velocity difference. The fitting formula for the velocity difference in Eq.(\ref{VFit}) only holds for $\eta=1\times 10^{8}~\text{T}~\text{m}^{-2}$, initial position = 100$~\mu \text{m}$.
  \item Time $T_{1}$ in Eq.(\ref{VFit}) is a variable, with an upper bound determined by both the maximum velocity difference and the mass of the nano-crystal. In this paper, the maximum velocity difference that can be achieved in stage $\text{\Romannum{1}}$ is about $1.4\times 10^{5}~\mu \text{m}~\text{s}^{-1}$. Therefore, the upper bound on $T_{1}$ is about 2.6 $\text{s}$, 26 $\text{s}$ and 260 $\text{s}$ for masses equal to $10^{-17}~\text{kg}$, $10^{-16}~\text{kg}$, $10^{-15}~\text{kg}$ respectively.
  \item As can be seen from Eq.(\ref{amplitude}), by decreasing the value of $\sqrt{A}$, we can get a larger superposition size, but correspondingly, we also need a longer time to close the wave packet trajectory.
  \item In Eq.(\ref{VFit}), the reason for using the velocity difference obtained in stage $\text{\Romannum{1}}$ to calculate the amplitude directly is that the velocity of the two wave packets corresponding to the linear fitting velocity difference in stage $\text{\Romannum{1}}$ is almost the same with the direction being opposite. The amplitude calculated from the velocity difference is equal to the sum of the amplitudes of the two wave packets, and gives the maximum superposition size.
\end{itemize}

	
	\section{Recovering spin coherence}\label{SpinCoherence}

	The SGI splits the two wave packets in the superposition state. First, they lose their spin coherence and then recombine to recover the spin coherence. We first use the definition of the spin coherence given in Ref.~\cite{Schwinger1988} to calculate the expression of the spin coherence in the case of our magnetic field profile and then study what experimental conditions are needed to recover the spin coherence. Heisenberg's equation of motion is given by
	\begin{equation}\label{HeisenbergEquation}
		i \hbar \frac{d \hat{A}_{H}(t)}{d t}=\left[\hat{A}_{H}(t), \hat{H}_{H}(t)\right],
	\end{equation}
where $\hat{A}_{H}(t)$ and $\hat{H}_{H}(t)$ are the Hermitian and the Hamiltonian operator in the Heisenberg picture, respectively. Using Eq.(\ref{HeisenbergEquation},) we can get the equation of motion for the position 
	\begin{align}\label{Position}
		\frac{d \hat{\boldsymbol{r}}(t)}{d t}&=\frac{1}{i \hbar}\big[\hat{\boldsymbol{r}}(t), \hat{H}_{H}(t)\big] 
		=\frac{\boldsymbol{p}(t)}{m},
	\end{align}
and the equation of motion for the momentum
	\begin{align}\label{Momentum}
		\frac{d \hat{\boldsymbol{p}}(t)}{d t}&=\frac{1}{i \hbar}\big[\hat{\boldsymbol{p}}(t), \hat{H}_{H}(t)\big] \nonumber\\
		&=\frac{\chi_{m} m}{2 \mu_{0}} \nabla \big(\boldsymbol{B}(\boldsymbol{r}(t))^{2}\big) + \nabla \big(\boldsymbol{\mu}(t)\cdot \boldsymbol{B}(\boldsymbol{r}(t))\big).
	\end{align}
	By integrating Eq.(\ref{Position}) and (\ref{Momentum}), the formal solution of the evolution of the position and the momentum with time can be written as \cite{Schwinger1988}
	\begin{widetext}
	\begin{align}
		\boldsymbol{p}(t)&=\boldsymbol{p}_{0} + \int_{0}^{t}\left(\frac{\chi_{m} m}{2 \mu_{0}} \nabla \big(\boldsymbol{B}(\boldsymbol{r}(t^{\prime}))^{2}\big) - \nabla \big(\mu_{B}\boldsymbol{\sigma}\cdot \boldsymbol{B}(\boldsymbol{r}(t^{\prime}))a^{\dagger}(t^{\prime}) a(t^{\prime})\big)\right)d t^{\prime},\label{Momentum1}\\
		\boldsymbol{r}(t)&=\boldsymbol{r}_{0} + \frac{t}{m}\boldsymbol{p}_{0} + \frac{1}{m} \int_{0}^{t} (t-t^{\prime})\left(\frac{\chi_{m} m}{2 \mu_{0}} \nabla \big(\boldsymbol{B}(\boldsymbol{r}(t^{\prime}))^{2}\big) - \nabla \big(\mu_{B}\boldsymbol{\sigma}\cdot \boldsymbol{B}(\boldsymbol{r}(t^{\prime}))a^{\dagger}(t^{\prime}) a(t^{\prime})\big)\right)d t^{\prime},\label{Position1}
	\end{align}
\end{widetext}
	where $\boldsymbol{r}_{0}$ and $\boldsymbol{p}_{0}$ are the initial position and the momentum respectively, $\boldsymbol{\sigma}$ is the spin operator and $a^{\dagger}(t)$ and $a(t)$ are creation and annihilation operators with $\boldsymbol{S}(t)=\boldsymbol{\sigma}a^{\dagger}(t)a(t)$. Since the beams' trajectories are split by an inhomogeneous magnetic field along the z-direction, the wave packet motion along the z-direction is studied next. With a bit of rearrangement, the Eq.(\ref{Momentum1}) and (\ref{Position1}) becomes
	\begin{align}
		p_{z}(t)-p_{z}&= \Delta p_{z},\\
		z(t)-z_{0}-\frac{t}{m}p_{z}&=\Delta z,
	\end{align}
	where 
	\begin{align}
		\Delta p_{z}=& \int_{0}^{t}\left(\frac{\chi_{m} m}{2 \mu_{0}} \frac{\partial}{\partial z} B_{z}(t^{\prime})^{2} -\mu_{B}\sigma_{z}\frac{\partial}{\partial z}B_{z}(t^{\prime})\right)d t^{\prime},\\
		\Delta z=&\frac{1}{m} \int_{0}^{t} (t-t^{\prime})\left(\frac{\chi_{m} m}{2 \mu_{0}} \frac{\partial}{\partial z} B_{z}(t^{\prime})^{2} - \mu_{B}\sigma_{z}\frac{\partial}{\partial z}B_{z}(t^{\prime})\right)d t^{\prime},
	\end{align}
	are the variations in the z-component of the position and the momentum. $r_{z}$ and $p_{z}$ are the initial z-component of the position and the momentum, respectively. We take into account that the wave function of a massive particle localized in the position space at $t=0$ is a Gaussian wave packet
	\begin{equation}\label{InitialWaveFunction}
		\psi(z,0)=\left(\frac{1}{2 \pi \delta z^{2}}\right)^{1 / 4} e^{-\frac{z^{2}}{4 \delta z^{2}}},
	\end{equation}
	with a minimum uncertainty $\delta z\delta p_{z}=\hbar/2$. In Appendix-B, we study the evolution of the wave packet in presence of the non-linear magnetic field. We show that the expected
	value of the position of each arm of the interferometer coincides with the classical trajectories Eq.(\ref{Acceleration2}).

	 In this situation, the spin coherence can be written as \cite{Englert1988,Schwinger1988}
	\begin{align}\label{SpinCoherence1}
		\left\langle\hat{\sigma}_{x}(t)\right\rangle=\cos(\Phi(t))\exp\left(-\frac{1}{2}\left[\left(\frac{\Delta z}{\delta z}\right)^2+\left(\frac{\Delta p}{\delta p}\right)^2\right]\right),
	\end{align}
	where
	\begin{align}
		\Phi(t)=\frac{1}{\hbar}g\mu_{B}\int_{0}^{t}B_{z}(t^{\prime})d t^{\prime},
	\end{align}
	is the accumulated Larmor precession angle. We are evaluating the expectation value with respect to the spin state $\frac{1}{\sqrt{2}}(\ket{\uparrow}_{z}+\ket{\downarrow}_{z})$. If $\Phi(t)=2n\pi$ ($n$ is an integer), $\Delta z=0$ and $\Delta p=0$, then $\left\langle\hat{\sigma}_{x}(t)\right\rangle=1$, the spin coherence is completely restored. In real experiments, we cannot control the experimental conditions with arbitrary precision. Assuming that the experimental error is bounded by certain parameters,
	given by \cite{Englert1988}
	\begin{align}\label{TolerateDeviation}
		|\delta \Phi(t)| \leqslant \varepsilon_{1}, \quad\left|\frac{\Delta z(t)}{\delta z}\right| \leqslant \frac{\varepsilon_{2}}{\sqrt{2}}, \quad\left|\frac{\Delta p(t)}{\delta p}\right| \leqslant \frac{\varepsilon_{3}}{\sqrt{2}}, 
	\end{align}
	where $\varepsilon_{1}$, $\varepsilon_{2}$, $\varepsilon_{3}$ are much less than 1, and the specific value depends on our requirements  of the experimental accuracy. Taylor expansion of Eq.(\ref{SpinCoherence1}), and then by taking the first-order term gives
	\begin{align}\label{SpinCoherence2}
		\left\langle\hat{\sigma}_{x}(t)\right\rangle&\lesssim1-\frac{1}{2}\left(|\delta \Phi(t)|^{2}+\left|\frac{\Delta z(t)}{\delta z}\right|^{2} +\left|\frac{\Delta p(t)}{\delta p}\right|^{2}\right),\nonumber \\
		&\leqslant 1-\varepsilon^{2}.
	\end{align}
	Here we assumed $\varepsilon_{1}=\varepsilon_{2}=\varepsilon_{3}=\varepsilon$ for simplify. When $\varepsilon=0.1$, the confidence level for spin coherence is 99\%. Next, we need to make an approximation for Eq.(\ref{Acceleration2}), so that we can analytically solve the equation of motion, which is convenient for estimating the trajectory deviation caused by the imprecision of magnetic field control.

We model Eq.(\ref{Acceleration2}) with the following acceleration expression 
	\begin{align}\label{Acceleration3}
		\frac{\mathrm{d}^2 z(t)}{\mathrm{d}t^2}=a_{z}=\left(C_{\text{correction}}\frac{\chi_{m}}{\mu_{0}}B_{0}-S_{z}\frac{g e \hbar }{m m_{e}}\right)\eta z,
	\end{align}
where $C_{\text{correction}}$ is a dimensionless  correction factor. The value of this correction factor is related to $\eta$. When $\eta$ is taking different values, we need to adjust the correction factor to make the approximate trajectory as close to the exact trajectory as possible. 

In the present paper, we have considered the mass $m \sim 10^{-17}~ \text{kg}, 10^{-16}~\text{kg}$ and $10^{-15}~\text{kg}$, and then substitute the values of other physical quantities into Eq.(\ref{Acceleration3}), we can find that the coefficient $\left(C_{\text{correction}}\frac{\chi_{m}}{\mu_{0}}B_{0}-S_{z}\frac{g e \hbar }{m m_{e}}\right) < 0$, which gives rise to the Harmonic oscillator equation with a solution of Eq.(\ref{Acceleration3}) is
	\begin{align}\label{MotionEquationofSpatial}
		z(t)=z_{0}\cos(\sqrt{A}t),
	\end{align}
	where $A$ is the square of frequency
	\begin{align}\label{frequency}
		A=-\left(C_{\text{correction}}\frac{\chi_{m}}{\mu_{0}}B_{0}-S_{z}\frac{g e \hbar }{m m_{e}}\right)\eta > 0,
	\end{align}
	$z_{0}$ is the amplitude. Here we select the value of $\eta$ corresponding to the second stage of the experiment to calculate the minimum accuracy required to control the magnetic field. This is because the fluctuation in the magnetic field is inversely proportional to $A$. The greater the value of $A$, the higher the accuracy requirements for the magnetic field control. And the value of $A$ is proportional to $\eta$ (Eq.(\ref{frequency})). The larger the $\eta$, the larger $A$ is. The value of $\eta$ used in stage $\text{\Romannum{2}}$ of the experiment is the smallest, so the fluctuation in the magnetic field will be the lowest at this stage, see below the expressions Eq.(\ref{accuracy-z}) and Eq.(\ref{accuracy-p}).
	
\begin{figure}[htpb]
		\centering
		\includegraphics[scale=0.3]{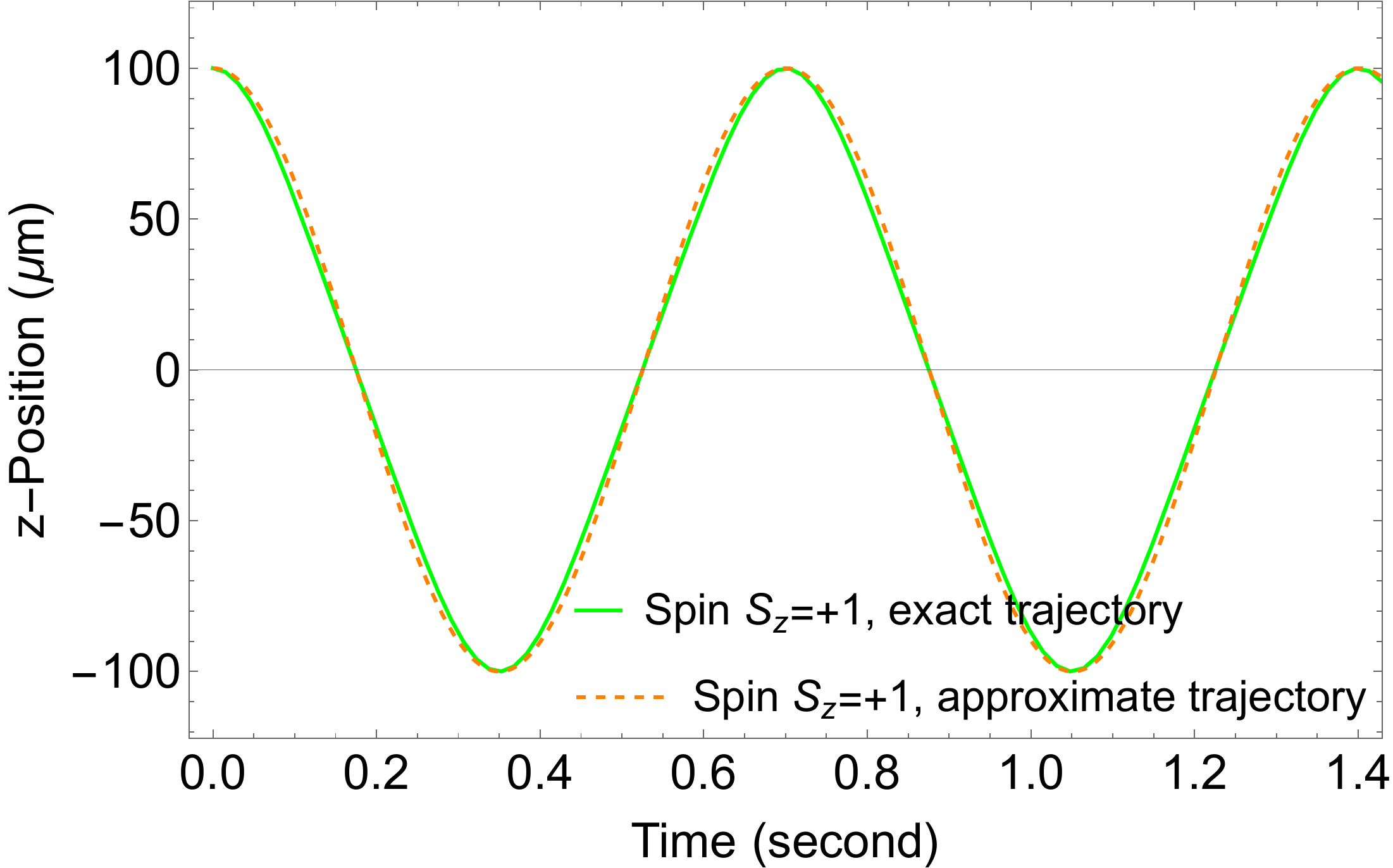}\\
		\vspace{0.3cm}
		\includegraphics[scale=0.3]{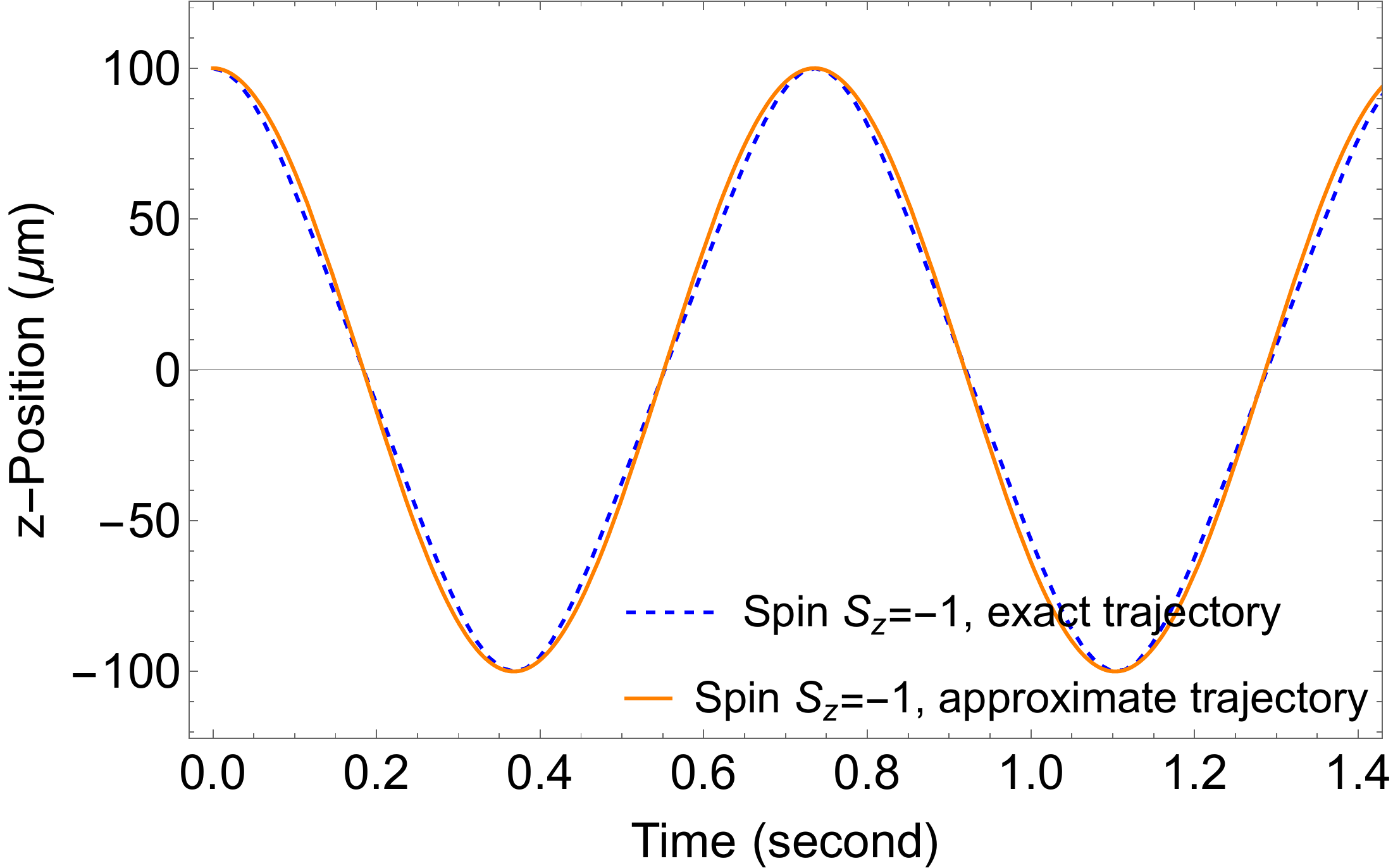}
		\caption{We have shown the comparison of the approximate and the exact trajectories of the two wave packets. Here $m=10^{-17}~ \text{kg}, \eta=1\times 10^{6}~\text{T}~\text{m}^{-2}$, $C_{\text{correction}}=27.3467$.}\label{CompareTrajectory}
	\end{figure}

For $m=10^{-17}~\text{kg}$, see Fig.~\ref{M17-FiveTypesFigures}, the value of $\eta=1\times 10^{6}~\text{T}~\text{m}^{-2}$ at stage II, and the corresponding correction factor $C_{\text{correction}}=27.3467$. The corrected approximate trajectory is compared with the exact trajectory as shown in Fig.(\ref{CompareTrajectory})~\footnote{We have performed a similar analysis for the stage I of the trajectory, where $\eta =10^{8}{\rm ~\text{T}~\text{m}^{-2}}$ for different masses, see Appendix \ref{MagneticFieldControl} for the discussions, and see Table~\ref{tabel2}. The constraints on $\eta $ for stage III will be very similar to that of stage I.}.
	
\begin{figure}[htpb]
    	\centering
    	\includegraphics[scale=0.3]{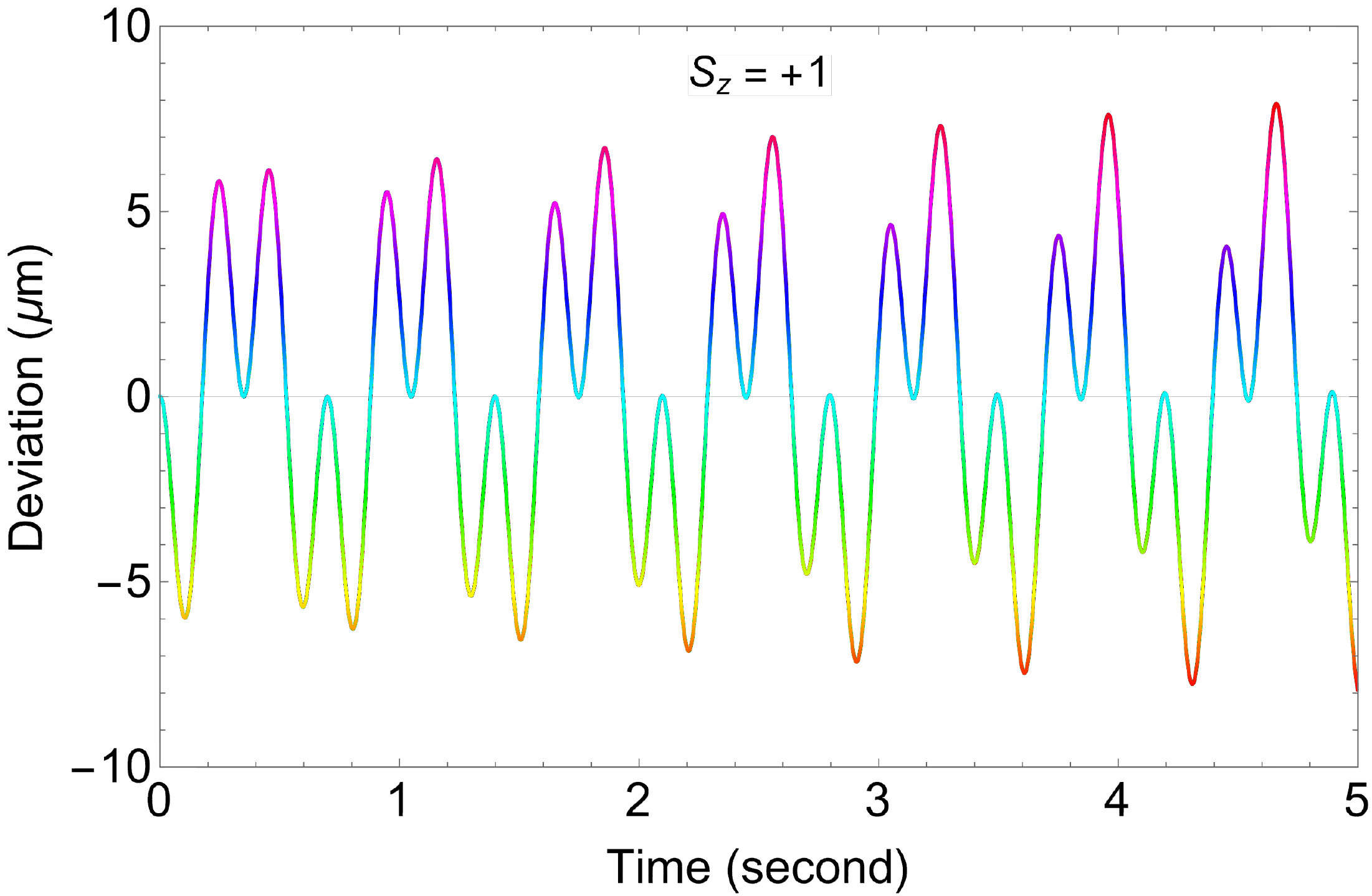}\\
    	\vspace{0.3cm}
    	\includegraphics[scale=0.3]{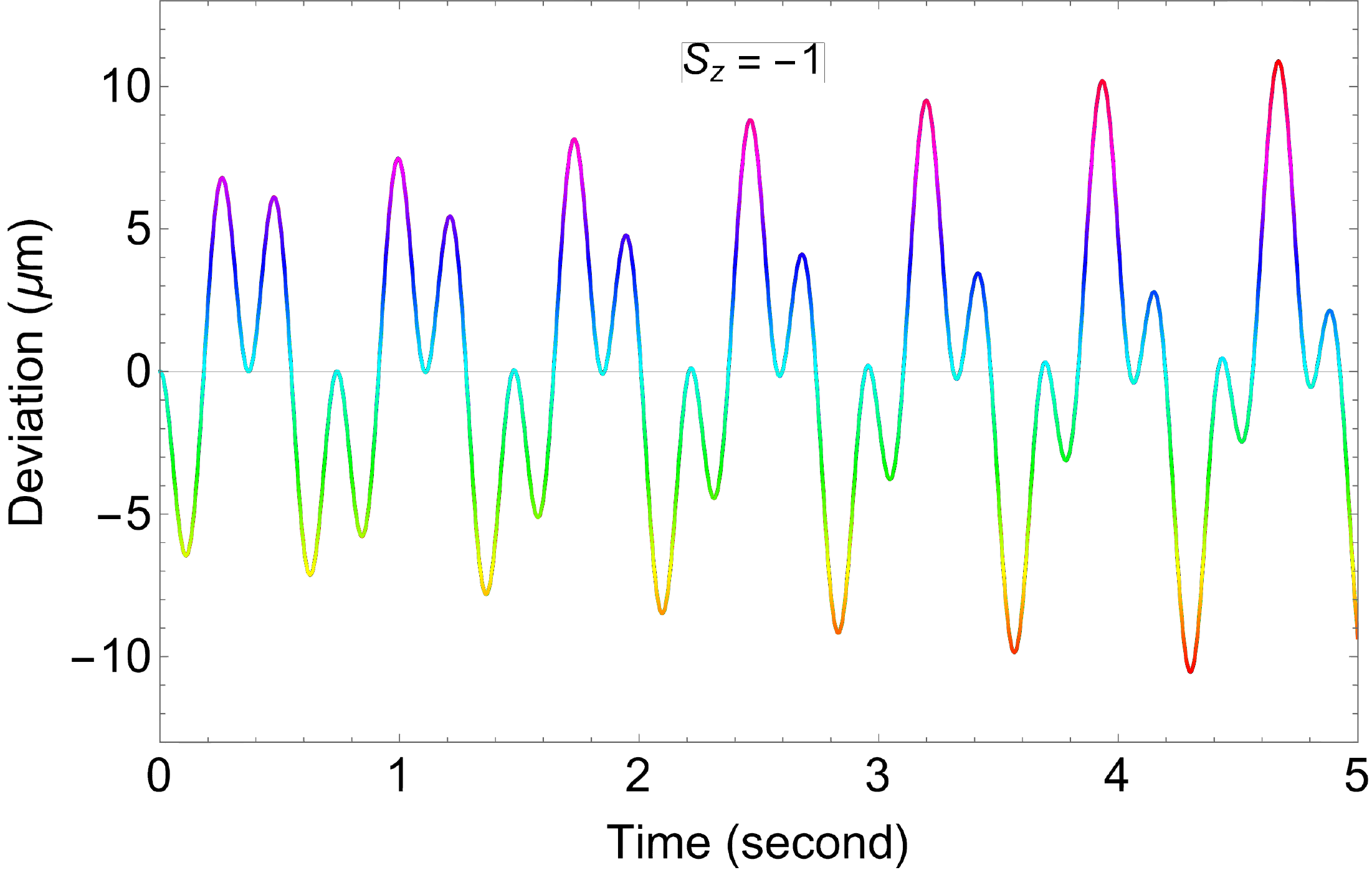}
    	\caption{The deviation between the approximate trajectory and the exact trajectory. Here $m=10^{-17} kg, \eta=1\times 10^{6} \text{T}~\text{m}^{-2}$, $C_{\text{correction}}=27.3467$.}\label{DeviationUpTo5SecondsEta10To6}
    \end{figure}

	 We can compare the approximated trajectory with regard to the exact trajectory, see Fig.~(\ref{CompareTrajectory}) for $m=10^{-17}~\text{kg}$ for $1.4$ seconds, and Fig.~(\ref{DeviationUpTo5SecondsEta10To6}) for a longer time period $5$ seconds for both the wavefunctions, i.e. up and down spin trajectories. Fig.(\ref{DeviationUpTo5SecondsEta10To6}) shows the deviation of the approximate trajectory from the exact trajectory. We can see that the maximum deviation between the trajectories within 4 seconds is less than 10 $\mu \text{m}$.

	
	 With the above mentioned approximations, we can now address the question of spin coherence and any fluctuations 
	 in the magnetic field, i.e. any deviation $\Delta \eta$ in the value of $\eta$ due to an inaccuracy in the control of the magnetic field. Then the deviation of the trajectory can be expressed as
	\begin{align}\label{delta_z}
		\Delta z(t)=&z_{0}-z_{0}\cos(\sqrt{A+\Delta A}t),\nonumber \\
		\leqslant&z_{0}\frac{\Delta A^{2}}{4A}t^{2}.
	\end{align}
	where 
	\begin{align}\label{delta_a}
		\Delta A=&-\left(C_{\text{correction}}\frac{\chi_{m}}{\mu_{0}}B_{0}-S_{z}\frac{g e \hbar }{m m_{e}}\right)\Delta \eta.
	\end{align}
	Here we have used $\sqrt{A}t=2n\pi$ (n is a positive integer), and then we have $\cos\left(\sqrt{A}t\right)=1$, $\sin\left(\sqrt{A}t\right)=0$. Combining Eq.(\ref{TolerateDeviation}), Eq.(\ref{frequency}), Eq.(\ref{delta_z}) and Eq.(\ref{delta_a}), we can get
	\begin{align}\label{accuracyrequiredZ}
		\left(\frac{\Delta \eta}{\eta}\right)_{z} \leqslant&\left(\frac{2\sqrt{2}\varepsilon}{z_{0}At^{2}}\delta z\right)^{\frac{1}{2}},
	\end{align}
	where the subscript $z$ represents the accuracy required to obtain from the positional uncertainty.
	
	Similarly, by using the last inequality about the momentum in Eq.(\ref{TolerateDeviation}), we can also obtain a requirement for the accuracy of the magnetic field. The momentum can be obtained by taking the derivative of Eq.(\ref{MotionEquationofSpatial})
	\begin{align}
		p_{z}(t)=m\frac{\mathrm{d} z}{\mathrm{d}t}=-mz_{0}\sqrt{A}\sin(\sqrt{A}t).
	\end{align}
	Since the initial momentum is zero, we can directly express the deviation of momentum as
	\begin{align}\label{delta_p}
		\Delta p_{z}(t)=&-mz_{0}\sqrt{A+\Delta A}\sin(\sqrt{A+\Delta A}t),\nonumber \\
		\leqslant&mz_{0}\left(\sqrt{A}+\frac{\Delta A}{2\sqrt{A}}\right)\sin\left(\left(\sqrt{A}+\frac{\Delta A}{2\sqrt{A}}\right)t\right),\nonumber \\
		=&\frac{1}{4}\frac{mz_{0}t}{A}\left(\Delta A+A\right)^{2}-\frac{1}{4}mz_{0}At.
	\end{align}
	Here we have also used $\cos\left(\sqrt{A}t\right)=1$ and $\sin\left(\sqrt{A}t\right)=0$. Combining Eq.(\ref{TolerateDeviation}), Eq.(\ref{frequency}), Eq.(\ref{delta_a}) and Eq.(\ref{delta_p}) we can get
	\begin{align}\label{accuracyrequiredP}
		\left(\frac{\Delta \eta}{\eta}\right)_{p_{z}} \leqslant&\left(\frac{2\sqrt{2}\varepsilon}{mz_{0}At}\delta p_{z}+1\right)^{\frac{1}{2}}-1,
	\end{align}
	where the subscript $p_{z}$ represents the accuracy required from the momentum uncertainty.
	
	Using the approximate solution of the wave packet trajectory (Eq.(\ref{MotionEquationofSpatial})), when time $t$ satisfies $\sqrt{A}t=2n\pi$, the variation of the trajectory can be written as
	\begin{align}\label{deltaz_t}
		(\delta z(t))^{2}=(\delta z_{0})^{2}+\left(\frac{t}{m}\right)^{2}(\delta p_{z})^{2}.
	\end{align}
	If we require that the wave packet does not spread significantly in time t, then the last term in Eq.(\ref{deltaz_t}) needs to be satisfied~\footnote{See the analysis of Appendix~\ref{SOWP}, where we have analysed the spread in the wave packet. Fig.(\ref{EvolutionOfWavePacket}) shows the evolution of the probability density of the wave packet for one of the spins for $m=10^{-17}~\text{kg}$ for $\eta =10^{6}~{\rm T~m^{-2}}$. The numerical values will not alter much for different values of $\eta$. Note that the expectation value of 
the wave packet position with the classical trajectory is shown in Fig.(\ref{ExpectationValueAndClassicalTrajectory}). We can see from Fig.~(\ref{UncertaintyDeltaZDeltaP}) that the minimum uncertainty is not always followed throughout the trajectory, but it is satisfied at certain times. If we could manage to close the interference at those moments, we will be able to recover the spin coherence as desired in the text.}
	\begin{align}\label{delta-p-required}
    \frac{t}{m}\delta p_{z}\cong \delta z_{0}.
	\end{align}
	Combining Eq.(\ref{delta-p-required}) with the minimum uncertainty $\delta z=\delta p_{z}= (\hbar/2)^{\frac{1}{2}}$, we can get
	\begin{align}\label{DzDp}
		\delta z=\left(\frac{t\hbar}{2m}\right)^{\frac{1}{2}}, \quad \delta p_{z}=\left(\frac{m\hbar}{2t}\right)^{\frac{1}{2}}.
	\end{align}
	By substituting Eq.(\ref{DzDp}) into Eq.(\ref{accuracyrequiredZ}) and Eq.(\ref{accuracyrequiredP}), we can get
	\begin{align}
		\left(\frac{\Delta \eta}{\eta}\right)_{z} \leqslant&\left(\frac{2\varepsilon\hbar^{\frac{1}{2}}}{z_{0}Am^{\frac{1}{2}}t^{\frac{3}{2}}}\right)^{\frac{1}{2}},\label{accuracy-z}\\
		\left(\frac{\Delta \eta}{\eta}\right)_{p_{z}} \leqslant&\left(\frac{2\varepsilon\hbar^{\frac{1}{2}}}{z_{0}Am^{\frac{1}{2}}t^{\frac{3}{2}}}+1\right)^{\frac{1}{2}}-1.\label{accuracy-p}
	\end{align}
	Using Eq.(\ref{accuracy-z}) and Eq.(\ref{accuracy-p}), we obtain the accuracy required on the magnetic field under for masses to recover the spin coherence, as shown in the Table (\ref{tabel1}).
	\renewcommand\arraystretch{1.5}
	\begin{table}[h]
		\begin{tabular}{p{2cm} p{1.3cm} p{2.2cm} p{2.2cm}}
			\hline
			\hline
			Mass  &$S_{z}$ & $\left(\frac{\Delta \eta}{\eta}\right)_{z} \lesssim$& $\left(\frac{\Delta \eta}{\eta}\right)_{p_{z}}\lesssim$ \\ \hline
			\multirow{2}{2cm}{$10^{-17}$kg} & 1 & $3.7\times 10^{-4}$ & $6.9\times 10^{-8}$ \\ 
			                              & -1 & $3.8\times 10^{-4}$ & $7.1\times 10^{-8}$ \\ \hline
			\multirow{2}{*}{$10^{-16}$kg} & 1 &$2.1\times 10^{-4}$  & $2.2\times 10^{-8}$ \\  
			                              & -1 & $2.1\times 10^{-4}$ & $2.2\times 10^{-8}$ \\ \hline
			\multirow{2}{*}{$10^{-15}$kg} & 1 & $1.2\times 10^{-4}$ & $6.9\times 10^{-9}$ \\  
			                              & -1 & $1.2\times 10^{-4}$ & $7.0\times 10^{-9}$ \\ 
			                              \hline
			                              \hline
		\end{tabular}
	\caption{We show the constraints on the magnetic field accuracy required for different masses. The value of spin $S_{z}$ does not affect the order of the magnitude of accuracy required in $\eta$.  We have demanded that we recover the spin coherence up to $99\%$. This particular table constraints the magnetic field accuracy from the stage II of the trajectory. A similar constraint on $\eta$ for the Stage I part of  the trajectory can be found, see Table \ref{tabel2}.  }
	\label{tabel1}
	\end{table}

	Where we have used $\varepsilon=0.1$, which corresponding to recover $99\%$ spin coherence; $z_{0}=1\times 10^{-4}~\text{m}$, which is the initial centre position of wave packet; $t=\frac{2\pi}{\sqrt{A}}\approx 0.7 s$, which is duration of experimental stage; $\eta=1\times 10^{6}~\text{T}~\text{m}^{-2}$, which is the gradient parameter we have used.
	

	\section{Conclusion}

In this paper, we have provided a simple mechanism for creating a large spatial superposition with heavy masses and with embedded spin. We have shown that it is possible to achieve $\Delta Z \sim {\cal O}(10^{3})~{\rm \mu \text{m}}$ for $m=10^{-17}~\text{kg}$, $\Delta Z \sim ~{\cal O}(10^{2})~{\rm \mu \text{m}}$ for $m=10^{-16}~\text{kg}$, and $\Delta Z \sim {\cal O}(10)~{\rm \mu \text{m}}$ for $m= 10^{-15}~\text{kg}$ within $\sim 1.4$ seconds. There is indeed an order of magnitude gain in the splitting of the wave function compared to our earlier proposal~\cite{Marshman:2021wyk}, where we had taken only the gradient term in the magnetic field, and could not achieve such a large spatial superposition in a short time scale (within $1-1.5$ seconds). In this regard, catapulting the trajectory of the two wave packets has yielded a better result with a magnetic field gradient of order ${\cal O}(10^{2}- 10^{4})~{\rm Tm^{-1}}$. 

We highlighted that there are primarily three stages of the trajectory. First, we create a large velocity difference between the two wave packets, which experience differential spin-dependent forces. The anharmonic oscillations gradually increase the amplitude, and when the two trajectories meet at $z=0$, their velocity difference is large, and the trajectories catapult to achieve a large spatial splitting. We employ three different values of $\eta$ parameter which controls the magnetic field gradient, see Figs.(\ref{M17-FiveTypesFigures},\ref{M16-FiveTypesFigures},\ref{M15-FiveTypesFigures}). We have ensured that the interference is completed 
within ${\cal O}(1-1.5)$ second, where the wave function overlap is such that the position and the momentum match to interfere with the two paths. We have also analysed the conditions required to maintain the spin coherence. To achieve $99\%$ coherence, we have obtained the stringent bound on the magnetic field fluctuations. The most stringent condition on the fluctuation in the magnetic field arises from stage I, see Table \ref{tabel2}, and similarly for stage II, see Table \ref{tabel1}. We have also analysed the spreading of the wave function and showed that the wave packets evolve and do not satisfy the minimum uncertainty principle throughout the trajectory at every moment, but the largest $\delta z\delta p \leq 4\hbar$ (This restriction only holds when the initial conditions are $\delta z\sim 5\times 10^{-3}~\mu \text{m}$ and initial position $z\sim 5\times 10^{-2}~\mu \text{m}$.) and it oscillates with a period of roughly $0.5$ seconds where it satisfies the minimum uncertainty principle for $m=10^{-17}~\text{kg}$ and for $\eta=10^{6}~{\rm T~m^{-2}}$. However, the wave function's classical and quantum trajectories match extremely well, see Fig.~\ref{ExpectationValueAndClassicalTrajectory}.

Indeed, in all our analysis, the time duration of the spin coherence is an important factor for the experiment, but the spin coherence times are perpetually rising (approaching 1 second ~\cite{Bar-Gill,Abobeih}, even 30 seconds ~\cite{Muhonen,Farfurnik}) adapting these to nano-crystals remains an open problem, but there is no fundamental constraints~\cite{Knowles}. The spatial coherence times can be made $100$ seconds, see ~\cite{Bose:2017nin,vandeKamp:2020rqh,Toros:2020dbf}. There are indeed other challenges, but achievable pressures, temperatures, distances from other sources and fluctuations~\cite{vandeKamp:2020rqh}. For example, a decoherence rate below 0.1 Hz is achievable for diamond spheres of masses $10^{-14}$ kg. This is expected \cite{Bose:2017nin,vandeKamp:2020rqh} for internal temperatures of 0.15~$\text{K}$, an environmental temperature of 1~$\text{K}$ and the environmental gas number density of $10^{-8}~{\rm m}^{-3}$.
In addition to these, we will need to take into account the effect of the rotation of the diamond \cite{Japha} and the excitation of phonons \cite{Henkel2021} on the spin coherence. However, we will study these effects separately.

{\bf Acknowledgements}: We would like to thank Yoni Japha, Ron Folman and Ben Stickler for the discussions.
R. Z. is supported by China Scholarship Council (CSC) fellowship. R. J. M. is supported  by the Australian Research Council (ARC) under the Centre of Excellence for Quantum Computation and Communication Technology (CE170100012). AM's research is funded by the Netherlands Organisation for Science and Research (NWO) grant number 680-91-119.  SB would like to acknowledge EPSRC Grant Nos. EP/N031105/1 and EP/S000267/1.

	\begin{appendices}
			
	\section{Spin coherence and magnetic field control for $\eta=1\times 10^{8}~\text{T}~\text{m}^{-2}$}\label{MagneticFieldControl}

\begin{figure}[h]
	\centering
	\includegraphics[scale=0.19]{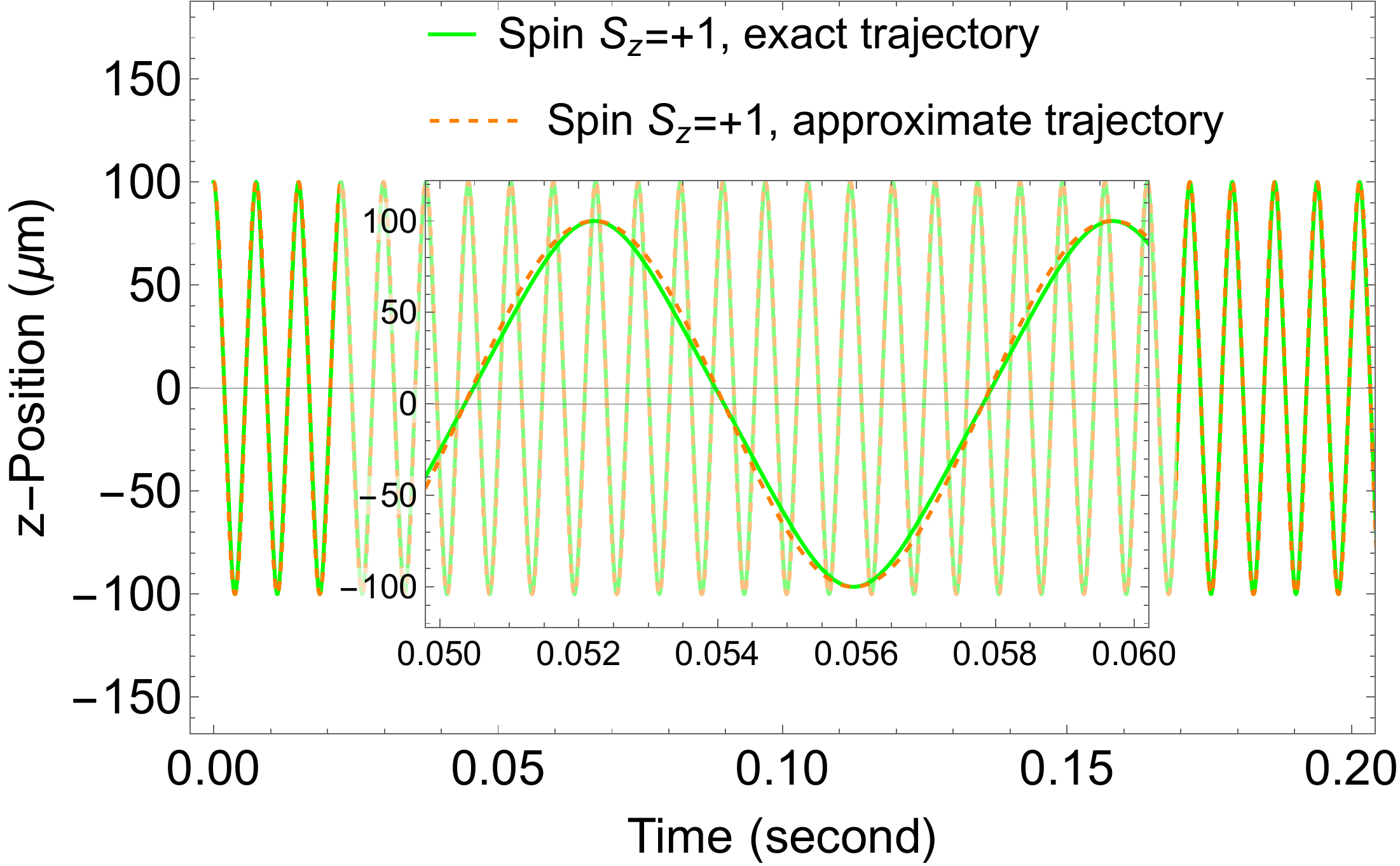}
	\includegraphics[scale=0.19]{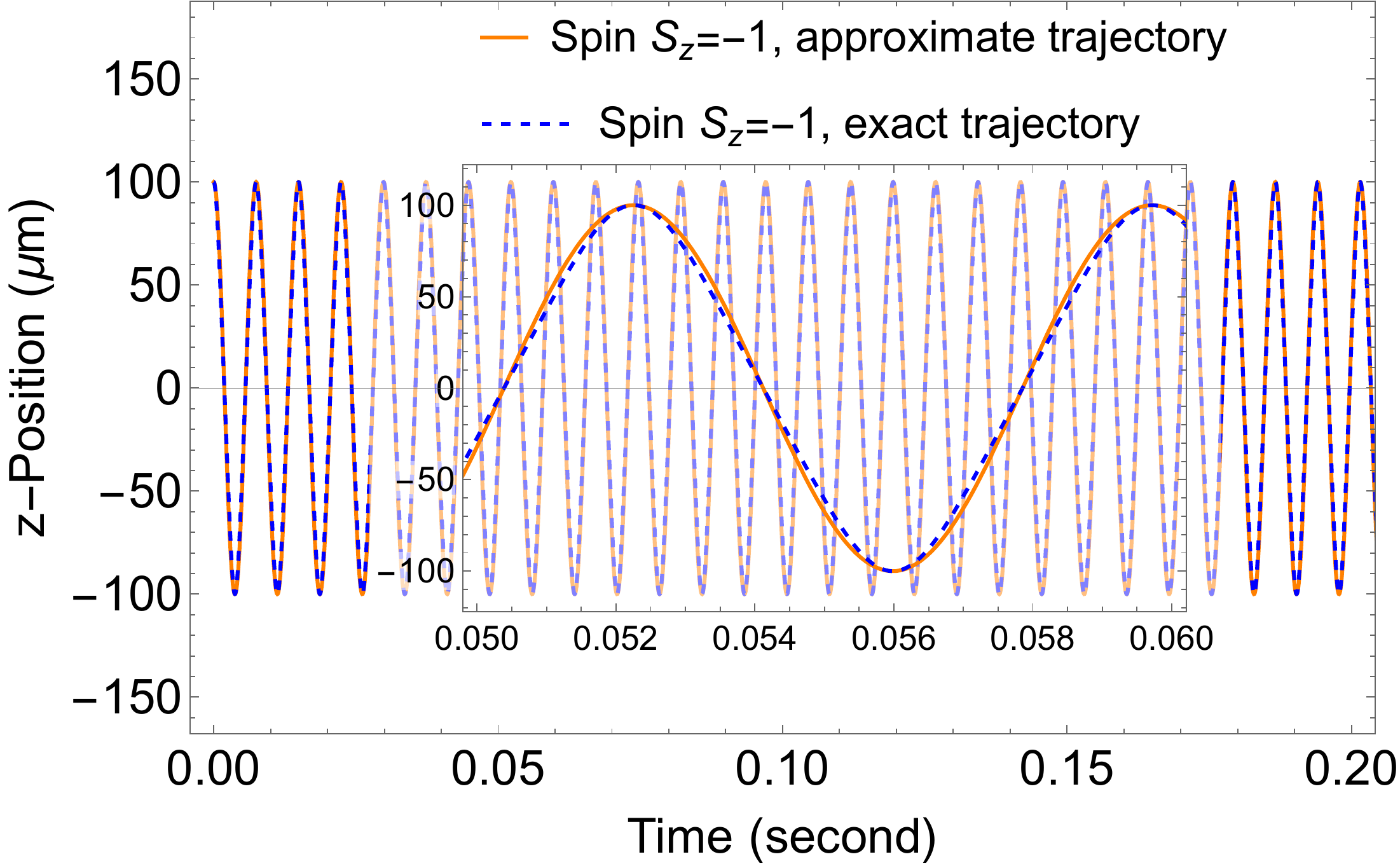}
	\caption{Comparison of approximate and exact trajectories of two wave packets. Here $m=10^{-17}~\text{kg}, \eta=1\times 10^{8} ~\text{T}~\text{m}^{-2}$, $C_{\text{correction}}=2526.82$, $p_{z}=0$.}\label{CompareTrajectory2}
\end{figure}

	According to the analysis in Sec.(\ref{SpinCoherence}), we can calculate the accuracy required for the magnetic field control under the new magnetic field gradient by directly replacing the corresponding values of $\eta$ and the correction factor. Now we set $\eta=1\times 10^{8}~\text{T}~\text{m}^{-2}$, and the corresponding correction factor $C_{\text{correction}} = 2526.82$. Using Eq.(\ref{Acceleration2}) and Eq.(\ref{MotionEquationofSpatial}), we can compare the approximate and the exact trajectories of the wave packets as shown in Fig.(\ref{CompareTrajectory2}).

\begin{figure}[h]
		\centering
		\includegraphics[scale=0.19]{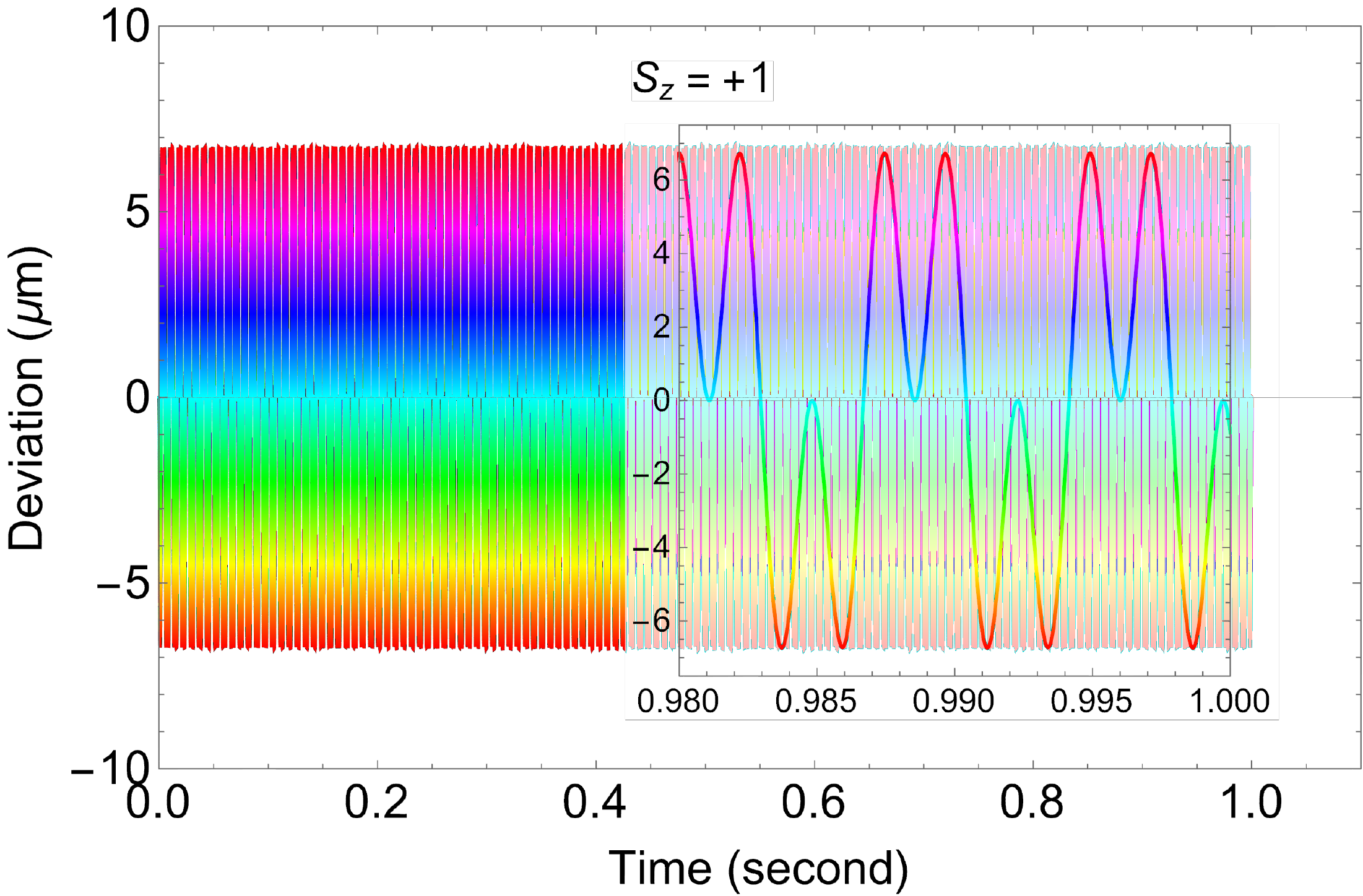}
		\includegraphics[scale=0.19]{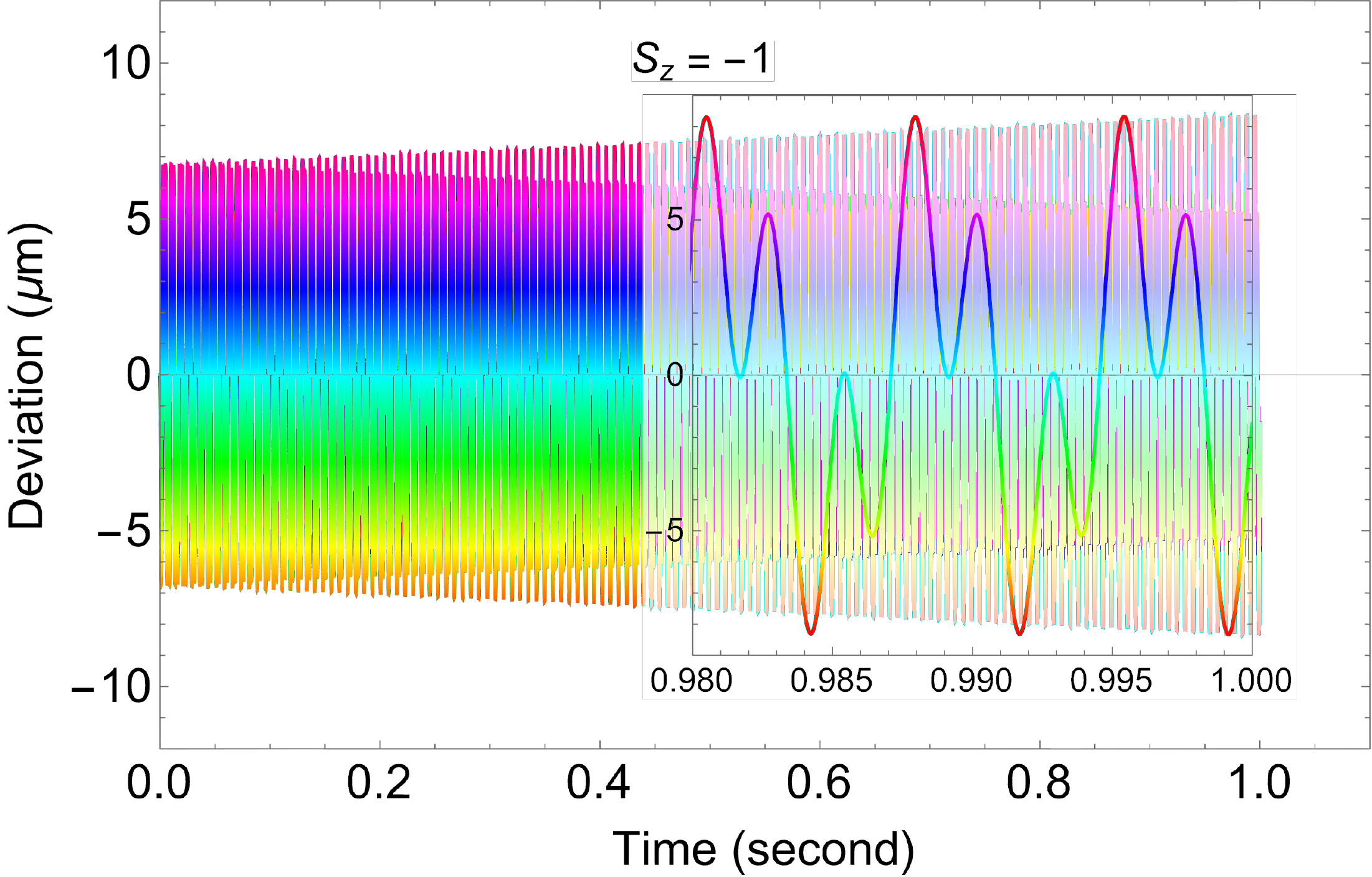}
		\caption{The deviation between the approximate trajectory and the exact trajectory for $S_\text{z}=\pm 1$. Here $m=10^{-17}~\text{kg}, \eta=1\times 10^{8} ~\text{T}~\text{m}^{-2}$, $C_{\text{correction}}=2526.82$, $p_{z}=0$.}\label{DeviationUpTo1SecondsEta10To8}
\end{figure}	

\begin{table}[h]
	\centering
	\begin{tabular}{p{2cm} p{1.3cm} p{2.2cm} p{2.2cm}}
		\hline
		\hline
		Mass  & $S_{z}$ & $\left(\frac{\Delta \eta}{\eta}\right)_{z} \lesssim$& $\left(\frac{\Delta \eta}{\eta}\right)_{p_{z}}\lesssim$ \\ \hline
		\multirow{2}{2cm}{$10^{-17}$kg} & 1 & $9.3\times 10^{-6}$ & $4.3\times 10^{-11}$ \\ 
		& -1 & $9.3\times 10^{-6}$ & $4.3\times 10^{-11}$ \\ \hline
		\multirow{2}{*}{$10^{-16}$kg} & 1 &$5.2\times 10^{-6}$  & $1.4\times 10^{-11}$ \\ 
		& -1 & $5.2\times 10^{-6}$ & $1.4\times 10^{-11}$ \\ \hline
		\multirow{2}{*}{$10^{-15}$kg} & 1 & $2.9\times 10^{-6}$ & $4.3\times 10^{-12}$ \\  
		& -1 & $2.9\times 10^{-6}$ & $4.3\times 10^{-12}$ \\ 
		\hline
		\hline
	\end{tabular}
	\caption{ We have shown the constraints on the magnetic field accuracy for different masses during stage I. The bounds on $\eta$ is based on recovering the spin coherence $99\%$. The value of spin $S_{z}$ does not affect the order of magnitude of accuracy.}
	\label{tabel2}
\end{table}		

Fig.(\ref{DeviationUpTo1SecondsEta10To8}) shows the deviation of the approximate trajectory from the exact trajectory. We can see that the maximum deviation between trajectories within 1 second is less than 9 $\mu \text{m}$.		
		
Using Eq.(\ref{accuracy-z}) and Eq.(\ref{accuracy-p}), we can get $\delta \eta/\eta$ in the magnetic field fluctuation for different masses, as shown in Table (\ref{tabel2}).
\renewcommand\arraystretch{1.5}

Where we have used $\varepsilon=0.1$, which is corresponding to maintain 99\% spin coherence; $z_{0}=1\times 10^{-4} ~\text{m}$, which is the initial centre position of wave packet; $t=\frac{60\pi}{\sqrt{A}}\approx 0.2 ~\text{s}$, which is duration of experimental stage; $\eta=1\times 10^{8}~\text{T}~\text{m}^{-2}$, which is the gradient parameter we used.		

\section{Spreading of the wave packet}\label{SOWP}

In Sec.\ref{SpinCoherence}, we assume that the wave packet is always kept to a minimum uncertainty when calculating the accuracy required of the magnetic field control to restore spin coherence. In this appendix, we study the evolution of the wave packet width in the quartic potentials, providing a theoretical basis for our hypothesis in Sec.\ref{SpinCoherence}.

\begin{figure}[h]
	\centering
	\includegraphics[scale=0.36]{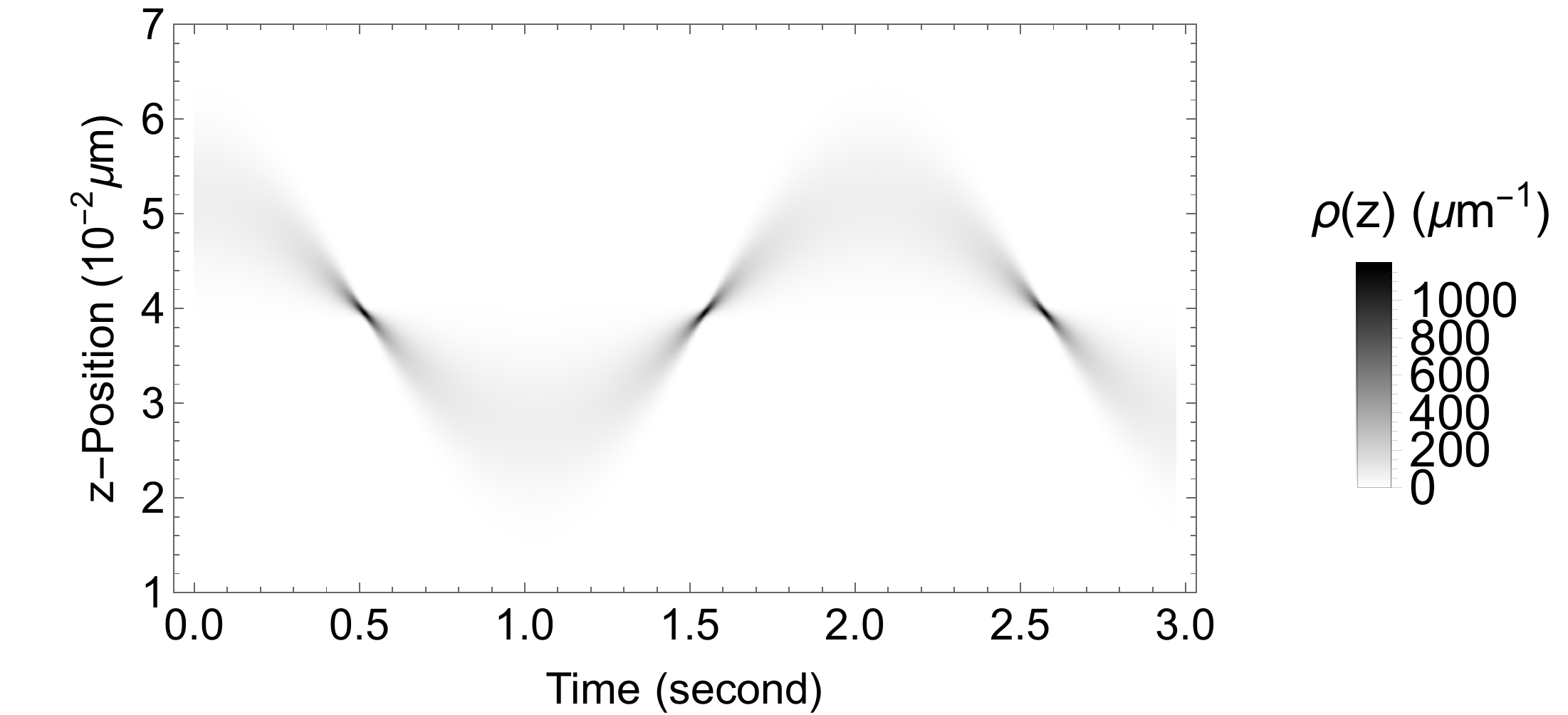}
	\caption{We show the evolution of the wave packet for $S_{\text{z}}=-1$. The shadow in the figure corresponds to the probability density that the NV centre is located at a certain spatial location at a certain time. The darker the color of the shadow, the greater the value of the probability density. Here initial width of wave packet $\delta_{z}\approx 5\times 10^{-3}\mu \text{m}$, $m=10^{-17} \text{kg}, \eta=1\times 10^{6} \text{T}~\text{m}^{-2}$.}\label{EvolutionOfWavePacket}
\end{figure}

The Schr\"odinger equation is 
\begin{equation}
	i \hbar \frac{d}{d t}|\Psi(t)\rangle=\hat{H}|\Psi(t)\rangle.
\end{equation}
By substituting the specific forms of Hamiltonian (Eq.(\ref{Hamitonian})) and magnetic field (Eq.(\ref{magneticfield})) into the Schr\"odinger equation and making the initial state as the Gaussian wave packet shown in Eq.(\ref{InitialWaveFunction}), we can use the Trotter expansion method to numerically calculate the evolution of the wave packet \cite{Schmied2020} as shown in Fig.(\ref{EvolutionOfWavePacket}).

\begin{figure}[h]
	\centering
	\includegraphics[scale=0.193]{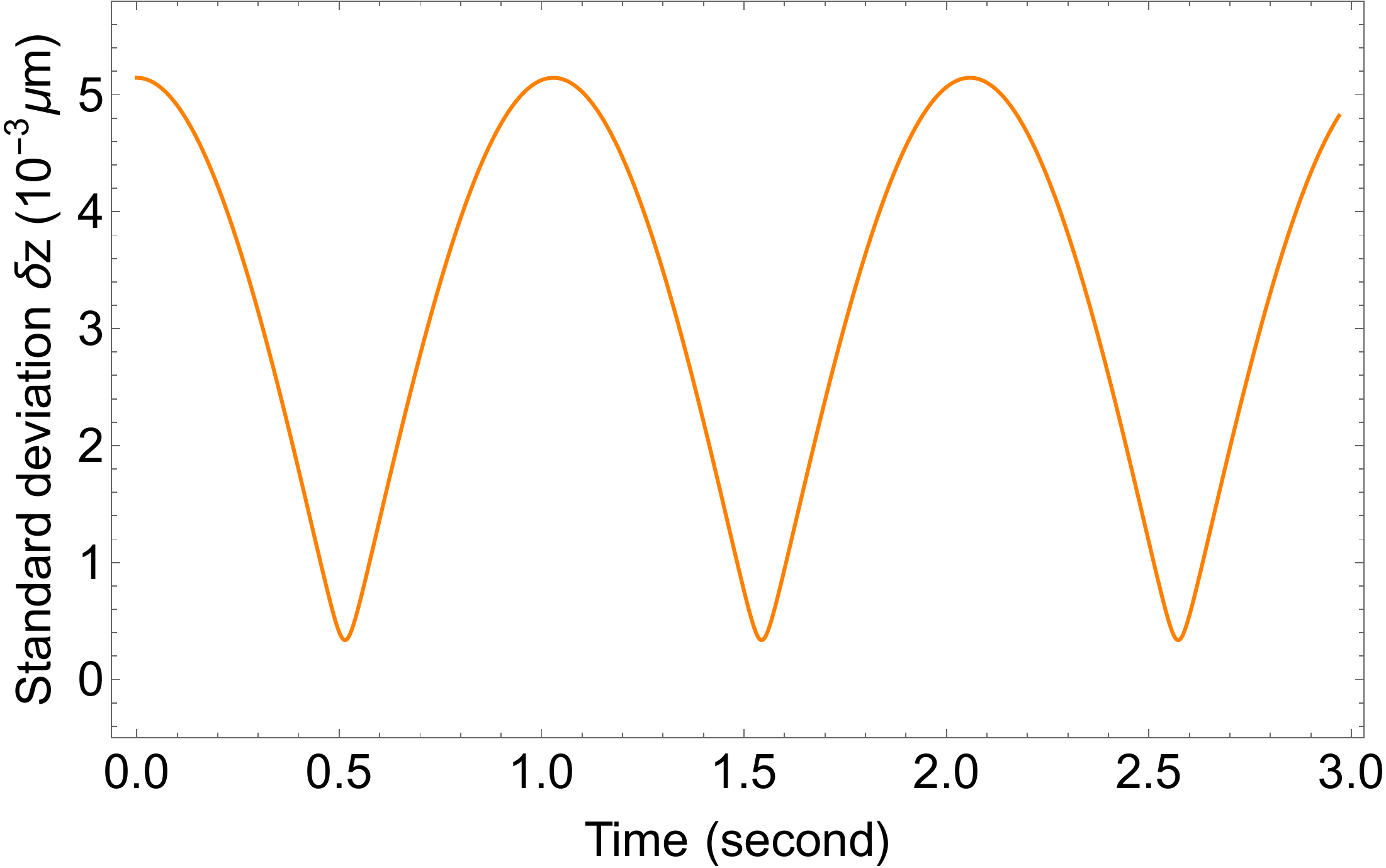}\hspace{-0.1cm}
	\includegraphics[scale=0.193]{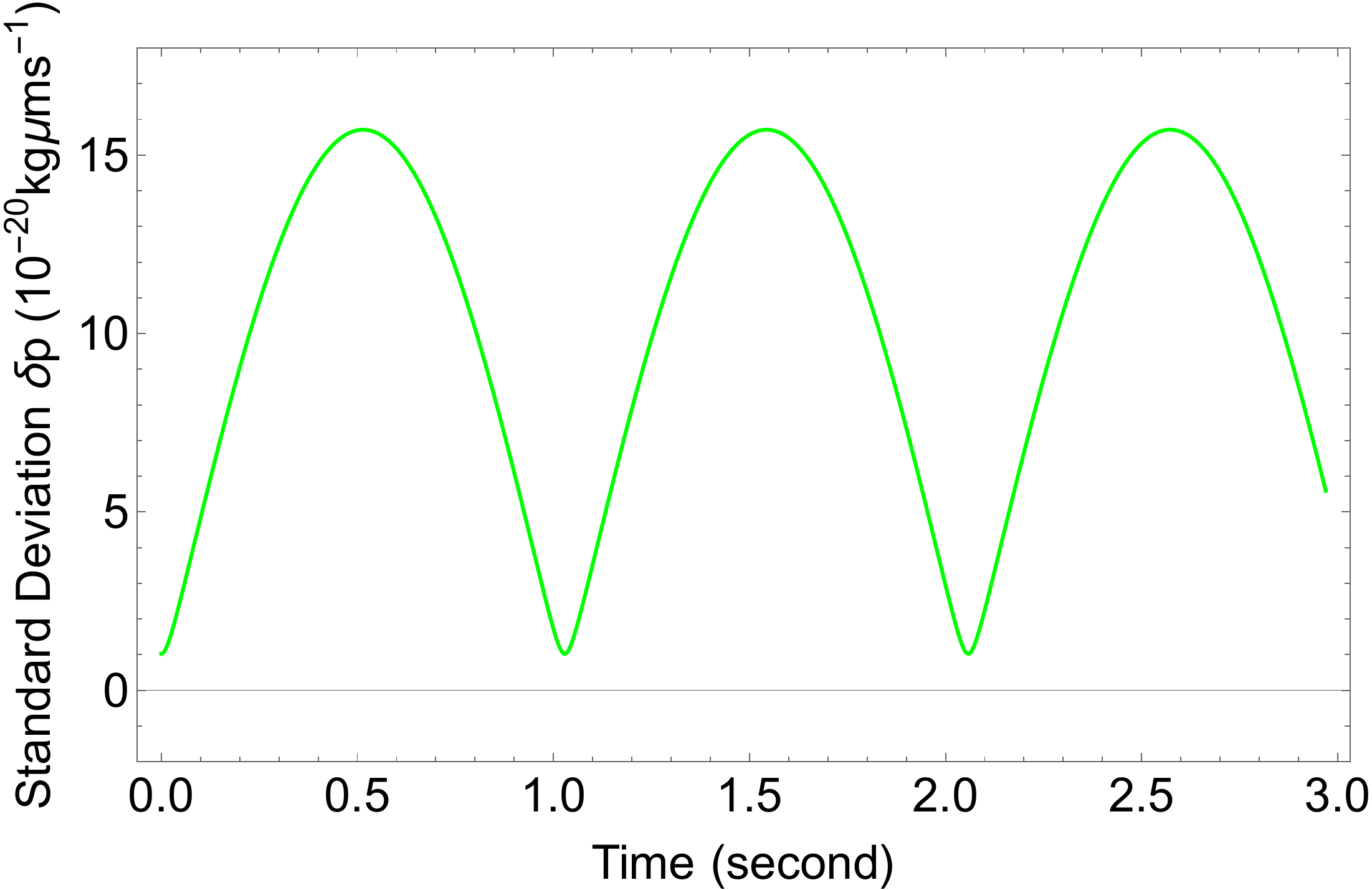}
	\caption{We have shown the standard deviation for the z-position and momentum for $S_{\text{z}}=-1$, which $\delta z=\left(\left\langle z^{2}\right\rangle-\left\langle z \right\rangle^{2}\right)^{\frac{1}{2}}$, $\delta p=\left(\left\langle p^{2}\right\rangle-\left\langle p\right\rangle^{2}\right)^{\frac{1}{2}}$. The standard deviation of the wave packet position and momentum shows a periodic oscillation behavior in the quartic potentials. Here $m=10^{-17} ~\text{kg}, \eta=1\times 10^{6} ~\text{T}~\text{m}^{-2}$.}\label{StandardDeviation}
\end{figure}

\begin{figure}[H]
	\centering
	\includegraphics[scale=0.25]{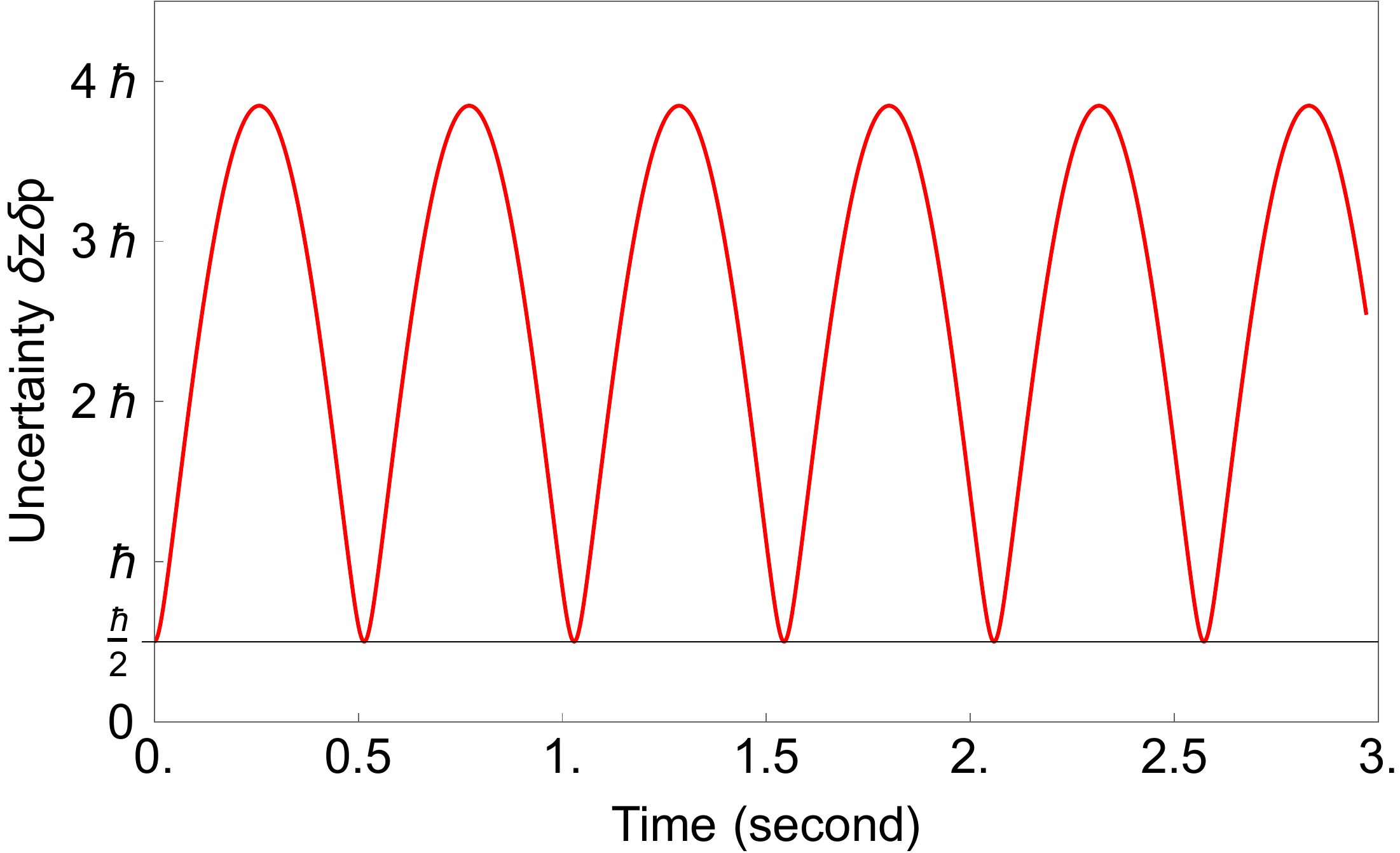}
	\caption{We have shown the uncertainty of the position and the momentum of the wave packet for $S_{\text{z}}=-1$.The uncertainty changes periodically with time and satisfies the uncertainty principle $\delta_{z}\delta_{p}\ge\hbar/2$. Here $m=10^{-17} ~\text{kg}, \eta=1\times 10^{6} ~\text{T}~\text{m}^{-2}$.}\label{UncertaintyDeltaZDeltaP}
\end{figure}

\begin{figure}[h]
	\centering
	\includegraphics[scale=0.25]{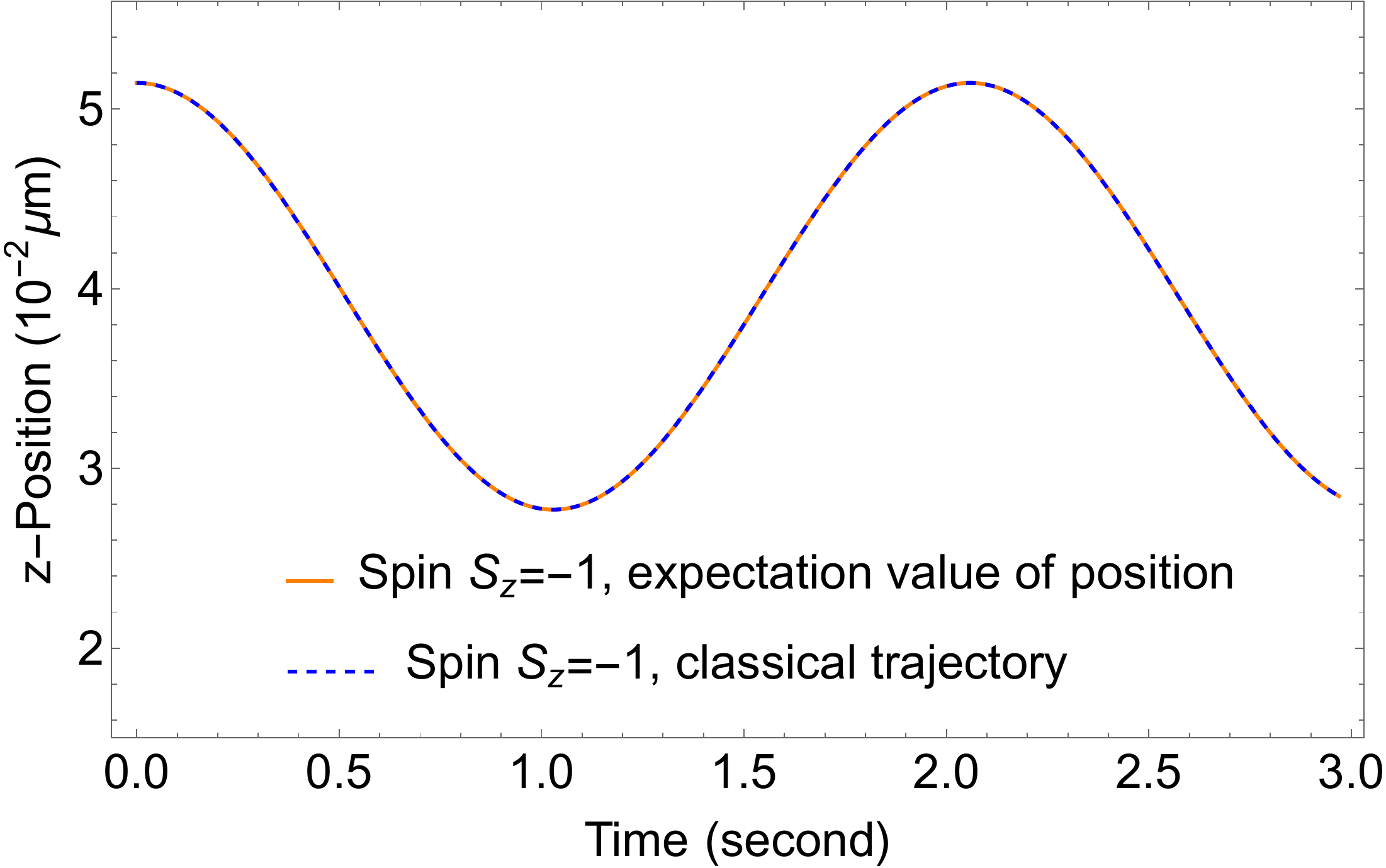}
	\caption{We have shown the comparison between the expectation value of the NV centre's position and the classical trajectory of the NV centre. The two trajectories coincide. Here $m=10^{-17} ~\text{kg}, \eta=1\times 10^{6}~ \text{T}~\text{m}^{-2}$.}\label{ExpectationValueAndClassicalTrajectory}
\end{figure}

It can be seen from Fig.(\ref{EvolutionOfWavePacket}) that the spreading of the wave packet in the quartic potential exhibits a periodic oscillation behaviour. For the purpose of illustration, we will take a single value of $\eta=10^{6}~{\rm T~m^{-2}}$ and for mass $m=10^{-17}~\text{kg}$. A similar analysis can be revised for different values of $\eta$ but the physical properties will not alter much.

The uncertainty of the position and momentum of the wave packet is shown in Fig.(\ref{StandardDeviation}). The product of the uncertainty of position and momentum satisfies the uncertainty principle, as shown in Fig.(\ref{UncertaintyDeltaZDeltaP}). As can be seen from Fig.(\ref{UncertaintyDeltaZDeltaP}), the value of $\delta_{z}\delta_{p}$ changes periodically over time and returns to the minimum uncertainty at the end of a period, which means that our assumptions in Sec.(\ref{SpinCoherence}) are reasonable. 

In order to verify the correctness of the numerical results, we compare the expectation value of the wave packet position with the classical trajectory as shown in Fig.(\ref{ExpectationValueAndClassicalTrajectory}).

As can be seen from Fig.(\ref{ExpectationValueAndClassicalTrajectory}), the expectation value of the wave packet position coincides with the classical trajectory (Eq.(\ref{Acceleration2})), which means that our numerical calculation of the evolution of the wave packet is correct.

\section{Trajectories for masses $10^{-16} ~\text{kg}$ and $10^{-15} ~\text{kg}$}\label{CatapultingProcess}

\begin{figure*}[h]
	\centering
	\begin{subfigure}{0.325\linewidth}
		\centering
		\includegraphics[width=0.9\linewidth]{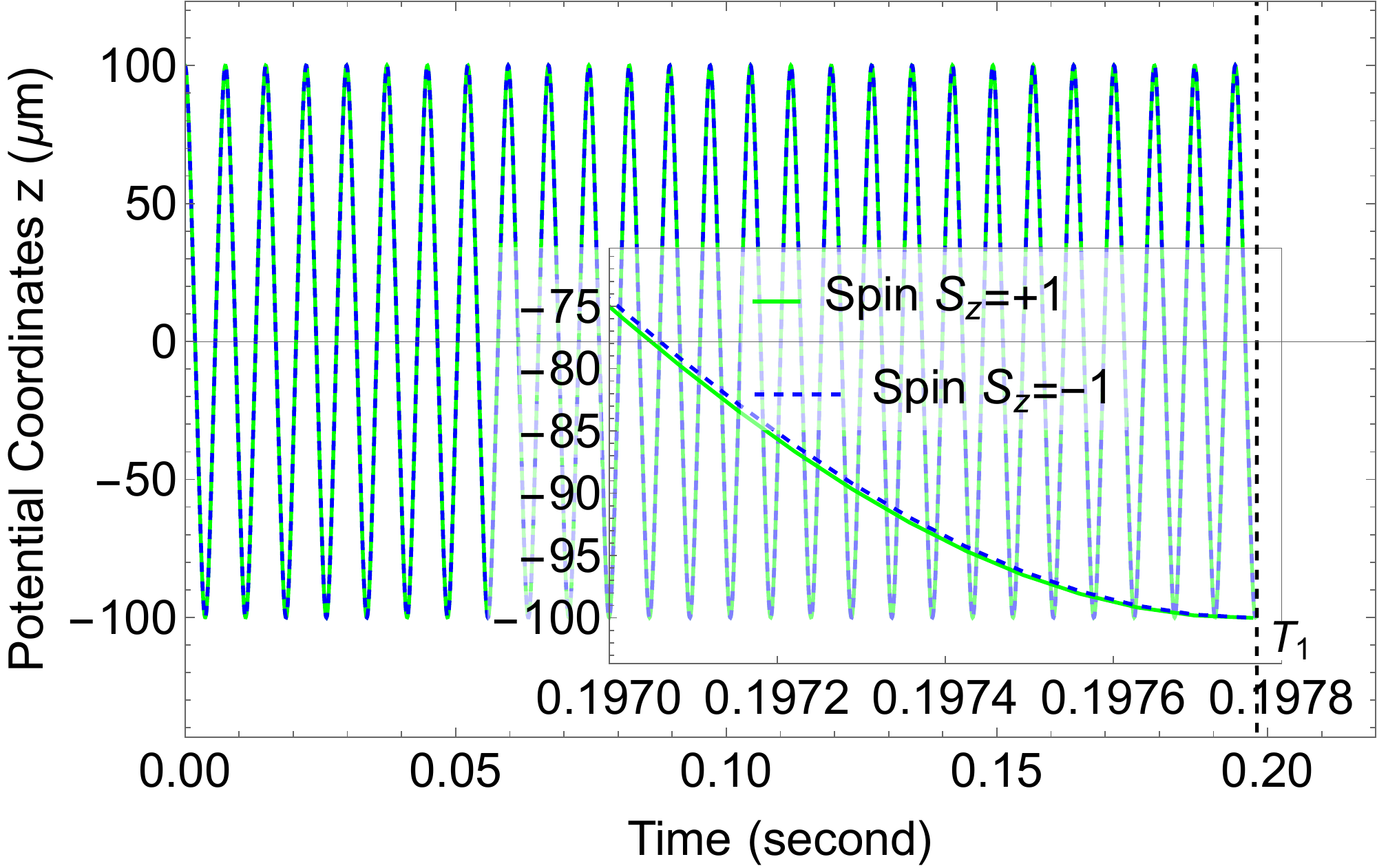}
		\caption{Potential coordinates - stage \Romannum{1}}\label{PotentialCoordinates-Mass16-1}
	\end{subfigure}
	\centering
	\begin{subfigure}{0.325\linewidth}
		\centering
		\includegraphics[width=0.9\linewidth]{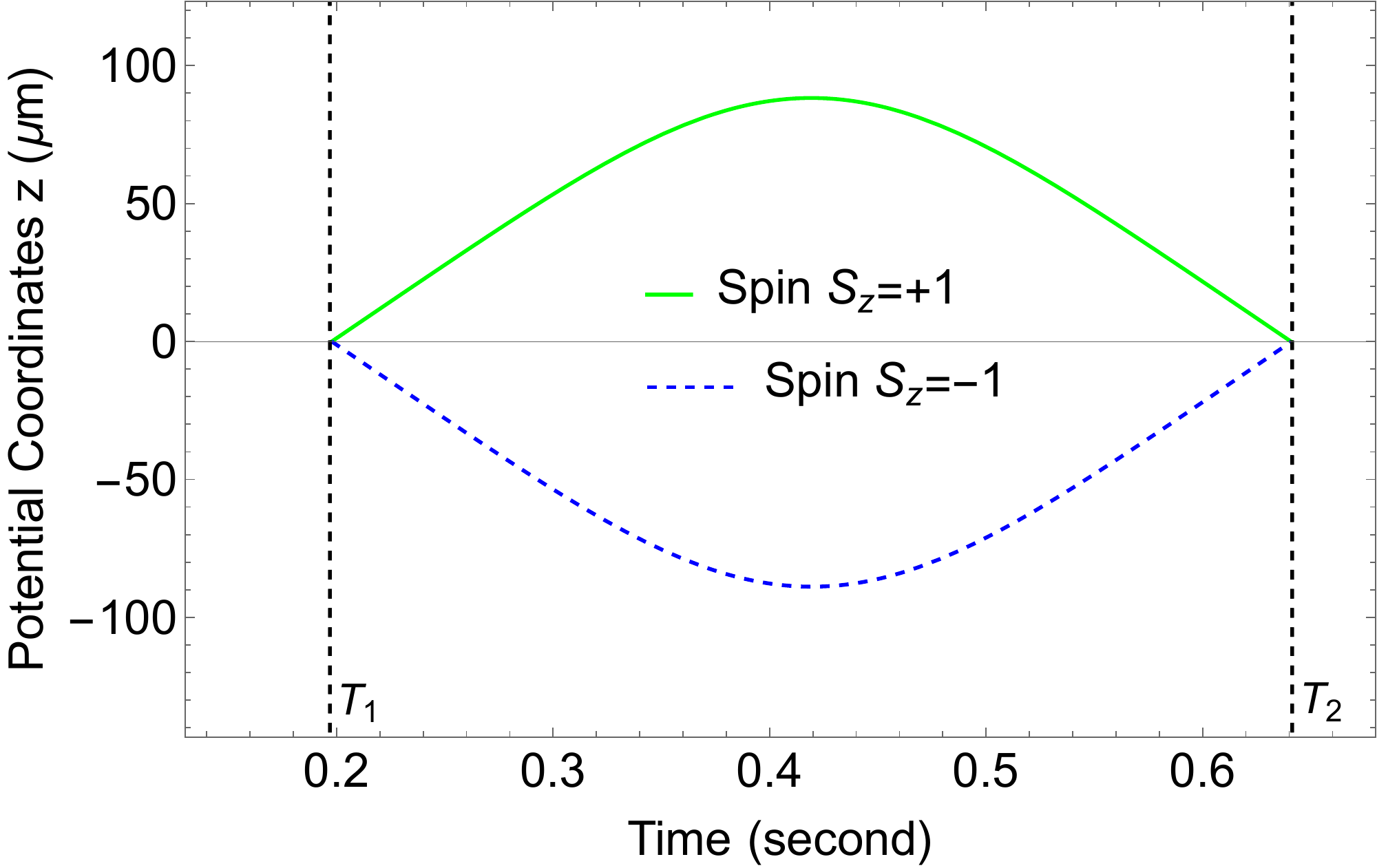}
		\caption{Potential coordinates - stage \Romannum{2}}
	\end{subfigure}
	\centering
	\begin{subfigure}{0.325\linewidth}
		\centering
		\includegraphics[width=0.9\linewidth]{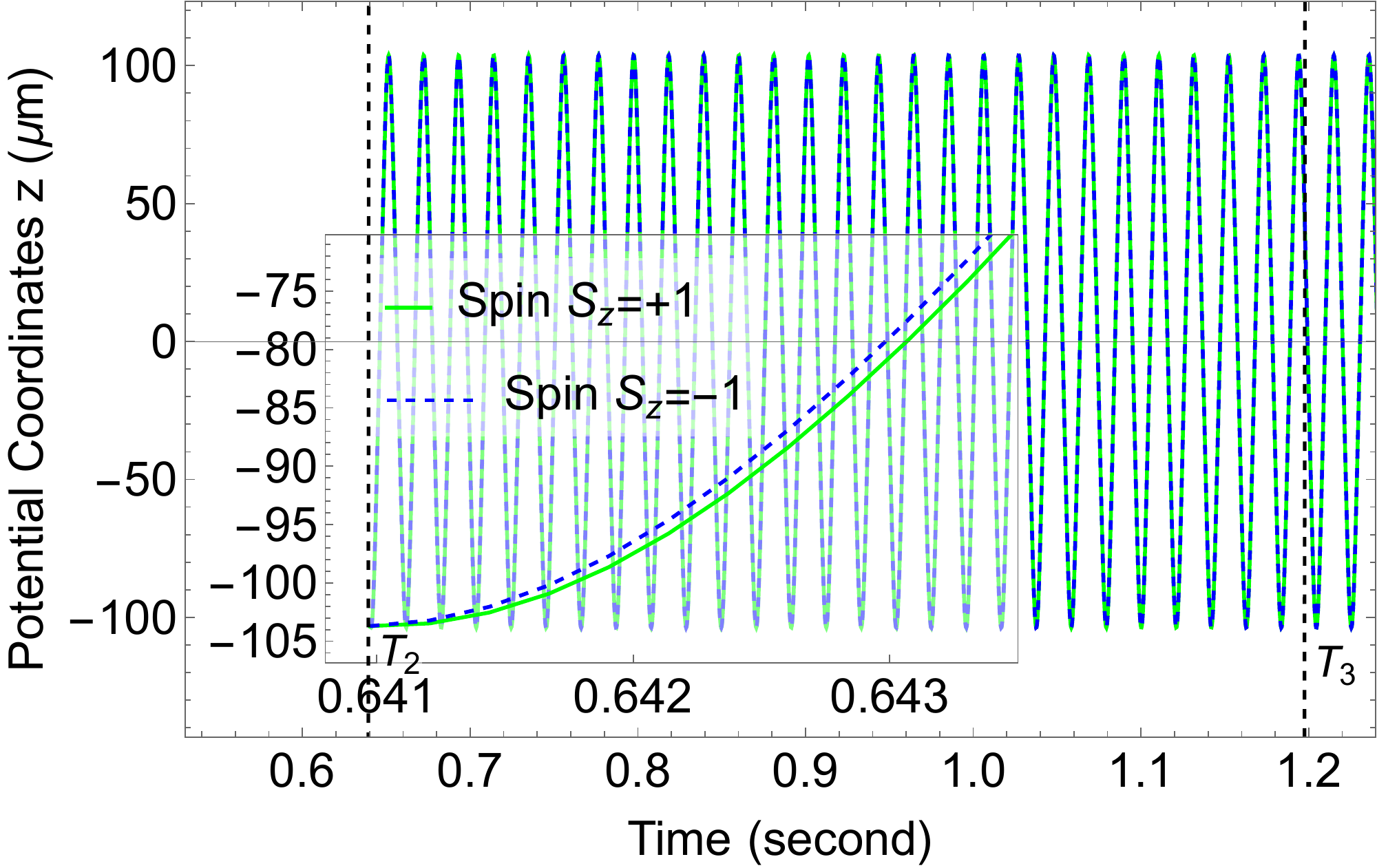}
		\caption{Potential coordinates - stage \Romannum{3}}
	\end{subfigure}\\
    \vspace{0.5cm}
	\centering
	\begin{subfigure}{0.325\linewidth}
		\centering
		\includegraphics[width=0.9\linewidth]{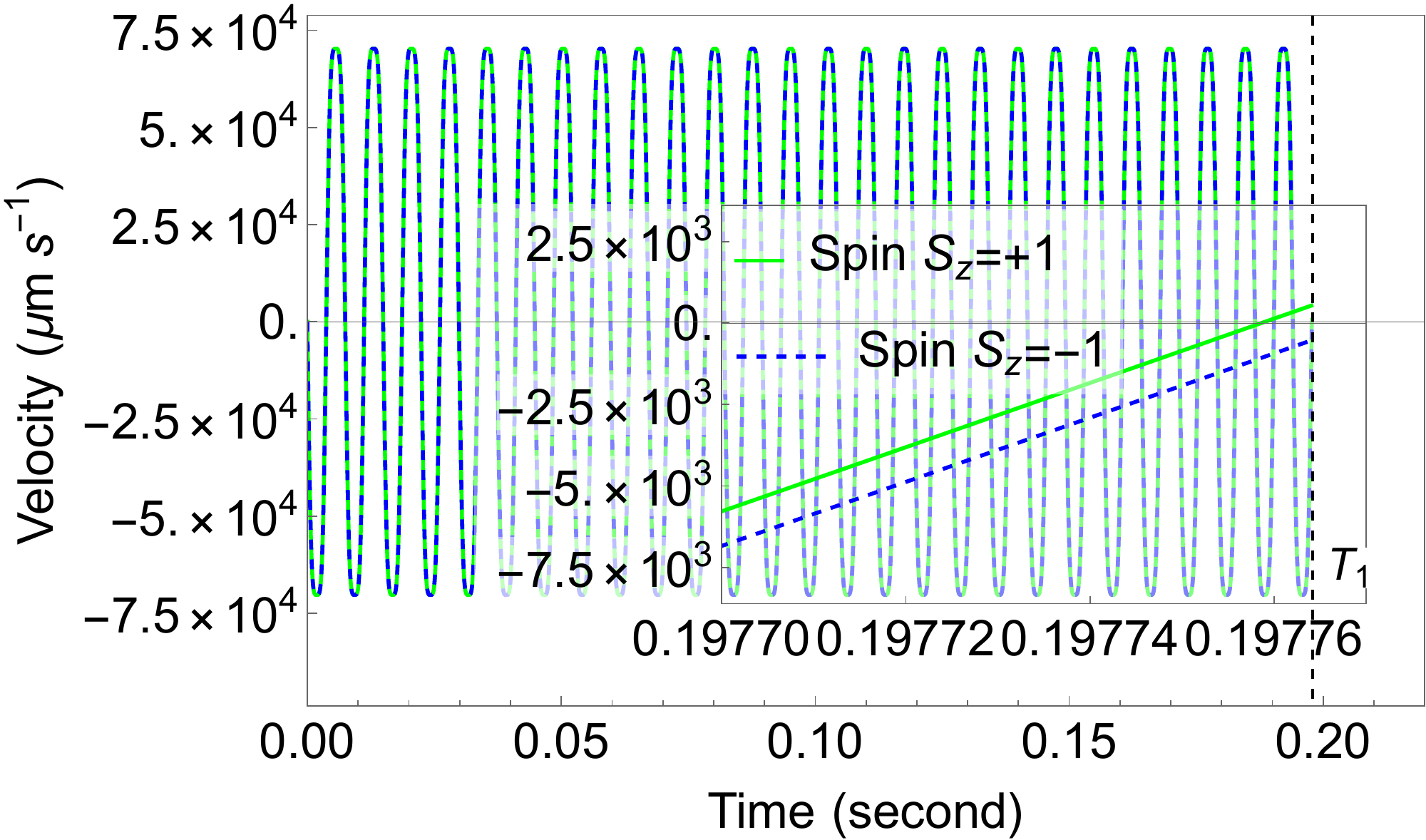}
		\caption{Velocity - stage \Romannum{1}}\label{Velocity-Mass16-1}
	\end{subfigure}
	\centering
	\begin{subfigure}{0.325\linewidth}
		\centering
		\includegraphics[width=0.9\linewidth]{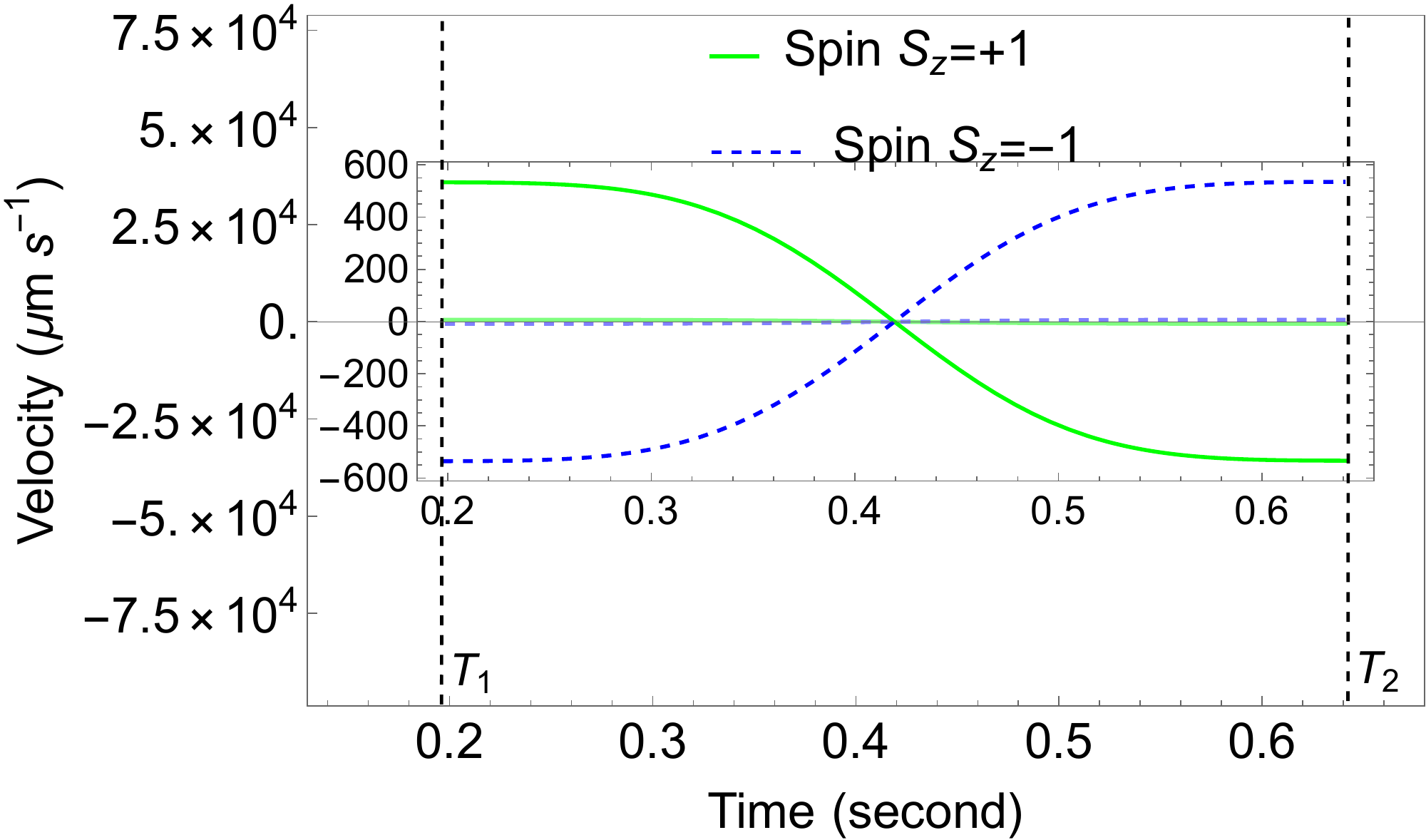}
		\caption{Velocity - stage \Romannum{2}}
	\end{subfigure}
	\centering
	\begin{subfigure}{0.325\linewidth}
		\centering
		\includegraphics[width=0.9\linewidth]{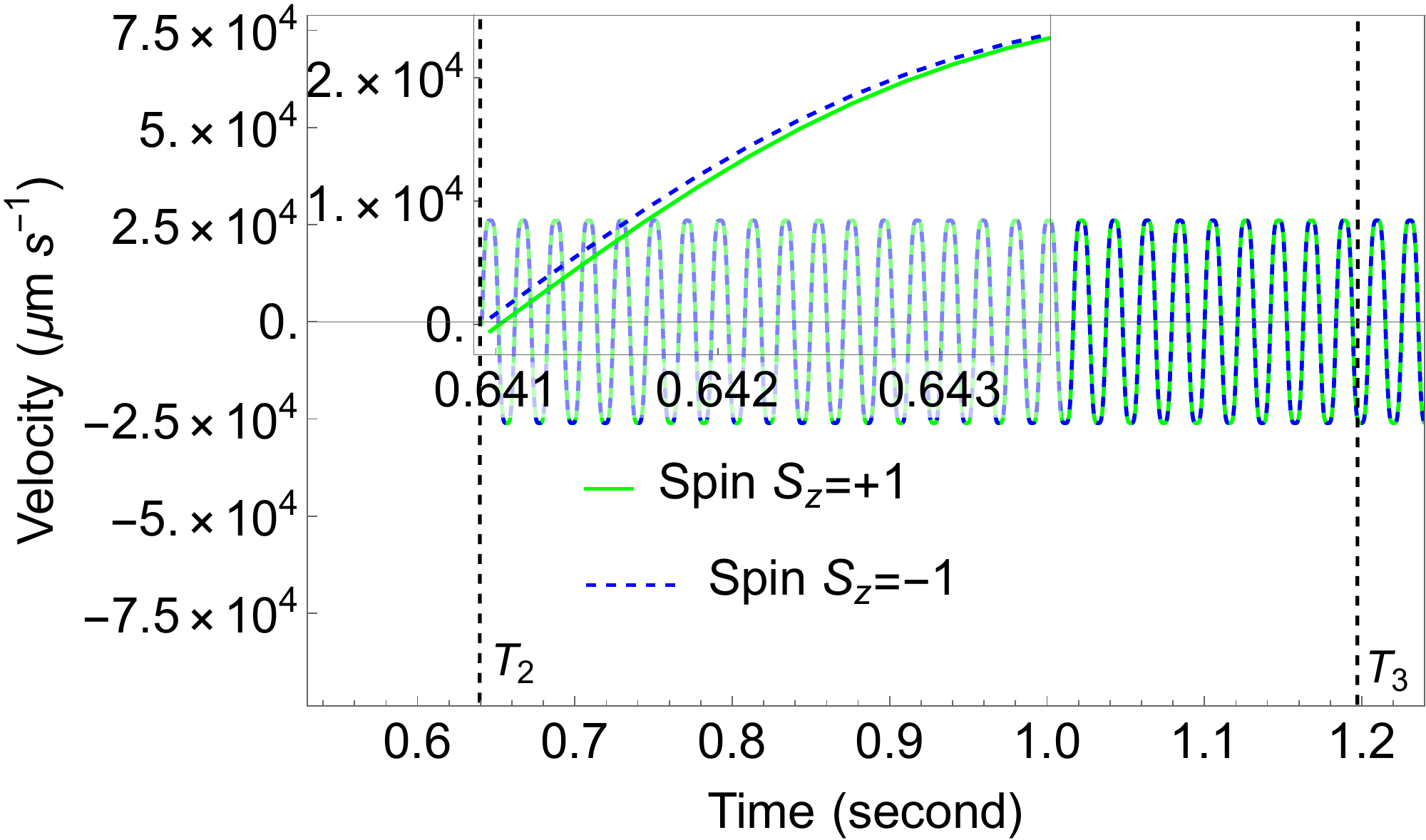}
		\caption{Velocity - stage \Romannum{3}}
	\end{subfigure}\\
    \vspace{0.5cm}
	\centering
	\begin{subfigure}{0.325\linewidth}
		\centering
		\includegraphics[width=0.9\linewidth]{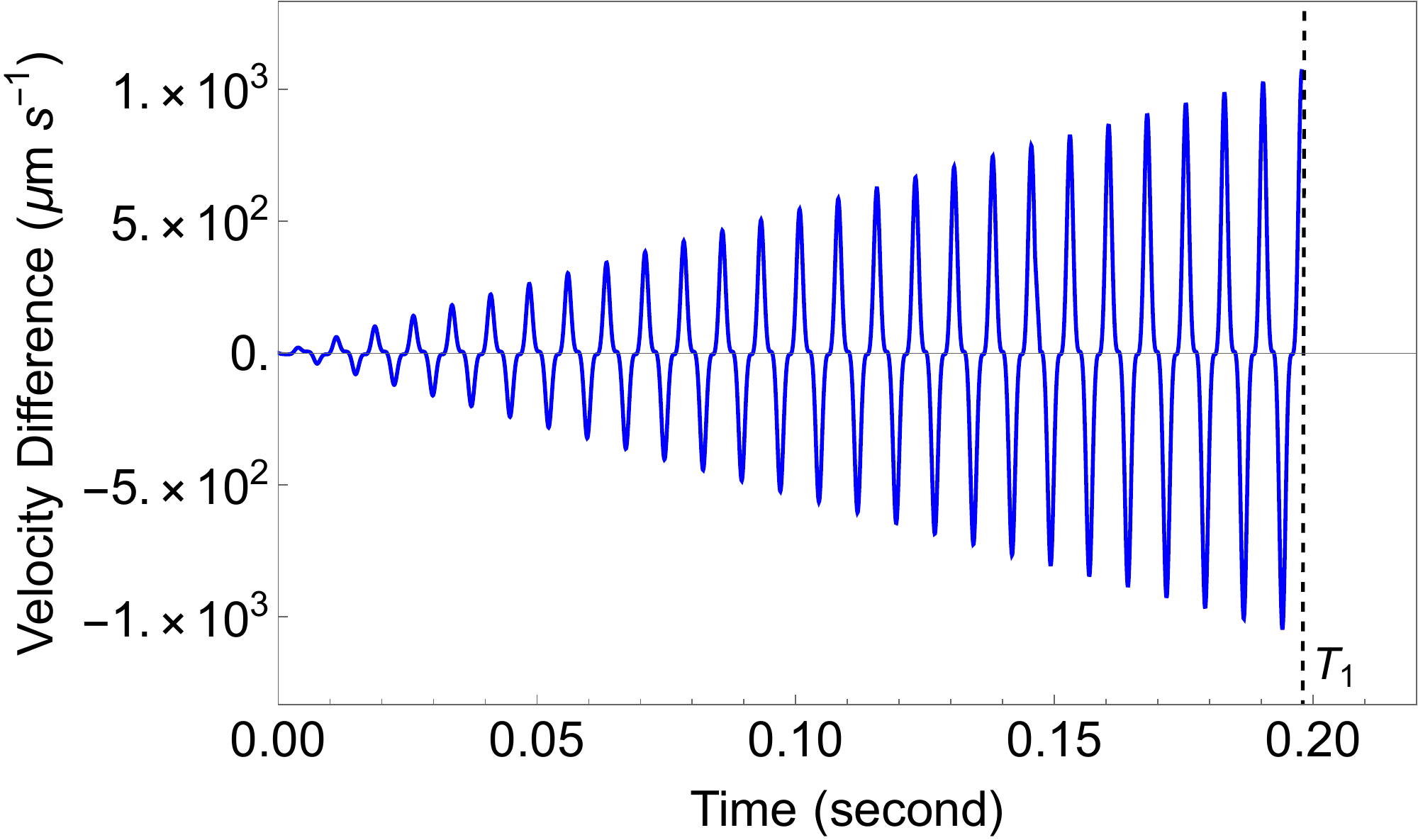}
		\caption{Velocity difference - stage \Romannum{1}}
	\end{subfigure}
	\centering
	\begin{subfigure}{0.325\linewidth}
		\centering
		\includegraphics[width=0.9\linewidth]{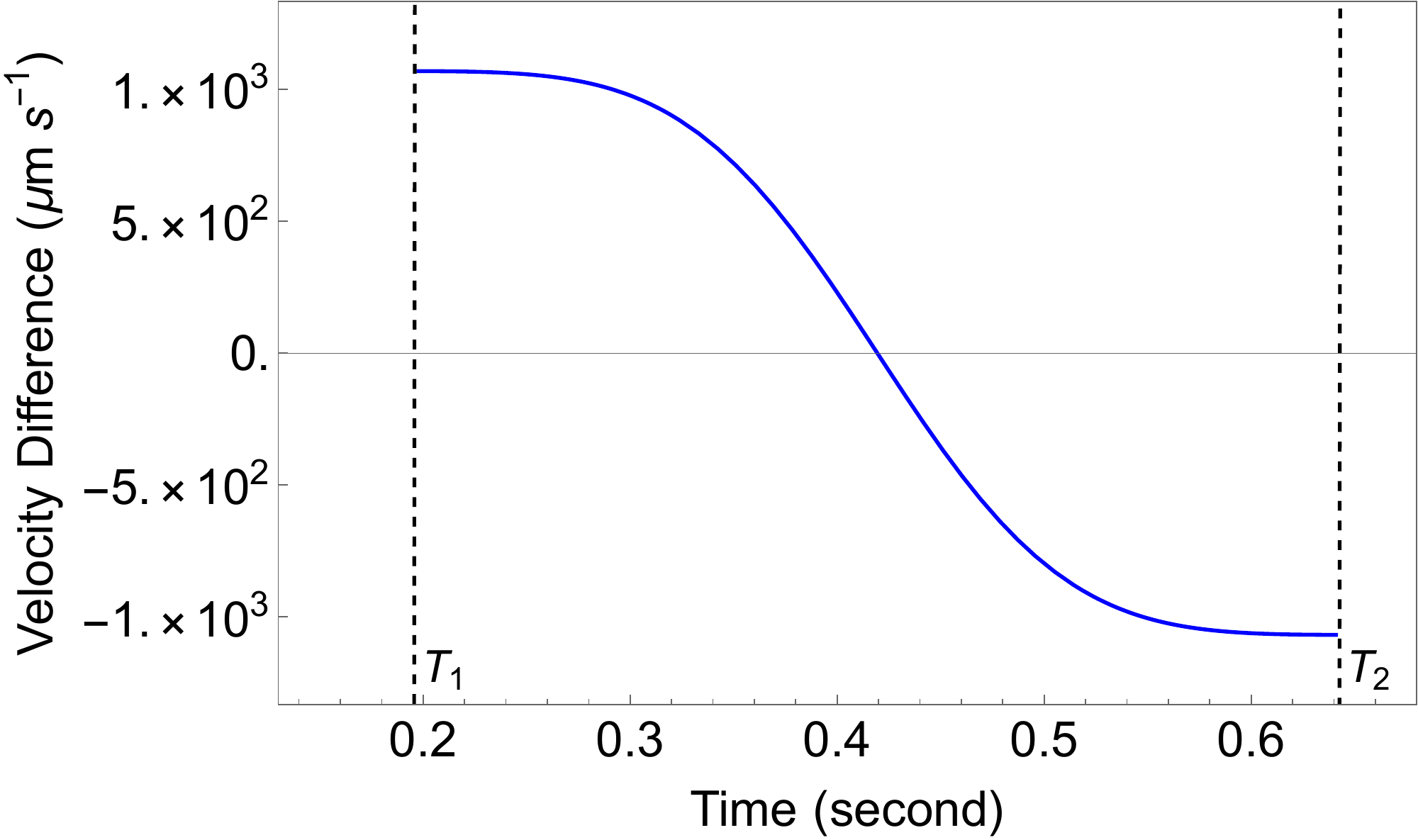}
		\caption{Velocity difference - stage \Romannum{2}}
	\end{subfigure}
	\centering
	\begin{subfigure}{0.325\linewidth}
		\centering
		\includegraphics[width=0.9\linewidth]{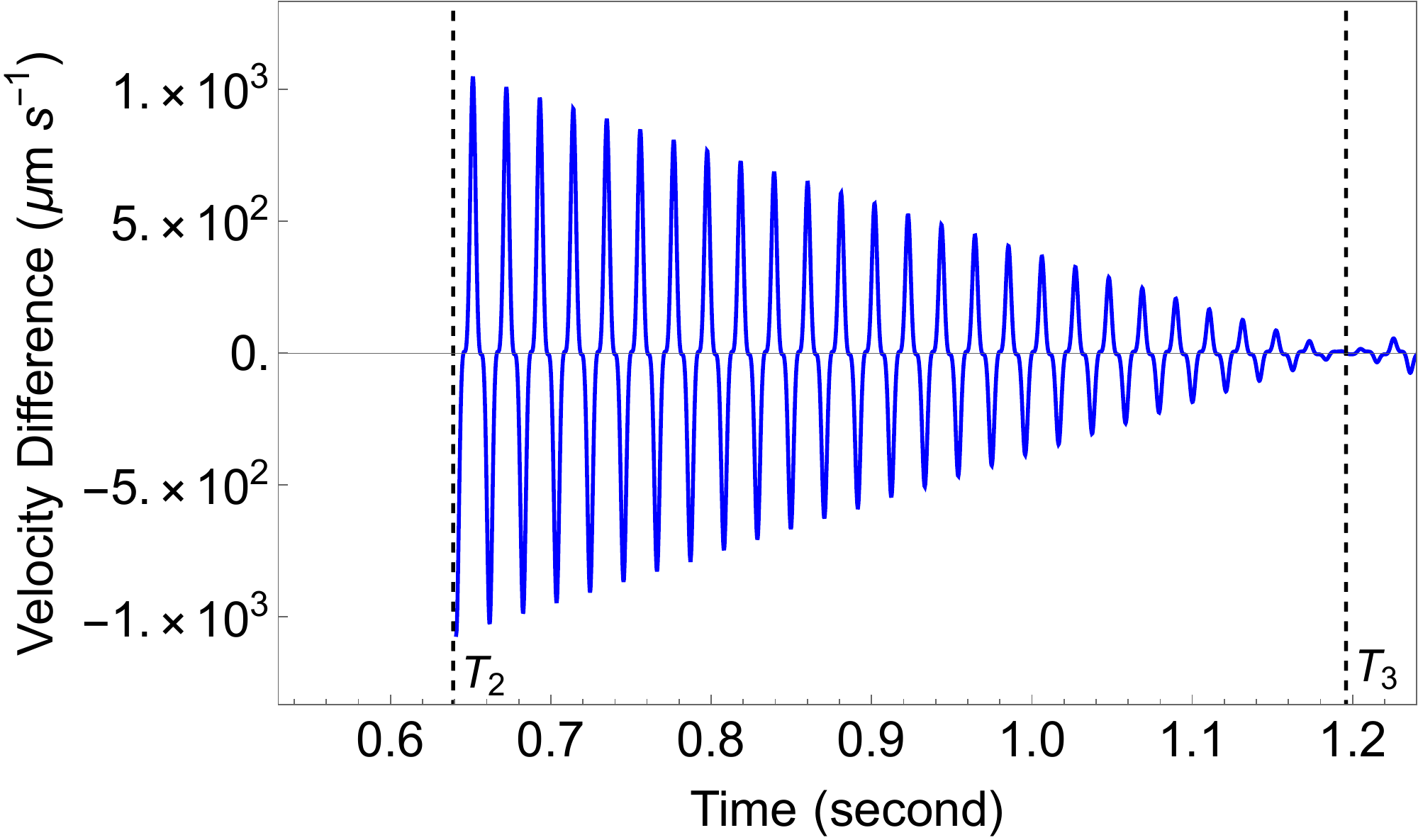}
		\caption{Velocity difference - stage \Romannum{3}}
	\end{subfigure}\\
    \vspace{0.5cm}
	\centering
	\begin{subfigure}{0.325\linewidth}
		\centering
		\includegraphics[width=0.9\linewidth]{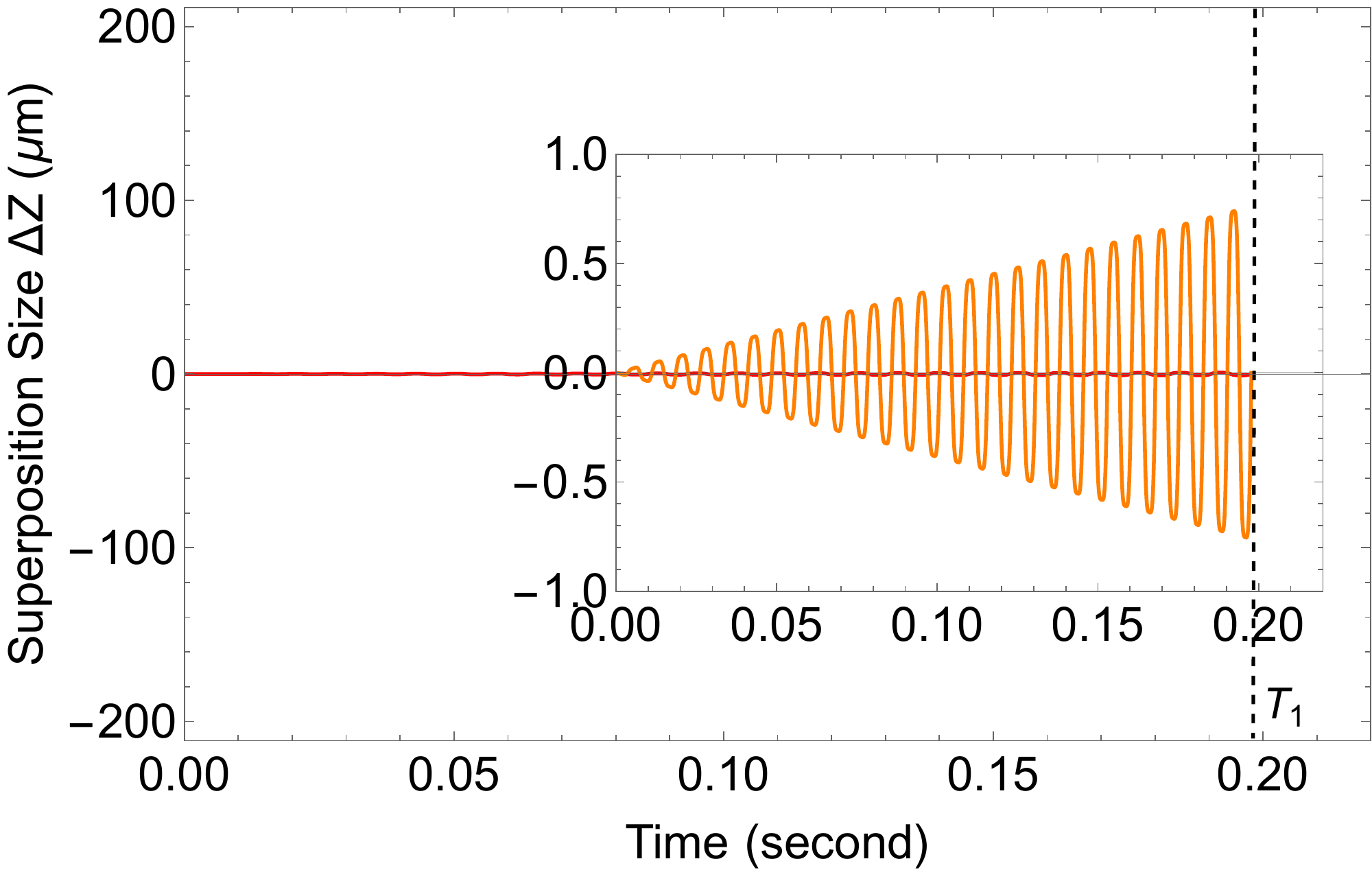}
		\caption{Superposition size - stage \Romannum{1}}
	\end{subfigure}
	\centering
	\begin{subfigure}{0.325\linewidth}
		\centering
		\includegraphics[width=0.9\linewidth]{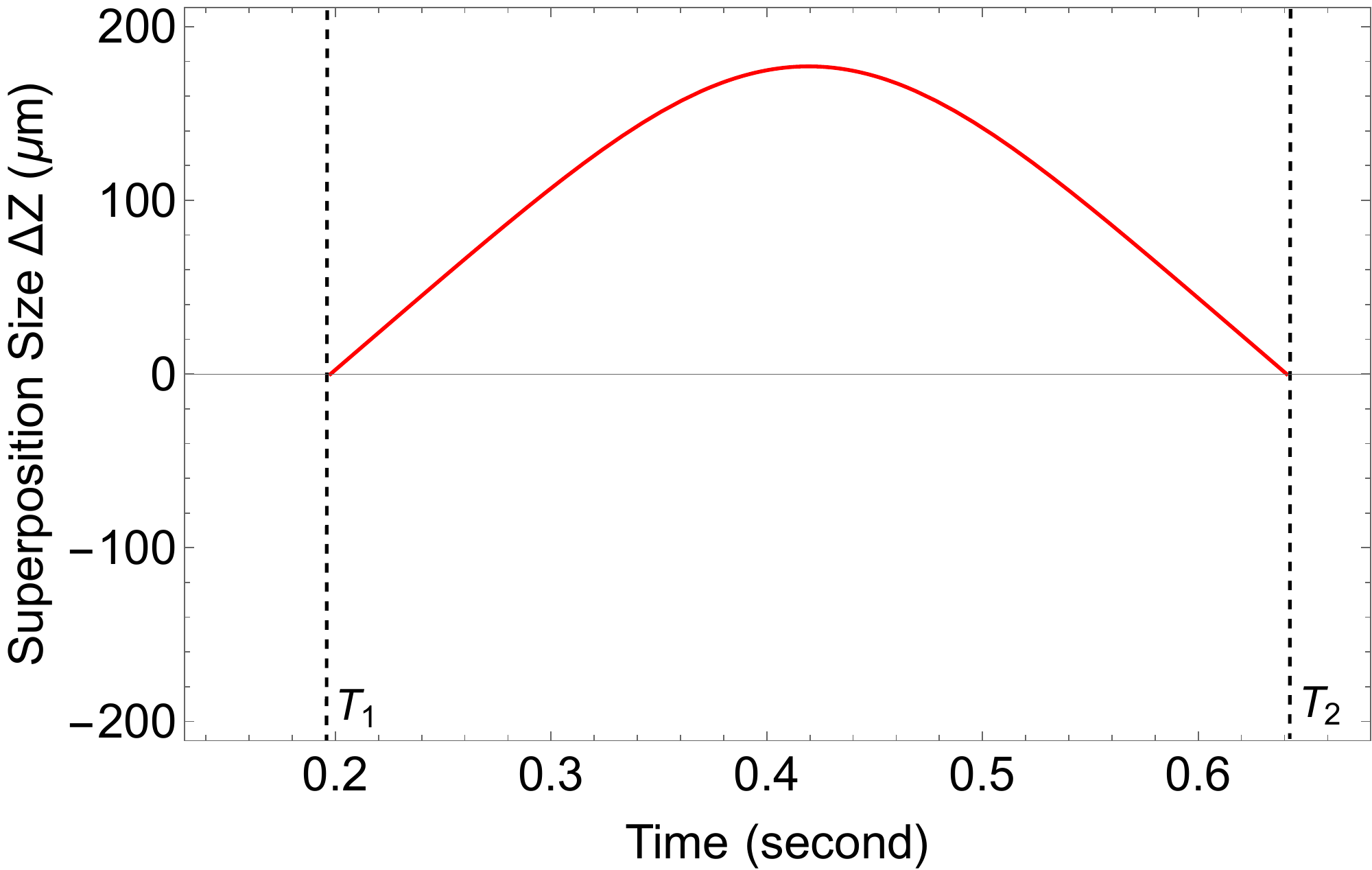}
		\caption{Superposition size - stage \Romannum{2}}\label{SuperpositionSize-Mass16-2}
	\end{subfigure}
	\centering
	\begin{subfigure}{0.325\linewidth}
		\centering
		\includegraphics[width=0.9\linewidth]{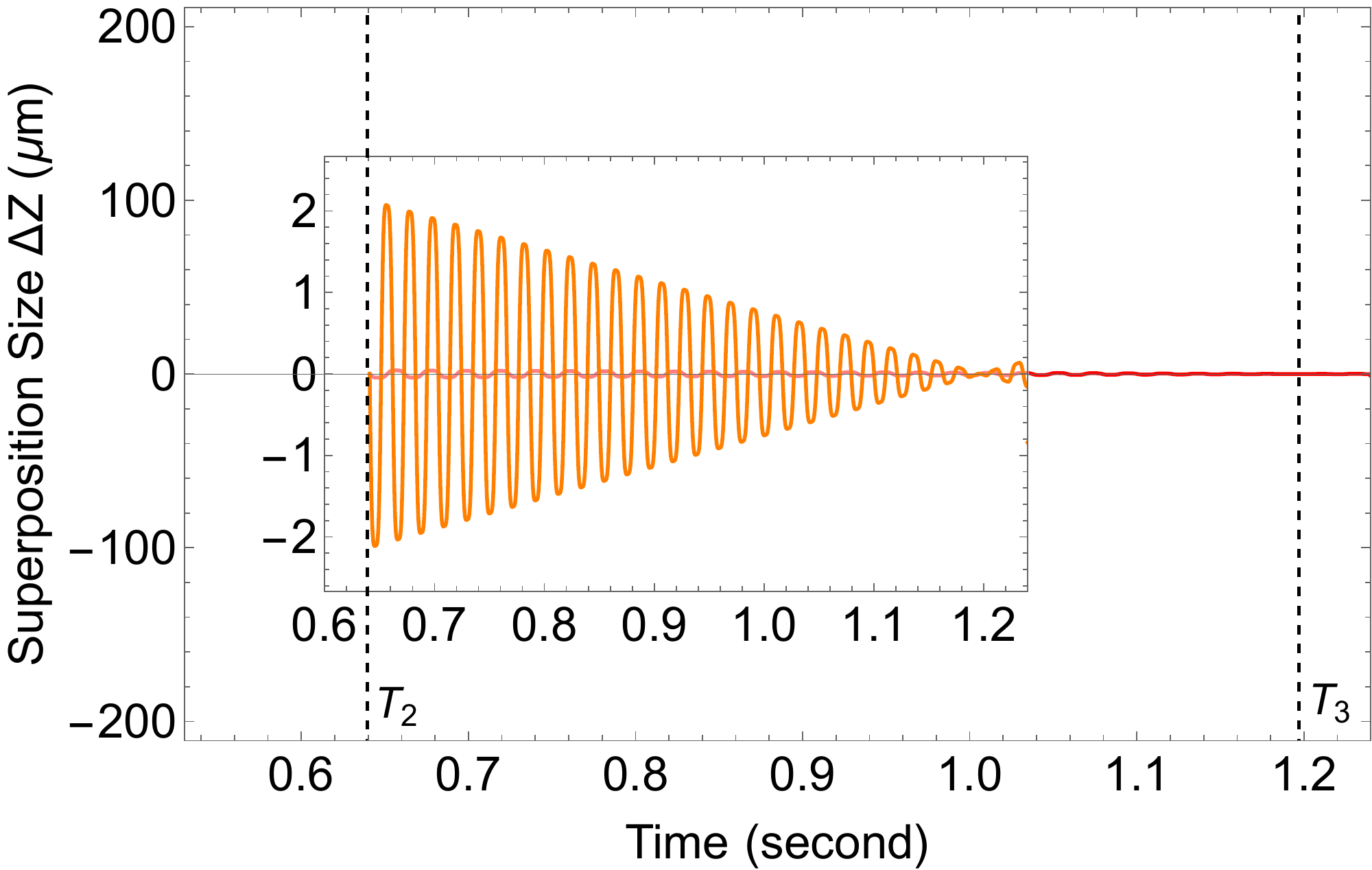}
		\caption{Superposition size - stage \Romannum{3}}\label{SuperpositionSize-Mass16-3}
	\end{subfigure}
	\caption{We show the dynamical aspects for the  mass $m= 10^{-16}$ kg, the magnetic field coordinates (potential coordinates) experienced, the velocities, the velocity differences, and the superposition size during the three experimental stages. We set different values of $\eta$ and the initial position of the wave packet in the magnetic field at different stages. Stage \Romannum{1}, $\eta=1\times 10^{8} ~\text{T}~\text{m}^{-2}$, with initial coordinate $z=100~\mu \text{m}$. Stage \Romannum{2}, $\eta=9\times 10^{5}~\text{T}~\text{m}^{-2}$, with initial coordinate $z=0~\mu \text{m}$. Stage \Romannum{3}, $\eta=3.445\times 10^{7}~\text{T}~\text{m}^{-2}$, with initial coordinate $z=-103.7~\mu \text{m}$. Times $T_{1}$ and $T_{2}$ are determined by constraining the moment when the superposition size is zero (with an accuracy of $z=10^{-6}~\mu \text{m}$). Time $T_{3}$ is the moment when the velocity difference between the two wave packets and the superposition size are zero.}\label{M16-FiveTypesFigures}
\end{figure*}

\begin{figure*}[h]
	\centering
	\begin{subfigure}{0.325\linewidth}
		\centering
		\includegraphics[width=0.9\linewidth]{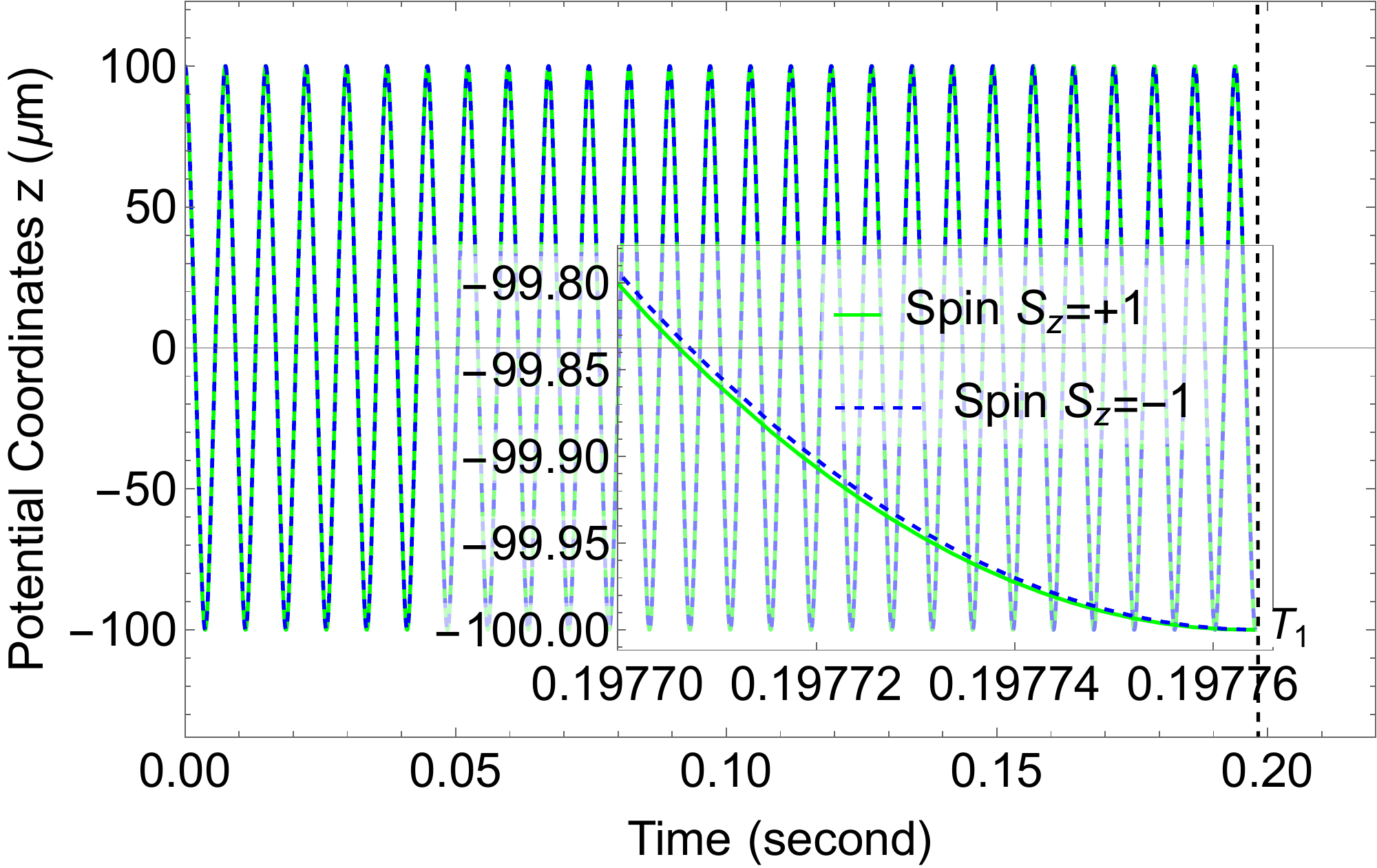}
		\caption{Potential coordinates - stage \Romannum{1}}\label{PotentialCoordinates-Mass15-1}
	\end{subfigure}
	\centering
	\begin{subfigure}{0.325\linewidth}
		\centering
		\includegraphics[width=0.9\linewidth]{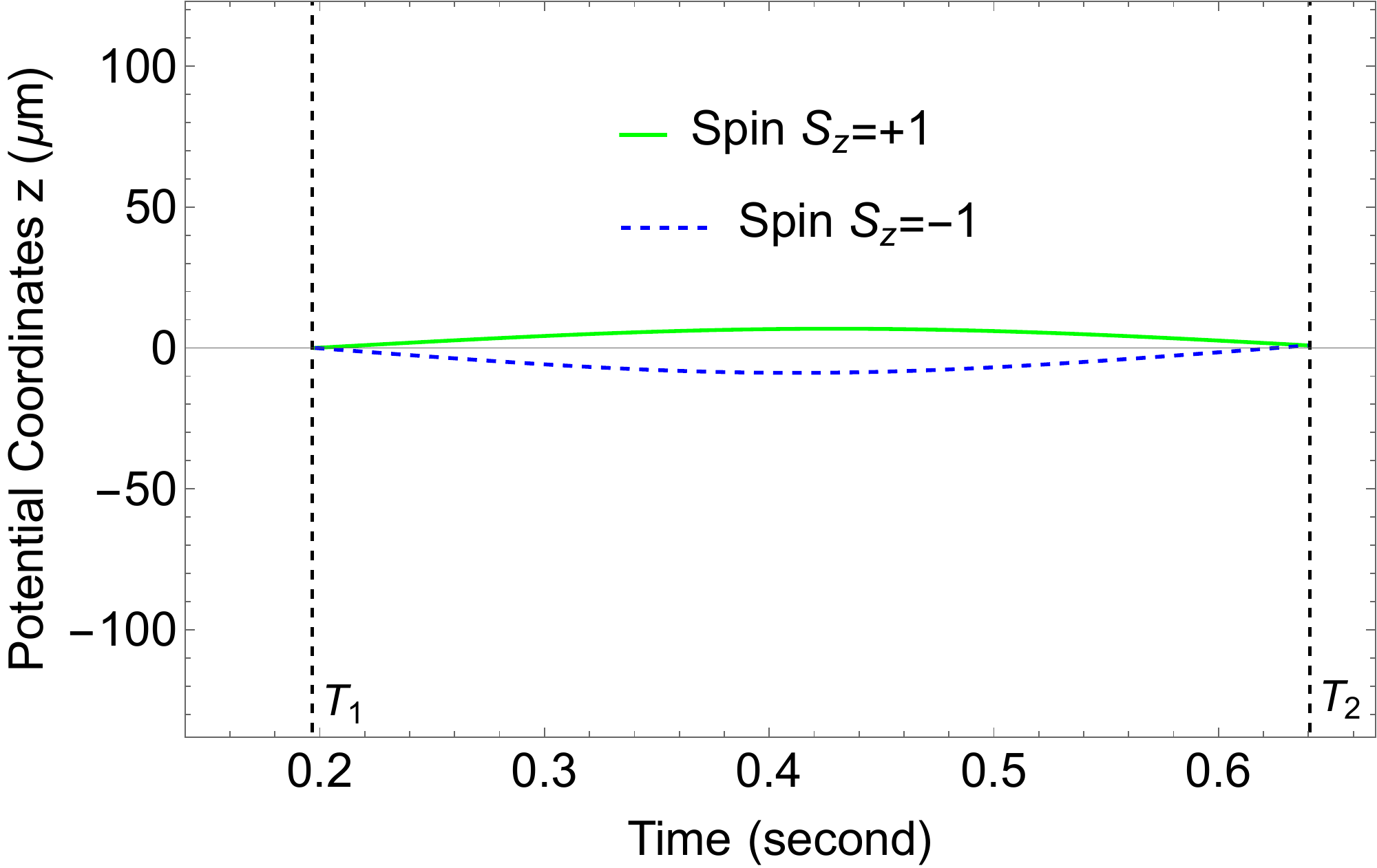}
		\caption{Potential coordinates - stage \Romannum{2}}
	\end{subfigure}
	\centering
	\begin{subfigure}{0.325\linewidth}
		\centering
		\includegraphics[width=0.9\linewidth]{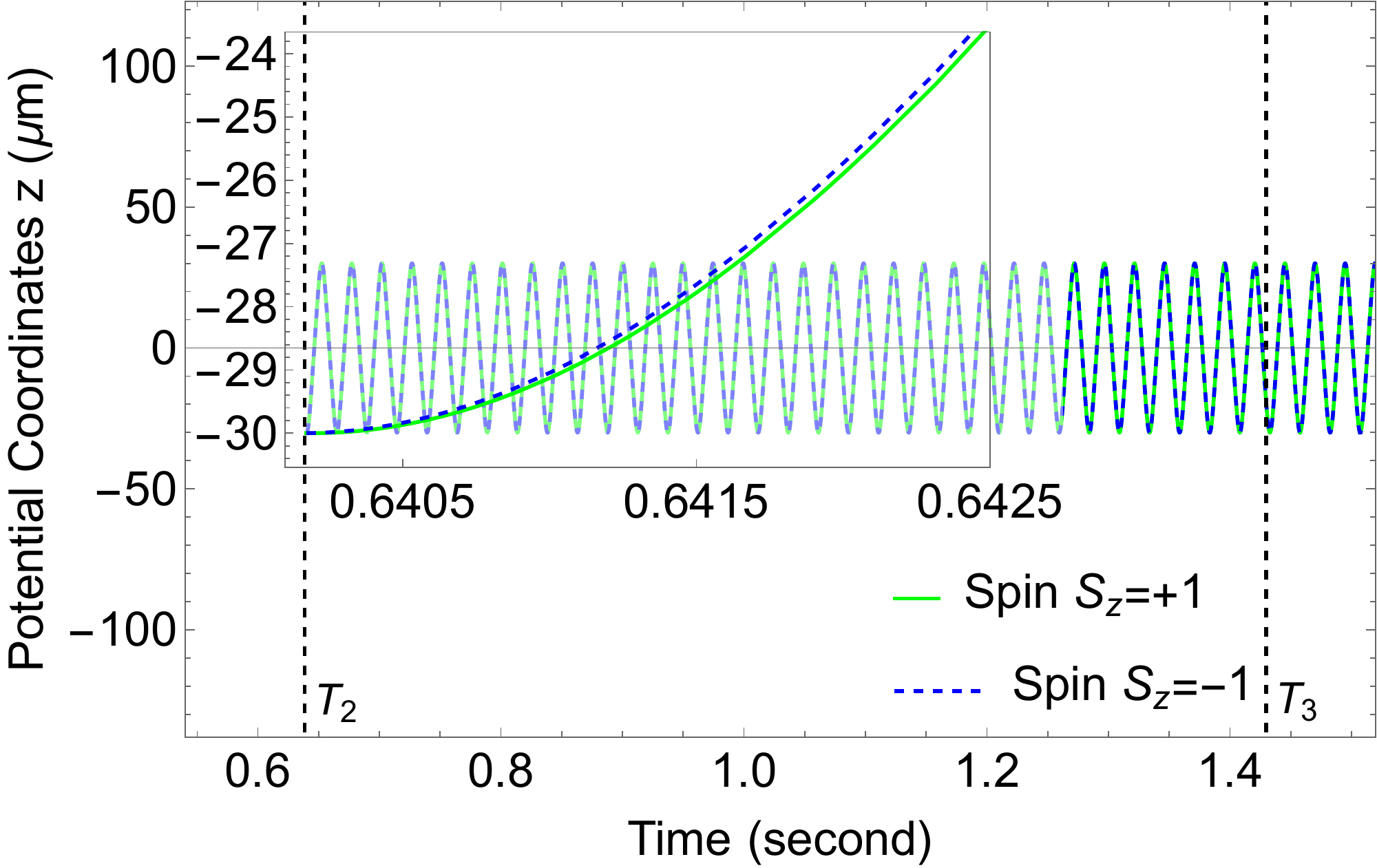}
		\caption{Potential coordinates - stage \Romannum{3}}
	\end{subfigure}\\
	\vspace{0.5cm}
	\centering
	\begin{subfigure}{0.325\linewidth}
		\centering
		\includegraphics[width=0.9\linewidth]{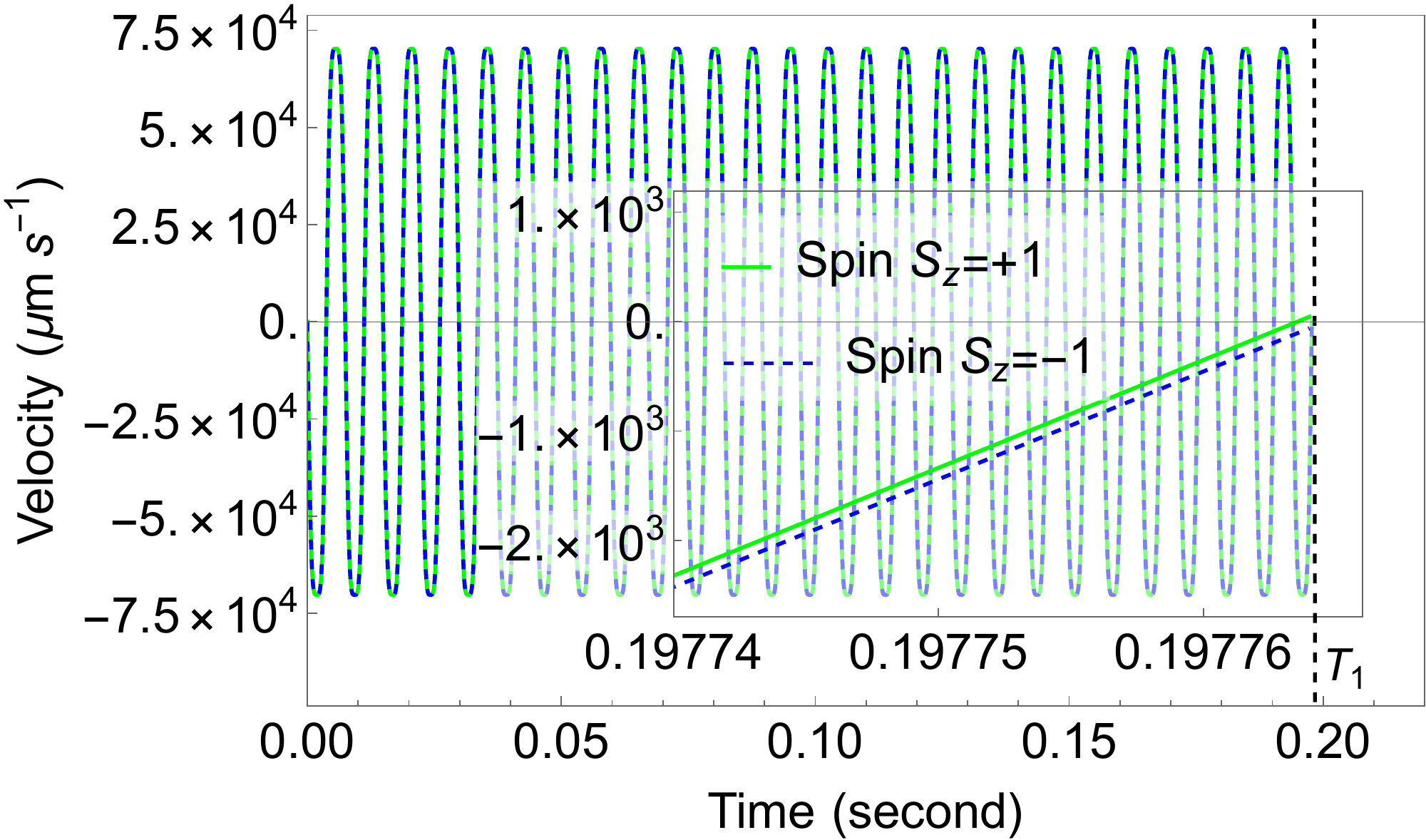}
		\caption{Velocity - stage \Romannum{1}}\label{Velocity-Mass15-1}
	\end{subfigure}
	\centering
	\begin{subfigure}{0.325\linewidth}
		\centering
		\includegraphics[width=0.9\linewidth]{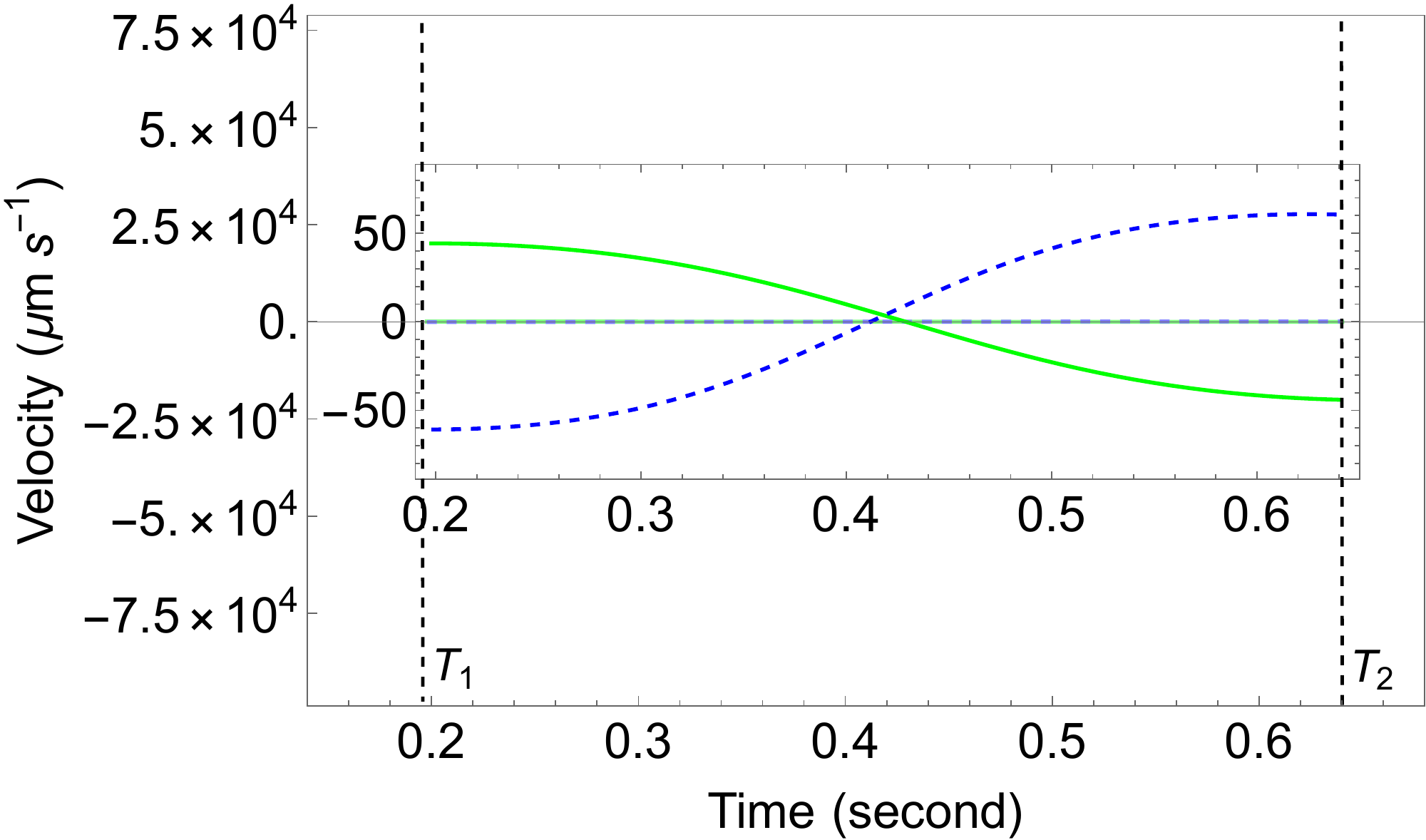}
		\caption{Velocity - stage \Romannum{2}}
	\end{subfigure}
	\centering
	\begin{subfigure}{0.325\linewidth}
		\centering
		\includegraphics[width=0.9\linewidth]{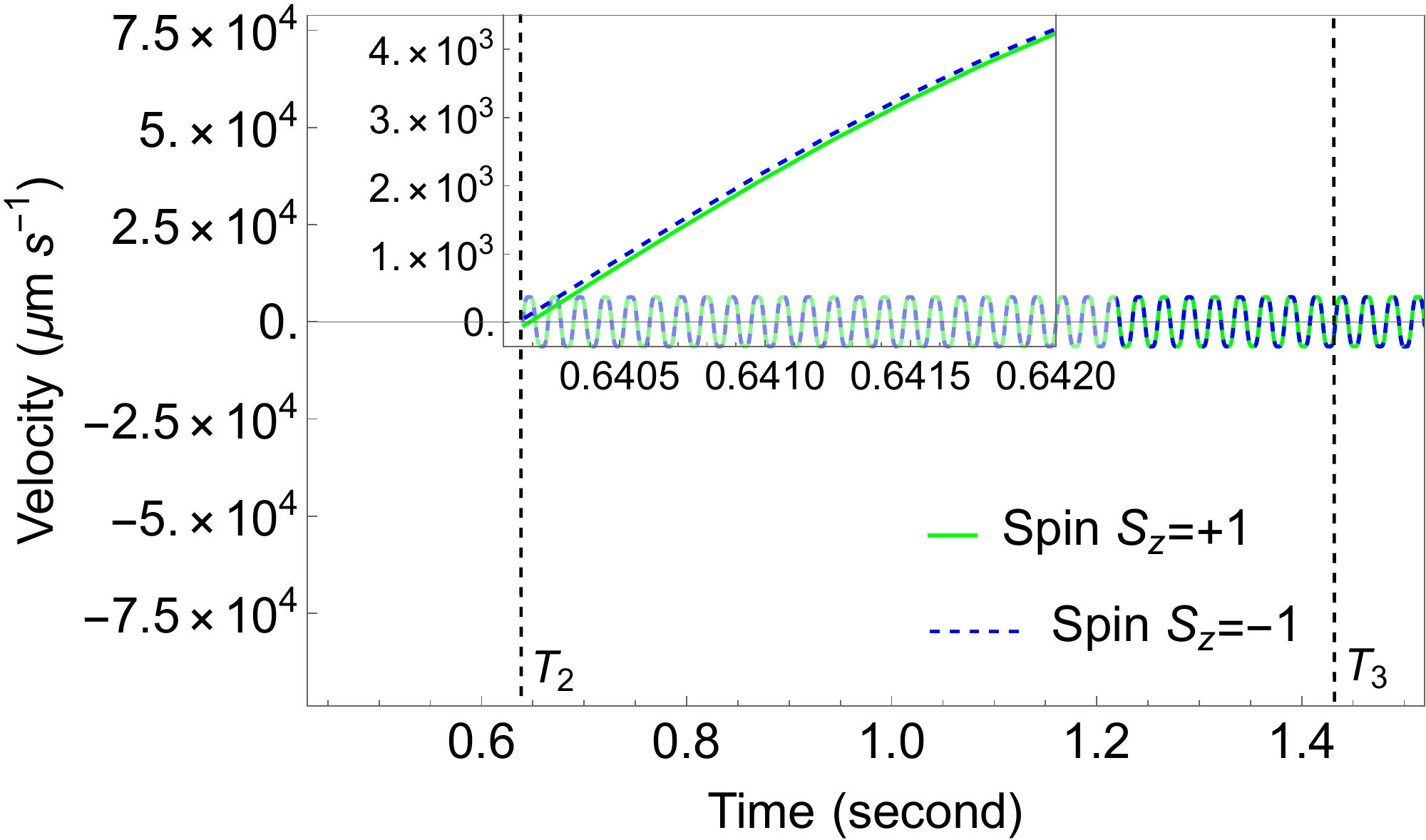}
		\caption{Velocity - stage \Romannum{3}}
	\end{subfigure}\\
	\vspace{0.5cm}
	\centering
	\begin{subfigure}{0.325\linewidth}
		\centering
		\includegraphics[width=0.9\linewidth]{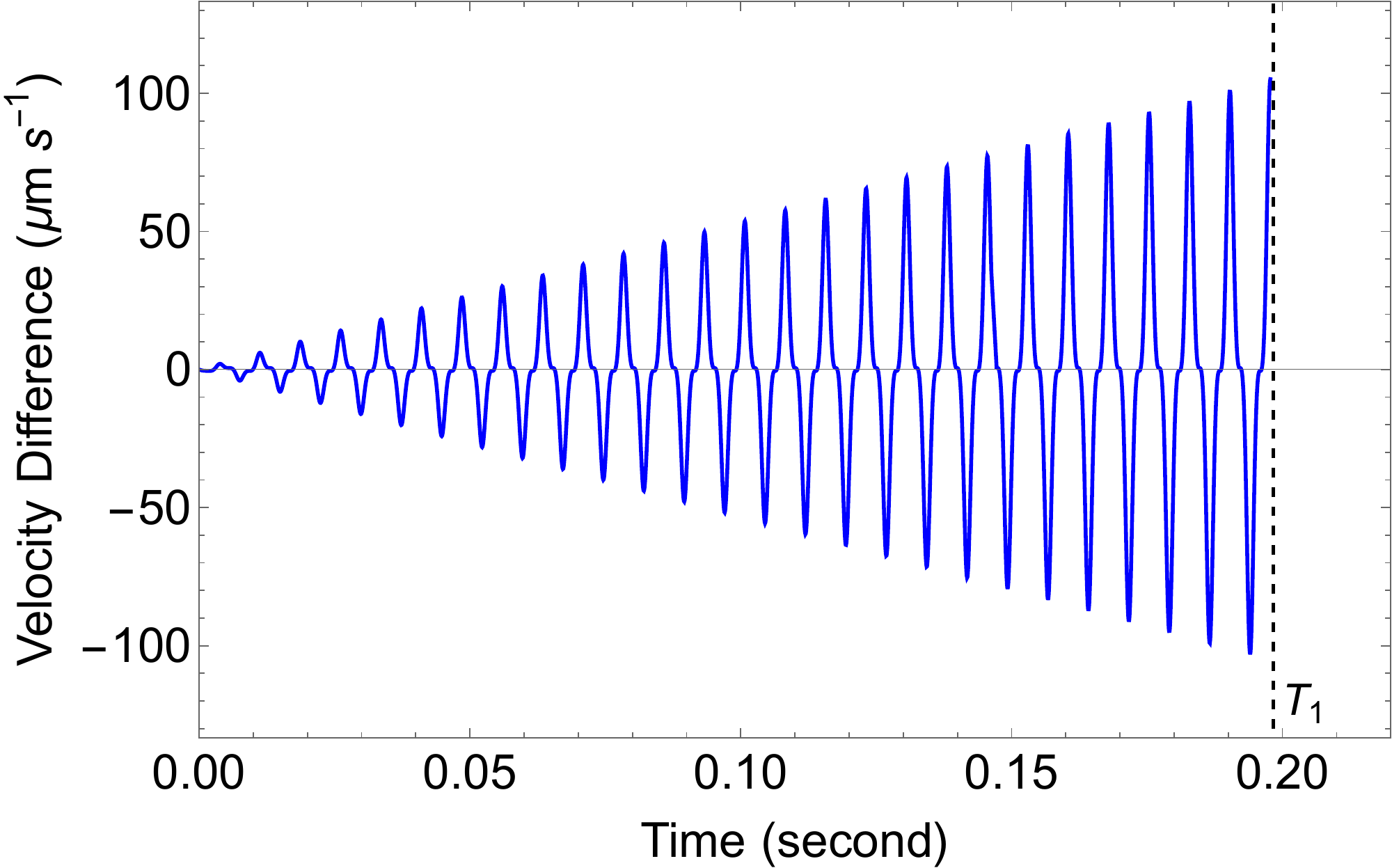}
		\caption{Velocity difference - stage \Romannum{1}}
	\end{subfigure}
	\centering
	\begin{subfigure}{0.325\linewidth}
		\centering
		\includegraphics[width=0.9\linewidth]{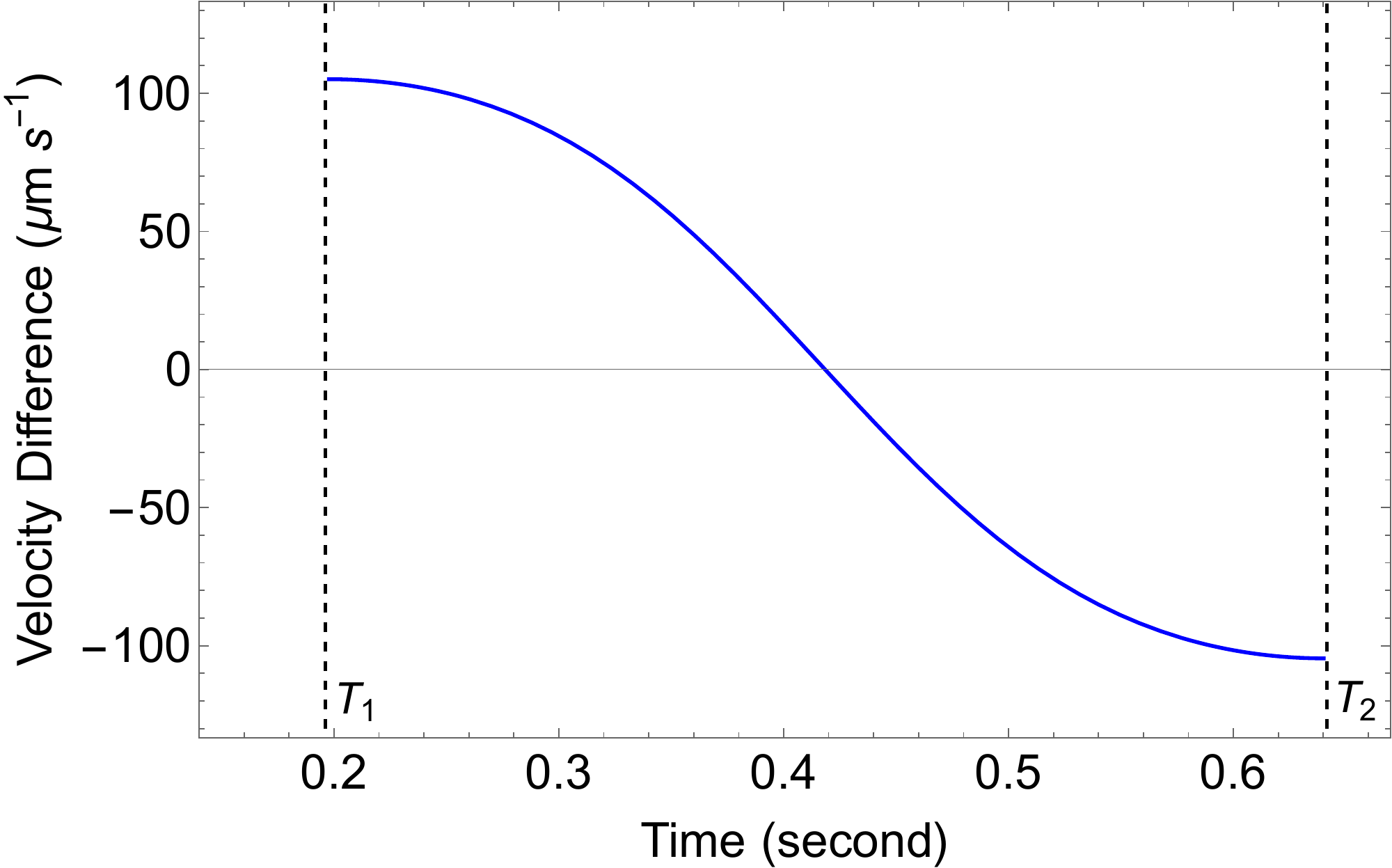}
		\caption{Velocity difference - stage \Romannum{2}}
	\end{subfigure}
	\centering
	\begin{subfigure}{0.325\linewidth}
		\centering
		\includegraphics[width=0.9\linewidth]{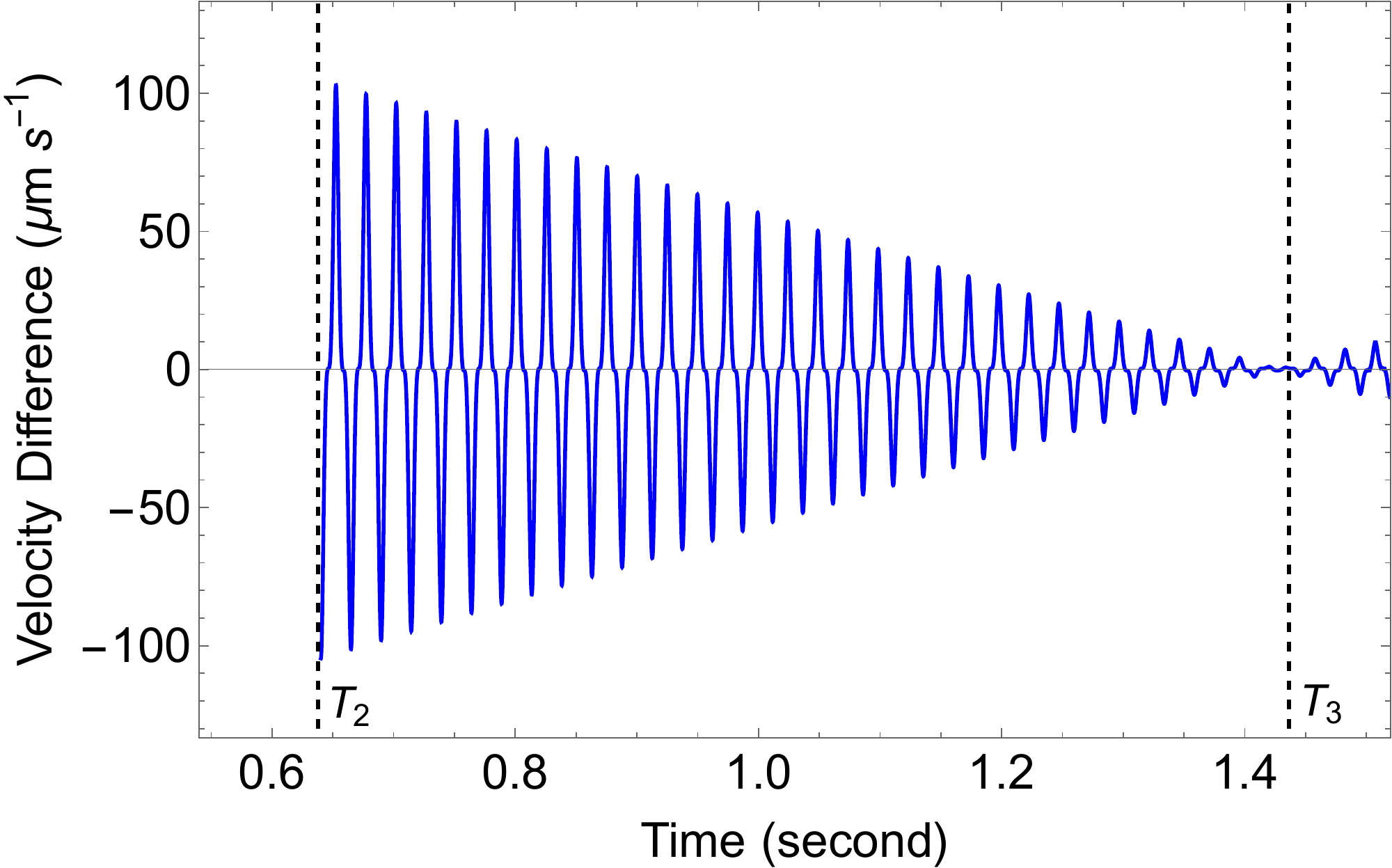}
		\caption{Velocity difference - stage \Romannum{3}}
	\end{subfigure}\\
	\vspace{0.5cm}
	\centering
	\begin{subfigure}{0.325\linewidth}
		\centering
		\includegraphics[width=0.9\linewidth]{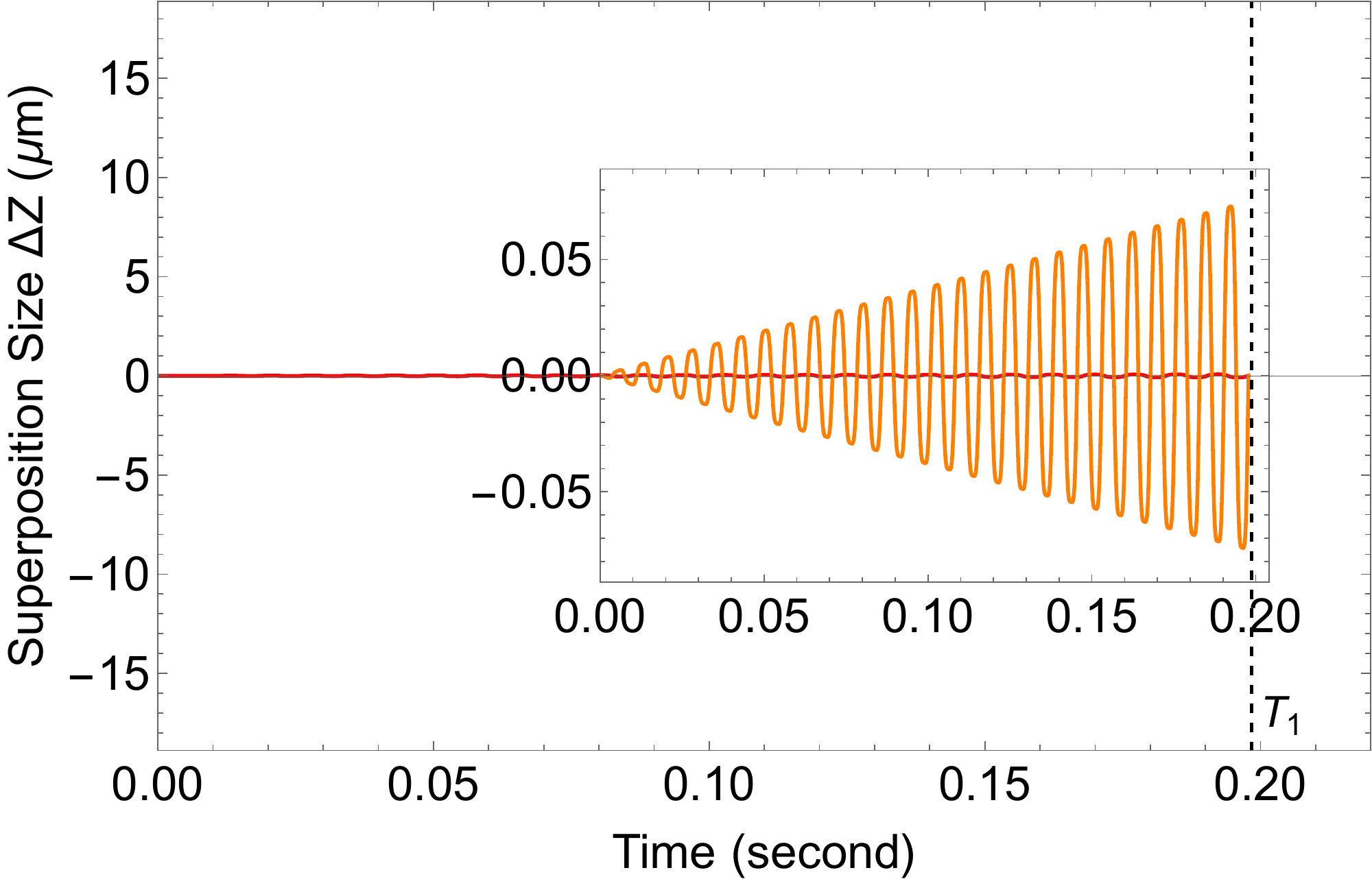}
		\caption{Superposition size - stage \Romannum{1}}
	\end{subfigure}
	\centering
	\begin{subfigure}{0.325\linewidth}
		\centering
		\includegraphics[width=0.9\linewidth]{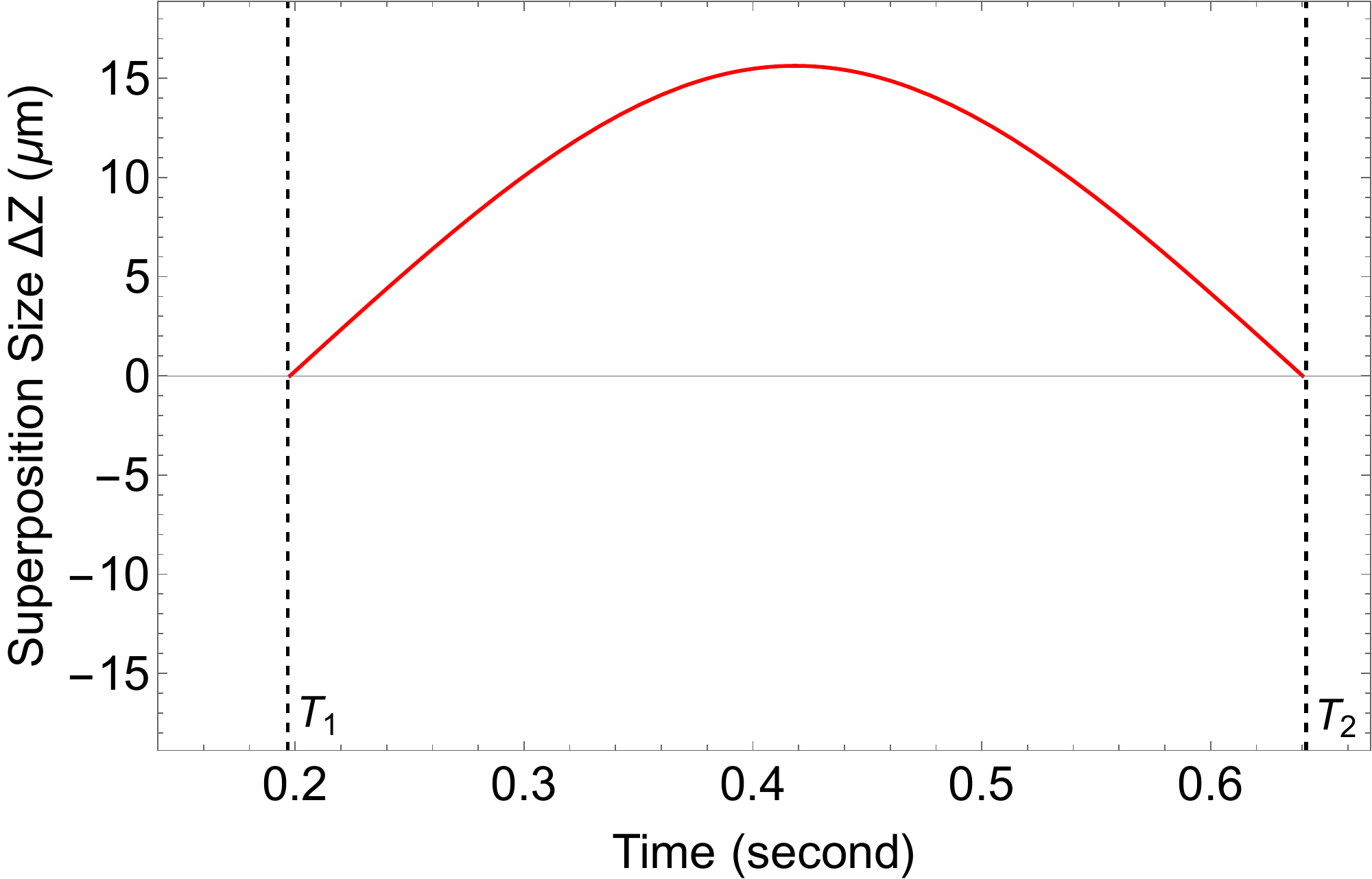}
		\caption{Superposition size - stage \Romannum{2}}\label{SuperpositionSize-Mass15-2}
	\end{subfigure}
	\centering
	\begin{subfigure}{0.325\linewidth}
		\centering
		\includegraphics[width=0.9\linewidth]{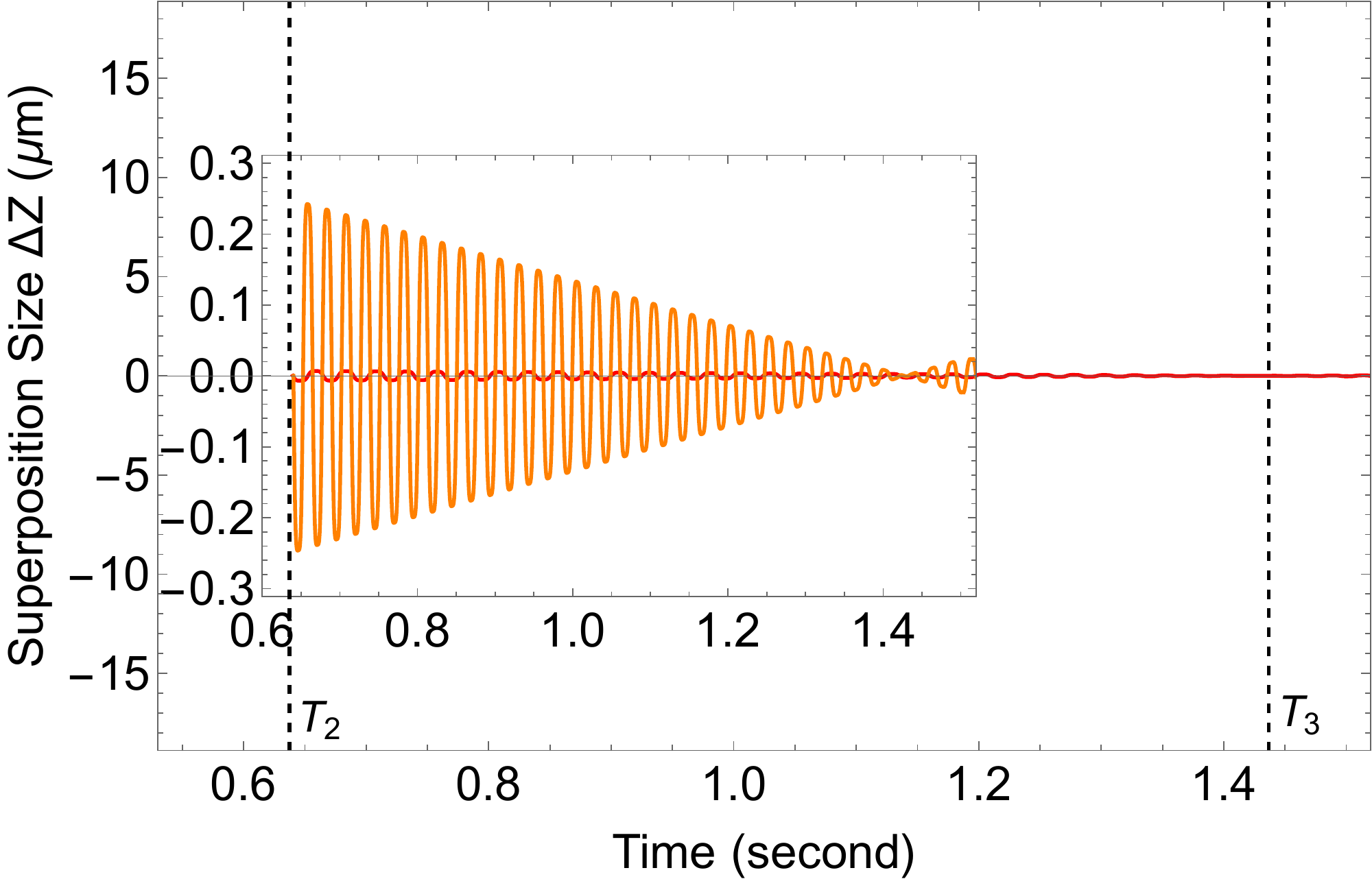}
		\caption{Superposition size - stage \Romannum{3}}\label{SuperpositionSize-Mass15-3}
	\end{subfigure}
	\caption{We show the dynamics for the mass $m=10^{-15}$~kg, the magnetic field coordinates (potential coordinates) experienced, the velocities, the velocity differences, and the superposition size during the three experimental stages. We set different values of $\eta$ and the initial position of the wave packet in the magnetic field at different stages. Stage \Romannum{1}, $\eta=1\times 10^{8} ~\text{T}~\text{m}^{-2}$, corresponds to  the coordinate $z=100~\mu \text{m}$. Stage \Romannum{2}, $\eta=6\times 10^{6} ~\text{T}~\text{m}^{-2}$, corresponding to the initial coordinate $z=0 ~\mu \text{m}$. Stage \Romannum{3}, $\eta=1\times 10^{8} ~\text{T}~\text{m}^{-2}$, with an initial coordinate $z=-30~\mu \text{m}$. Times $T_{1}$ and $T_{2}$ are determined by constraining the moment when the superposition size is zero (with an accuracy of $z=10^{-6}~\mu \text{m}$). Time $T_{3}$ is the moment when the velocity difference between two wave packets and the superposition size are zero.}\label{M15-FiveTypesFigures}
\end{figure*}

The time evolution of the nano-crystals for different masses in a non-linear magnetic field shows a very similar pattern as that of $m=10^{-17}$kg. The main difference between these evolutions is that the maximum velocity difference between the wave packets with the opposite spin orientations is inversely proportional to the mass for the same magnetic field gradient parameter $\eta$ (Fig.(\ref{ScalingBehavior1})). The difference in velocity between the wave packets then determines the superposition size that we can obtain in the same time span. The magnetic field used to control the motion of the wave packets in the second and third stages will change accordingly for different masses and the difference in velocity between the wave packets at the end of the first stage. We have shown the numerical results of specific parameters  and the evolution of the nano-crystals  in Fig.\ref{M16-FiveTypesFigures} and Fig.\ref{M15-FiveTypesFigures}. An important point to note is that for the heaviest mass $m=10^{-15}$kg we can obtain the spatial superposition size of $15$ micron, which is the required valued we require for testing the quantum nature of gravity in a lab by including the Casimir screening, see~\cite{vandeKamp:2020rqh}. The simple scaling of the superposition size $\Delta Z_{0}$ is given by Eq.~(\ref{amplitude}).

\end{appendices}

\end{document}